\documentclass[a4paper,reqno]{amsart}
\usepackage[english]{babel}
\usepackage{amssymb,graphicx}
\usepackage{xcolor}
\usepackage{hyperref}

\numberwithin{equation}{section}

\newcommand{\assign}{:=}

\newcommand{\exterior}{\wedge}
\newcommand{\infixand}{\text{ and }}
\newcommand{\infixor}{\text{ or }}
\newcommand{\nobracket}{}
\newcommand{\nospace}{}
\newcommand{\tmem}[1]{{\em #1\/}}
\newcommand{\tmop}[1]{\ensuremath{\operatorname{#1}}}

\newcommand{\tmtextit}[1]{\text{{\itshape{#1}}}}
\newtheorem{theorem}{Theorem}[section]
\newtheorem{corollary}[theorem]{Corollary}
\newtheorem{lemma}[theorem]{Lemma}
{\theoremstyle{remark}\newtheorem{remark}[theorem]{Remark}}

\usepackage{a4}
\usepackage{enumitem}
\allowdisplaybreaks[4]
\begin{document}

\title{Stochastic quantization of $\lambda \phi_2^4$-theory in 2-d Moyal
space}

\author{Chunqiu Song}
\address{Chunqiu Song: Institut f{\"u}r Analysis und Numerik, Universit{\"a}t
M{\"u}nester}
\email{chunqiu.song@uni-muenster.de}

\author{Hendrik Weber}
\address{Hendrik Weber: Institut f{\"u}r Analysis und Numerik, Universit{\"a}t
M{\"u}nster}
\email{hendrik.weber@uni-muenster.de}

\author{Raimar Wulkenhaar}
\address{Raimar Wulkenhaar: Mathematisches Institut, Universit{\"a}t M{\"u}nster}
\email{raimar@math.uni-muenster.de}

\keywords{stochastic quantization, non-commutative quantum field theory, Moyal product}

\subjclass[2010]{ 60H15, 35B45, 81T08, 81T75}

\begin{abstract}
  There is strong evidence for the conjecture that the $\lambda \phi^4$ QFT-model 
  on 4-dimensional non-commutative Moyal space can be non-perturbatively
  constructed. As preparation, in this paper we construct the 2-dimensional
  case with the method of stochastic quantization. We show the local
  well-posedness and global well-posedness of the stochastic quantization
  equation, leading to a construction of the Moyal $\lambda \phi^4_2$ measure
  for any non-negative coupling constant $\lambda$.
\end{abstract}

{\maketitle}

{\tableofcontents}

\section{Introduction}

The quantum field theoretical model that we study in this paper appeared at
the end of the last century in string theory with D-branes. In presence of a
magnetic field on the branes, the field theory limit of string theory has an
effective description in terms of a non-commutative $\star$-product
{\cite{Schomerus:1999ug,Seiberg:1999vs}}. The perturbative expansion of field
theories with $\star$-product is organized by ribbon graphs, which are
analogues of Feynman graphs and can be planar or non-planar. Planar graphs
show the usual divergences (related to products of distributions)
{\cite{Filk:1996dm}} of QFT. Non-planar graphs are superficially finite but
get a large amplitude near exceptional momenta, which produces intractable
problems when inserted as subgraphs (UV/IR-mixing, {\cite{Minwalla:1999px}}).

An investigation {\cite{Grosse:2003aj}} of the renormalization group flow in
the $\lambda \phi^4$-model on non-commutative Moyal space led one of us with
H.~Grosse to the identification of another marginal coupling in this model:
the frequency $\Omega$ of an harmonic oscillator potential. The resulting
action functional\footnote{In the literature this is sometimes called
Grosse--Wulkenhaar model.} reads
\begin{equation}
  S [\phi] = \frac{1}{(8\pi)^{d/2}}
  \int_{\mathbb{R}^d}  \left( \frac{Z}{2} \phi (x) \left( - \Delta
  + M^2 + 4 \Omega^2 \| \Theta^{- 1} x\|^2 \right) \phi (x) +
  \frac{Z^2 \lambda}{4} \phi^{\star 4} (x)  \right) d \nospace x.
  \label{GW-action}
\end{equation}
Here $\phi^{\star 4} (x) \assign (\phi \star \phi \star \phi \star \phi) (x)$
and $\star$ denotes the Moyal product on $\mathbb{R}^d$ ($d$ even), which
involves a skew-symmetric constant $d \times d$-matrix $\Theta$. We give more
details in Section \ref{1section}. In the renormalization group (RG) spirit
due to Wilson {\cite{Wilson:1973jj}}, the fields $\phi$ decompose into modes
depending on a scale $\Lambda$, and also the parameters $Z, M^2, \lambda,
\Omega$ depend on $\Lambda$. The RG flow of effective actions in $\Lambda$ has
been analyzed for $d = 2$ in {\cite{Grosse:2003nw}} (where $Z = 1$ and
$\lambda$ is constant) and $d = 4$ in {\cite{Grosse:2004yu}} and shown to be
consistent as formal power series in $\lambda$ (for $d = 2$) and $\lambda
(\Lambda_R)$ (for $d = 4$). The result has been reconfirmed in several other
renormalization schemes; we refer to {\cite{Rivasseau:2007qda}} for a review.
For $d = 2$ there are paths $\Omega (\Lambda)$ along which this frequency can
be removed for $\Lambda \to \infty$. This is not possible in $d = 4$ (in
agreement with UV/IR-mixing); here $\lim_{\Lambda \to \infty} \Omega (\Lambda)
= 1$, and the ratio $\frac{\lambda (\Lambda)}{\Omega^2 (\Lambda)}$ is
RG-constant up to $\mathcal{O} (\lambda)$ {\cite{Grosse:2004by}}. Therefore,
and in sharp distinction to the usual $\lambda \phi^4_4$-model where $\lambda
(\Lambda)$ develops a Landau pole at finite $\Lambda_0$ (reflecting marginal
triviality {\cite{Aizenman:2019yuo}}), the RG-flow of $\lambda$ of the model
(\ref{GW-action}) in $d = 4$ stays one-loop bounded over all scales
(asymptotic safety).

The asymptotic safety result has initiated a research program that aims at
establishing existence of moments/cumulants of the measure
\begin{align}
  \mu(d\phi)= \frac{1}{\mathcal{Z}} \exp\Big(-  S[\phi]\Big)  d\phi,
\label{Euclidean}
\end{align}
with $S[\phi]$ given in (\ref{GW-action}), beyond formal power series. This
article is part of that program. A key insight was the suggestion of
{\cite{Disertori:2006uy}} to place oneself at the RG-fixed point $\Omega
\equiv 1$, which is preserved over all scales. In this setting,
{\cite{Disertori:2006uy}} proved that $\lambda$ remains RG-constant up to
3-loop order. The reason for this remarkable stability was discovered in
{\cite{Disertori:2006nq}}: there is a Ward identity which can be employed to
prove that $\lambda$ is RG-constant to all orders in perturbation theory for
$\Omega \equiv 1$.

There are two research directions along which the rigorous construction of
(\ref{GW-action}) was pursued. This article opens a third direction. First,
Rivasseau developed the loop vertex expansion {\cite{Rivasseau:2007fr}} as a
new framework to Borel resum the series, later extended to a multi-scale loop
vertex expansion {\cite{Gurau:2013oqa}}. With these tools, Zhituo Wang
succeeded in constructing the $d = 2$-dimensional model
(\ref{GW-action})+(\ref{Euclidean}) at
$\Omega = 1$ and proved that the logarithm of the partition function
$\mathcal{Z}$ is the
Borel sum of the perturbation series, analytic in $\lambda$ in a cardioid
domain {\cite{Wang:2018sed}}.

On the other hand, building on {\cite{Disertori:2006nq}}, one of us with H.
Grosse established in {\cite{Grosse:2012uv}} a hierarchy of non-perturbative
Dyson-Schwinger equations for cumulants of (\ref{Euclidean}).
The hierarchy starts with a non-linear integral equation
for the planar two-point function alone, which was solved for $d = 2$ with E.
Panzer in {\cite{Panzer:2018tvy}} and for $d = 4$ with H. Grosse and A. Hock
in {\cite{Grosse:2019qps}} (with a main step in {\cite{Grosse:2019jnv}}). The
solutions are concrete integrals of classical special functions in which
$\lambda$ is a parameter. All planar correlation functions are obtained from
the planar 2-point function by a combinatorial recipe {\cite{deJong:2019oez}}.
The other equations of the hierarchy follow a recursion in the Euler
characteristic $\chi = 2 - 2 g - n$ of a genus-$g$ Riemann surface with $n$
boundary components. Intuitively, correlation functions of topology $(g, n)$
resum all Feynman ribbon graphs that can be drawn on a genus-$g$ Riemann
surface, with the external lines of the graph ending at the $n$ boundary
components. But in fact one never expands into graphs; the equations are exact
in $\lambda$ and can in principle be solved recursively in decreasing Euler
characteristic.

However, to really construct the model in this way one needs to sum over all
genera $g \in \mathbb{Z}_{\geq 0}$ of Riemann surfaces. This sum cannot
converge; it is expected to be Borel summable, but proving the assumptions of
Borel summability seems hopeless. We therefore propose a new strategy to
construct the non-commutative QFT-model (\ref{GW-action})+(\ref{Euclidean}).

This strategy builds on recent spectacular achievements in the SPDE approach
to sub-critical quantum field theories. The method of stochastic quantization
was proposed by Parisi and Wu {\cite{parisi1981perturbation}} to study gauge
fields without gauge fixing. The key idea is to study a Euclidean field
theory, formally given by (\ref{Euclidean}),
through the Langevin dynamics (again formally) given by
\begin{equation}
  d \phi = - \nabla S (\phi) d \nospace t + \sqrt{2} d \nospace W (t).
\end{equation}
This should define a Markov process with (\ref{Euclidean}) as equilibrium
measure. For example, the standard $\lambda \phi^4$-theory
\begin{equation*}
  S (\phi) = \int \left( \frac{1}{2} | \nabla \phi |^2 + \frac{r}{2} \phi^2 +
    \frac{\lambda}{4} \phi^4 \right) d \nospace x
\end{equation*}
leads to the stochastic PDE
\begin{equation}
  \partial_t \phi = \Delta \phi - r \phi - \lambda \phi^3 + \xi \label{Phi4}
\end{equation}
where $\xi$ is a space-time white noise. Equation (\ref{Phi4}) has to be
renormalized analogously to the ``static'' field theory. Early mathematical
works on (\ref{Phi4}) (e.g. {\cite{jona1985stochastic}} and
{\cite{albeverio1991stochastic}}) established the existence (and sometimes
uniqueness) of probabilistically weak solutions to (\ref{Phi4}) in the case $d
= 2$. Da Prato and Debussche {\cite{da2003strong}} observed that using a
simple transformation, probabilistically strong solutions could be
constructed. In {\cite{mourrat2017global}} and  {\cite{tsatsoulis2018spectral}}
it was observed that the non-linear damping term could be used to derive
strong a priori estimates that can in turn be leveraged to yield an SPDE-based
construction of the Euclidean field theory.

The theory of this equation has seen drastic developments since Hairer's
introduction of regularity structures {\cite{hairer2014theory}} and
Gubinelli-Imkeller-Perkowski's work on paracontrolled distributions
{\cite{gubinelli2015paracontrolled}}. The theory of these singular stochastic
PDEs is now well-developed and is able to cover all sub-critical dimensions $d
< 4$ (see  {\cite{MR4210726}}, {\cite{MR3719541}}, {\cite{MR4252872}},
{\cite{MR4164267}}, {\cite{MR4586861}}, {\cite{duch2023parabolic}},
{\cite{esquivel2024priori}}).

In this paper we adapt the SPDE techniques to quantum fields on
non-commutative spaces and completely settle the $d = 2$-dimensional case of
(\ref{GW-action})+(\ref{Euclidean}) at $\Omega = 1$. In the recent work
{\cite{chandra2023stochastic}}, a different but related non-commutative
variant of $\phi^4$-model was constructed using stochastic PDEs. Their model is
defined over a $d$-dimensional torus and they work with the standard Besov
spaces which are common in the theory of singular SPDEs. Due to the presence
of the harmonic oscillator potential $4\Omega^2 \| \Theta^{- 1}
x\|^2$ in (\ref{GW-action}), our model does not have translation invariance.
The correlation function of the free field part is explicit but complicated
(see formula (\ref{Corre})), which makes working with spatial variables
impractical. Instead we work with the matrix basis (see discussion in Section
\ref{1section}), in which the action takes the form
\[ S [\phi] = \frac{\theta}{4} \left( \sum_{m, n \geqslant 0} \frac{1}{2}  \left(
   M^2 + \frac{4}{\theta} (m + n + 1) \right) | \phi_{m \nospace n} |^2 +
   \sum_{m, n, k, l \geqslant 0} \frac{\lambda}{4} : \phi_{m \nospace n}
   \phi_{n \nospace k} \phi_{k \nospace l} \phi_{l \nospace n} : \right) . \]
We show that the EQFT can be realized as the invariant measure of the
stochastic quantization equation
\[ \partial_t \phi_{mn} = - A_{m \nospace n} \phi_{m \nospace n} - 
   \frac{\theta \lambda}{4} \sum_{k, l} : \phi_{m \nospace k} \phi_{k \nospace l}
   \phi_{l \nospace n} : + \dot{B}_t^{(m \nospace n)}
 \]
where $A_{m \nospace n} \assign m+n+1+ \frac{\theta M^2}{4}$
and $B_t^{(m \nospace n)} = \overline{B_t^{(n \nospace m)}}$
are complex Brownian motions such that $\mathbb{E} [\dot{B}_t^{(m \nospace n)}
\dot{B}_s^{(k \nospace l)}] = 2 \delta (t - s) \delta_{m \nospace l} \delta_{n
\nospace k}$.
The definition of various Wick products can be found in Section
\ref{1section}.

We study these equations using the Da Prato-Debussche trick, which
means that we view $\phi$ as perturbation of the non-interactive
(i.e. Gaussian) stationary solutions of the equations (see section
\ref{DPDT})
\[
\partial_t z_{mn} = - A_{m \nospace n} z_{m \nospace n} + \dot{B}_t^{(m
\nospace n)},\]
 and study the remainder $v_{m \nospace n} (t) \assign \phi_{m
\nospace n} (t) - z_{m \nospace n} (t)$ for all $m, n \in \mathbb{N}$ in the
matrix valued function space
\[ K^{\beta}_T \assign \{ (c_{m n} (t))_{t \in [0, T]} | \sup_{t \in [0, T]}
   t^{\beta} \| c (t) \|_{H^{\beta}} + \sup_{t \in [0, T]} \| c (t) \|_{H^0} <
   \infty \} \]
where
\[ \| c (t) \|_{H^{\beta}} \assign \left( \sum_{m, n=0}^\infty A_{m \nospace n}^{2
   \beta} | c_{m \nospace n} |^2 \right)^{\frac{1}{2}} . \]
Our first main theorem is the local well-posedness.

\begin{theorem}
  For any initial value $v (0) \in H^0$, \ there exists a random time
  $T$, which depends on the initial data $\| v (0)\|_{H^0}$ and $z$,
  such that the renormalized remainder equation
  \[ \partial_t v_{m \nospace n} = - A_{m n} v_{m \nospace n} -
    \frac{\theta\lambda}{4}
    (v^3 + v^2 z + v \nospace z \nospace v + z \nospace v^2 + : z^2 :
    v + v : z^2 : + z \nospace v \nospace z + : z^3 :)_{m n}
  \]
  has a unique solution up to time $T$ in the space $K_T^{\frac{1}{2} -}$
  almost surely.
\end{theorem}

Here we use the notation $K_T^{\frac{1}{2} -}$ to denote $K_T^{\frac{1}{2} -\epsilon}$ for some small positive constant $\epsilon$.
The main difficulty in deriving this result is the $z \nospace v \nospace z$
term. This term corresponds to non-planar ribbon graphs and therefore does not
require a renormalization. However, it is still important to view the action
of both $z$ factors as a single operation to capture stochastic cancellations.
We found it most convenient to realize this by considering the random operator
$v \mapsto z \nospace v \nospace z$ acting on an $L^2$-based Hilbert space. In
this framework we are able to obtain the required estimates, but unfortunately
it requires to estimate 105 different diagrams, see Appendix \ref{Eappendix}.

We then show an a priori estimate for the equations to get global
well-posedness. It turns out that one step in a Da Prato-Debussche expansion is not
enough; we have to do the second order expansion: we let $y$ be the stationary solution (see section \ref{APE} for the explicit formula) to 
 \[
 \partial_t y_{m n} = - A_{m n} y_{m n} - \frac{\theta \lambda}{4} : z^3 :_{m n}
 \]
  and show an a priori estimate for the second order remainder $w \assign v - y$.

\begin{theorem}
  We have
  \[ \partial_t \| w \|_{H^0}^2 + \| w \|_{H^{\frac{1}{2}}}^2 +
    \frac{\theta \lambda}{4}
    \| w^2 \|_{H^0}^2 \leqslant C \nospace F [y, z] \]
  where $C$ is a positive constant and $F [y, z]$ (see formula
  (\ref{Function})) only depends on $y$ and $z$, and has time independent
  stochastic moments of all orders. Moreover
  \[ \| w \|_{H^0}^2 (t) \leqslant e^{- t} \| w \|_{H^0}^2 (0) + C \int_0^t
     e^{- (t - s)} F [y, z] (s) d \nospace s. \]
\end{theorem}

We use this statement to conclude the global existence for $v$.

\begin{theorem}
  The renormalized remainder equation
  \[ \partial_t v_{m \nospace n} = - A_{m n} v_{m \nospace n} - \frac{\theta
     \lambda}{4} (v^3 + v^2 z + v \nospace z \nospace v + z \nospace v^2 + : z^2 :
     v + v : z^2 : + z \nospace v \nospace z + : z^3 :)_{m n} \]
  can be solved on $[0, \infty)$ almost surely.
\end{theorem}

The invariant measure can be constructed using the Krylov--Bogoliubov
method, as in {\cite{tsatsoulis2018spectral}}.  Suppose $\mu_t$ is the
probability measure of the solution at time $t$, where $\mu$ is the law
of the initial value. The sequence of probability measures
\[
  \frac{1}{t} \int_0^t \mu_s d \nospace s
\]
has a weak limit, which as expected is an invariant measure of the process. We
have the following main theorem.

\begin{theorem}\label{thm:main}
 Suppose $\phi (0)=z(0)+v(0) \in H^{- \frac{1}{2} - \varepsilon}$ with any random field $v(0)\in H^0$ a.s., then there exists a
  sequence of time variables $t_k \rightarrow \infty$, such that the sequence
  of probability measures
  \[ \frac{1}{t_k} \int_0^{t_k} \mu_s d \nospace s \]
  has an invariant weak limit in $\mathcal{M}_1 \left( H^{- \frac{1}{2} - \varepsilon}
  \right)$.
\end{theorem}
\noindent
Here $\mathcal{M}_1 \left( H^{- \frac{1}{2} - \varepsilon} \right)$ is the set
of all probability measures in the Hilbert space $H^{- \frac{1}{2} - \varepsilon}$.

\begin{remark}
  The method of stochastic quantization
  allows us to construct the measure for any $\lambda \geqslant 0$, which is
  different from the result in {\cite{Wang:2018sed}} where Borel summability
  of renormalized perturbation series for $\lambda$ in a (complex) cardioid
  domain was proved.
\end{remark}

\begin{remark}
We do not prove uniqueness of the invariant measure for the stochastic quantization dynamics. The harmonic oscillator potential ($4\Omega^2||\Theta^{-1}x||^2$) is confining and plays a role analogous to a finite-volume condition. We expect that the strategy of Tsatsoulis and Weber (\cite{tsatsoulis2018spectral}) to establish uniqueness can be adapted to the present setting with essentially no new conceptual ingredient, but we do not pursue the details here.
 
\end{remark}

\begin{remark}
  One may ask whether the measure constructed above could instead be
  obtained by the classical finite-volume argument of Nelson. To
  explain why this argument does not carry over to the non-commutative
  model, let us first recall its two key ingredients in the usual
  finite-volume $\phi^4_2$ model. Let $\phi_N$ be a Fourier cutoff of
  the Gaussian free field on a fixed two-dimensional torus $\Lambda$,
  and set $c_N=\mathbb{E}[\phi_N(x)^2]\sim\log N$. First, the
  integrated fourth Wick power
\[
	V_N:=\int_\Lambda:\phi_N^4:(x)\,dx
\]
converges in $L^2$ to $V=\int_\Lambda:\phi^4:(x)\,dx$; one has	
	\begin{equation}\label{eq:into-Nelson1}
	\mathbb{E}\bigl[|V_N-V|^2\bigr]\lesssim \frac{\log^2N}{N}.
	\end{equation}
Second, the cutoff Wick polynomial satisfies
	\begin{equation}\label{eq:into-Nelson2}
	:\phi_N^4:(x)
	=\phi_N(x)^4-6c_N\phi_N(x)^2+3c_N^2
	=\bigl(\phi_N(x)^2-3c_N\bigr)^2-6c_N^2
	\geq-6c_N^2.
	\end{equation}
Consequently, $V_N\geq-C_\Lambda(\log N)^2$. Together with the  quantitative convergence estimate \eqref{eq:into-Nelson1} and Gaussian hypercontractivity, this yields Nelson's estimate
\[
	\sup_{N\geq1}\mathbb{E}\bigl[e^{-pV_N}\bigr]<\infty
	\qquad\text{for every }p<\infty.
\]
Neither  \eqref{eq:into-Nelson1} nor \eqref{eq:into-Nelson2} are available for the natural matrix cutoff. Let $z_N$ be the $(N+1)\times(N+1)$ cutoff of $z$,  write $:\operatorname{Tr}(z_N^4):$. Since $z_N$ is Hermitian,  we have
\[	
	\mathbb{E}[(:\operatorname{Tr}(z_N^4):)^2]
	\gtrsim
	\sum_{a,b,c,d=0}^N
	\frac1{A_{ab}A_{bc}A_{cd}A_{da}}
	\gtrsim\log N,
\]	
and hence $:\operatorname{Tr}(z_N^4):$ does not possess the required $L^2$ limit. 
	
	
\end{remark}

\subsection{Structure of the paper} In Section \ref{1section} we introduce
the model including the definition of the Moyal product and the matrix base.
In Section \ref{DPDT} the da Prato-Debussche remainder equation and its
constituents are defined, various terms are estimated in Section \ref{FPM}
while the fixed point argument leading to the local-in-time well-posedness
result is completed in Section \ref{LocalESQ}. Section \ref{APE} contains the
derivation of a priori bounds. A Krylov-Bogoliubov argument which leads to the
existence of an invariant measure is executed in Section \ref{EIM}. Various
background facts as well as technical calculations (including the stochastic
estimates of 105 diagrams) can be found in Appendices
\ref{Hyper}-\ref{Eappendix}. \

\subsection{Notation} \label{notation}
Throughout the paper, we write $\lesssim$ and $\sim$
for $\leqslant$ and $=$ up to some irrelevant constants. We collect all the spaces used in this paper  for reader's convenience. 

For $m,n\geq 0$, set
\[
A_{mn}
:=
m+n+1+\frac{\theta M^2}{4}.
\]
Define the following spaces
\begin{align*}
H^\alpha
&:=
\left\{
c=(c_{mn})_{m,n\geq 0}
\ \middle|\
\|c\|_{H^\alpha}
:=
\left(
\sum_{m,n=0}^{\infty}
A_{mn}^{2\alpha}|c_{mn}|^2
\right)^{\frac12}
<\infty
\right\},
\\
M^p
&:=
\left\{
c=(c_{mn})_{m,n\geq 0}
\ \middle|\
\|c\|_{M^p}
:=
\sup_{m,n\geq 0}
A_{mn}^{p}|c_{mn}|
<\infty
\right\},
\\
G^{a,b}
&:=
\left\{
L=(L^i_{jk})_{i,j,k\geq 0}
\ \middle|\
\|L\|_{G^{a,b}}
:=
\sup_{i,j,k\geq 0}
A_{ii}^{a}A_{jk}^{b}|L^i_{jk}|
<\infty
\right\}.
\end{align*}
For any Banach space $E$, we use the notation
\[
C_T E:=C([0,T];E),
\]
and equip this space with the uniform norm
\[
\|f\|_{C_T E}
:=
\sup_{t\in[0,T]}\|f(t)\|_E.
\]
In particular, $C_T H^\alpha$, $C_T M^p$, and $C_TG^{a,b}$ denote the
spaces of paths that are continuous, with respect to the corresponding
Banach-space norms, from $[0,T]$ into $H^\alpha$, $M^p$, and $G^{a,b}$,
respectively.

Finally, for $\beta>0$, define the weighted path space
\[
K_T^\beta
:=
\left\{
c\in C([0,T];H^0)\cap C((0,T];H^\beta)
\ \middle|\
\|c\|_{K_T^\beta}<\infty
\right\},
\]
where
\[
\|c\|_{K_T^\beta}
:=
\sup_{t\in[0,T]}\|c(t)\|_{H^0}
+
\sup_{t\in(0,T]}t^\beta\|c(t)\|_{H^\beta}.
\]

Some properties of these spaces are listed in Appendix \ref{spaceofmatrix}.

\subsection{Acknowledgements} CS and HW are funded by the European Union (ERC,
GE4SPDE, 101045082). CS, HW and RW are funded by the Deutsche
Forschungsgemeinschaft (DFG, German Research Foundation) under Germany's
Excellence Strategy EXC 2044 -390685587, Mathematics M{\"u}nster:
Dynamics-Geometry-Structure. CS would like to thank Fabian
H{\"o}fer for a discussion on compact embeddings.

\subsection{Declaration of Generative AI Use} After the mathematical results and proofs in this manuscript had been developed by the authors, ChatGPT was used during the revision process as an editorial and presentational aid. It was used to refine the English, and shorten or reorganize the exposition of some calculations, especially in Appendices D and E. 

\section{$\lambda \phi^4$ Model in 2-d Moyal Space}\label{1section}

Suppose that $d$ is an even integer, $\Theta$ is a $d \times d$ real
non-degenerate antisymmetric block diagonal matrix of the form
\[
\tmop{diag} (\Theta_1,\ldots, \Theta_{d / 2}), \qquad \text{where} \qquad \Theta_1 = \ldots = \Theta_{d / 2} = \left(
\begin{array}{cc}
  0 & \theta\\
  - \theta & 0
\end{array} \right)
\]
 with $\theta \in \mathbb{R}^+$. The Moyal product of two
complex-valued Schwartz functions $f, g \in \mathcal{S} (\mathbb{R}^d)$ is
defined by
\[ (f \star g) (x) \assign \int_{\mathbb{R}^d} \int_{\mathbb{R}^d} \frac{d
   \nospace k d \nospace y}{(2 \pi)^d} f \left( x + \frac{1}{2} \Theta k
   \right) g (x + y) e^{i \langle k, y \rangle} \]
which is again a Schwartz function. See Appendix \ref{Aappendix} for a collection of some
properties of the  Moyal product. The Euclidean action of the $\lambda \phi^4$
model in 2-d Moyal space (see {\cite{Grosse:2003nw}}) is  given by
\begin{equation}
  S_E [\phi] = \frac{1}{8\pi}
  \int_{\mathbb{R}^2} \left( \frac{1}{2} | \nabla \phi |^2 +
  \frac{2 \Omega^2}{\theta^2} | x |^2 \phi^2 + \frac{1}{2} M^2 \phi^2 +
  \frac{\lambda}{4} \phi^{\star 4} \right) d \nospace x \label{GWmodel}
\end{equation}
where we assume $\phi$ is a real function. Then the stochastic quantization
equation is
\[ \partial_t \phi = \Delta \phi - \frac{4 \Omega^2}{\theta^2} | x |^2 \phi -
  M^2 \phi - \lambda \phi^{\star 3} + \sqrt{2}\xi
\]
where $\xi$ is a space-time white noise. Its invariant measure is formally
given by Gibbs measure $\frac{1}{\mathcal{Z}} \exp (- S_E [\phi])$. The
covariance $\langle \phi (x) \phi (y) \rangle$ of the free field part
($\lambda = 0$) given by $\left( - \frac{1}{2} \Delta + \frac{2
\Omega^2}{\theta^2} | x |^2 + \frac{1}{2} M^2 \right)^{- 1}$ can be computed
explicitly and the result is given by the following formula
\begin{equation}
  \int_0^{+ \infty} \nospace \frac{\omega^{d / 2} {e^{- t \left( \frac{\omega
  d}{2} + \frac{1}{2} M^2 \right)}} }{\pi^{d / 2} \left( {1 - e^{- 2 \omega
  t}}  \right)^{d / 2}} \exp \left( - \frac{\omega \left( {1 + e^{- \omega t}}
  \right)^2 | x - y |^2 + \omega \left( {1 - e^{- \omega t}}  \right)^2 | x +
  y |^2}{4 \left( {1 - e^{- \omega t}}  \right) \left( {1 + e^{- \omega t}} 
  \right)} \right) d \nospace t \label{Corre}
\end{equation}
where $\omega = \frac{2 \Omega}{\theta}$. Working with this correlation
function is not easy, following the works {\cite{Grosse:2003nw}} we use the
matrix basis instead.

We restrict ourselves to the case $d = 2$ and $\Theta = \left(
\begin{array}{cc}
  0 & \theta\\
  - \theta & 0
\end{array} \right)$ with $\theta \in \mathbb{R}^+$. The matrix basis is an
orthonormal basis $\{ b_{m \nospace n} \}_{m, n = 0}^{+ \infty}$ of $L^2
(\mathbb{R}^2;\mathbb{C})$ such that if two Schwartz functions $f, g \in \mathcal{S}
(\mathbb{R}^2)$ are expanded in this basis as
\[ f (x) = \sum_{m, n = 0}^{\infty} f_{m \nospace n} b_{m \nospace n} (x),
   \quad g (x) = \sum_{m, n = 0}^{\infty} g_{m \nospace n} b_{m \nospace n}
   (x) \]
then the coefficients of Moyal product become the matrix product of
corresponding coefficients
\[ (f \star g) (x) = \sum_{m, n = 0}^{\infty} \left( \sum_{k = 0}^{\infty}
   f_{m \nospace k} g_{k \nospace n} \right) b_{m \nospace n} (x) . \]
A more detailed description of the matrix basis can be found in Appendix
\ref{Aappendix}.

In the matrix basis, the Euclidean action (\ref{GWmodel}) takes 
 the following form
\[ 
S [\phi] = \frac{\theta}{4} \sum_{m, n, k, l} \left( \frac{1}{2} \phi_{m
   \nospace n} G_{m \nospace n ; k \nospace l} \phi_{k \nospace l} +
   \frac{\lambda}{4} \phi_{m \nospace n} \phi_{n \nospace k} \phi_{k \nospace
   l} \phi_{l \nospace n} \right) \]
where the quantities $G_{m \nospace n ; k \nospace l}$ are given by
\begin{align*}
 G_{m \nospace n ; k \nospace l} &= \left( M^2 + \frac{2 (1 +
   \Omega^2)}{\theta} (m + n + 1) \right) \delta_{n \nospace k} \delta_{m
   \nospace l} - \\
 &- \frac{2 (1 - \Omega^2)}{\theta} \sqrt{(m + 1) (n + 1)} \delta_{n + 1, k}
   \delta_{m + 1, l} - \frac{2 (1 - \Omega^2)}{\theta} \sqrt{m \nospace n}
   \delta_{n - 1, k} \delta_{m - 1, l} . 
   \end{align*}
For simplicity of our treatment, we assume $\Omega = 1$ (see
\cite{Disertori:2006uy}), so that
\[
 G_{mn ; kl} = \left( M^2 + \frac{4}{\theta} (m + n + 1) \right) \delta_{n
   \nospace k} \delta_{m \nospace l}
\]
whose inverse can be obtained by solving the equations
\[
  \sum_{k, l = 0}^{\infty} G_{m \nospace n ; k \nospace l} \Delta_{l \nospace
   k ; s \nospace r} = \sum_{k, l = 0}^{\infty} \Delta_{n \nospace m ; \ell
   \nospace k} G_{k \nospace \ell ; r \nospace s} = \delta_{m \nospace r}
   \delta_{n \nospace s} \]
to get
\[ \Delta_{n \nospace m ; l \nospace k} = \frac{\delta_{m \nospace l}
    \delta_{\nospace n \nospace k}}{M^2 + \frac{4}{\theta}  (m + n + 1)} .
\]
Before going to the stochastic quantization equation, we need to make
clear how to renormalize those matrix powers. The definition of Wick
powers is different from the commutative models.  Below we will
decompose the field $\phi=z+v$ where the random field $z$ is Gaussian
and distributed according to the Gaussian non-interacting theory, see
equation \eqref{Free} below.
The second Wick power of $\phi$ is defined as
$$: \phi^2 \assign v^2 + v \nospace z + z \nospace v + : z^2 : =
\phi^2 - \langle z^2 \rangle$$ where
$: z^2 : = z^2 - \langle z^2 \rangle$ (notice $\langle z^2 \rangle$
provides divergence, the way to define these objects correctly is to
use the finite matrix cut-off to approximate the original one, see
appendix \ref{Dappendix} for more detailed explanations and
constructions). Note that we define $: z^2 :$ simply as the usual Wick
renormalization of its matrix product components, but this naive idea
does not apply to higher powers, as we can see from the third Wick
power. We define the third Wick power of the interactive field by
\[
  : \phi^3 : = v^3 + v^2 z + v \nospace z \nospace v + z \nospace v^2 + v :
z^2 : + z \nospace v \nospace z + : z^2 : v + : z^3 \assign \phi^3 - \langle
z^2 \rangle \phi - \phi \langle z^2 \rangle
\]
and here
$: z^3 : = z^3 - \langle z^2 \rangle z - z \langle z^2 \rangle$.
This differs from the commutative case as we do not
subtract the contraction between first and third free fields. For the
fourth Wick power, we have
\begin{align*}
  : \phi^4 : & =  v^4 + v^3 z + v^2 z \nospace v + v \nospace z \nospace v^2
  + z \nospace v^3 + v^2 : z^2 : + v : z^2 : v + : z^2 : v^2 + v \nospace z
  \nospace v \nospace z + z \nospace v^2 z + z \nospace v \nospace z \nospace
  v 
  \\
  &  \qquad +: z^3 : v + v : z^3 : + : z^2 : v \nospace z
  + z \nospace v : z^2 : +  : z^4 :
  \\
  & =  \phi^4 - \phi^2 \langle z^2 \rangle - \phi \langle z^2 \rangle \phi -
  \langle z^2 \rangle \phi^2 + \langle z^2 \rangle^2
\end{align*}
and here
$: z^4 : = z^4 - z^2 \langle z^2 \rangle - z \langle z^2 \rangle z -
\langle z^2 \rangle z^2 + \langle z^2 \rangle^2$, which is now very
different from the commutative case. The trace functional should be
renormalized differently from the powers of the matrix: taking an
example, the cyclic symmetry makes two $z$ in the $\mathrm{Tr}(zvz)$ adjacent
and hence produces a divergence. The correct renormalization of
$\mathrm{Tr}(\phi^4)$ is
\[
  :\mathrm{Tr}(\phi^4):=\mathrm{Tr}(\phi^4)-4\mathrm{Tr}
  (\langle z^2 \rangle\phi^2)+2\mathrm{Tr}(\langle
z^2 \rangle^2)
\]
which is different from $\mathrm{Tr}(:\phi^4:)$. Its
functional derivative with respect to $\phi_{ij}^{*}=\phi_{ji}$ is
clearly $4:\phi^3:_{ij}$. 

The correct renormalised stochastic quantization equation
should thus be
\[ \partial_t \phi_{mn}=
  -\frac{\partial S[\phi]}{\partial \phi_{nm}}+\dot{B}_t^{(m \nospace n)}
\]  
In short summary, we only need to subtract contractions of adjacent free field components.

\section{Da Prato-Debussche Trick}\label{DPDT}

The stochastic quantization equation in matrix basis is formally given by the system of SDEs
\begin{align}\notag
    \partial_t \phi_{mn} & =  - \frac{\theta}{4}\bigg(
            \sum_{k, l} G_{m \nospace n ; k
     \nospace l} \phi_{k \nospace l} +\lambda : \sum_{k, l}
     \phi_{m \nospace k} \phi_{k \nospace l} \phi_{l \nospace n}: \bigg)+
     \dot{B}_t^{(m \nospace n)}
     \\
     \label{stochasticquantizationeq}
     & =  -  A_{m \nospace n} 
     \phi_{m \nospace n} - \frac{\theta \lambda}{4}  : \sum_{k, l} \phi_{m \nospace
     k} \phi_{k \nospace l} \phi_{l \nospace n} :+ \dot{B}_t^{(m \nospace n)}
 \end{align}
 where  $A_{m \nospace n} \assign m+n+1+ \frac{\theta M^2}{4}$ and 
 $B_t^{(m \nospace n)} = \overline{B_t^{(n \nospace m)}}$ are complex
Brownian motions such that $B_t^{(m \nospace n)}$ only correlates with
$B_t^{(n \nospace m)}$, or equivalently we can write 
\[
\mathbb{E} [\dot{B}_t^{(m \nospace
n)} \dot{B}_s^{(k \nospace l)}] = 2 \delta (t - s) \delta_{m \nospace l}
\delta_{n \nospace k}.
\] 
Since $\phi (t, x) = \sum_{m, n = 0}^{\infty} \phi_{m
\nospace n} (t) b_{m \nospace n} (x)$ is a real field,  $\phi_{m \nospace
n} (t)$ is a Hermitian matrix valued function, that is $\phi_{m \nospace n}
(t) = \overline{\phi_{n \nospace m} (t)}$.

In order to give a meaning to this system of  renormalised SDEs
we use the Da Prato-Debussche trick {\cite{da2003strong}}, which means that  we
regard equation (\ref{stochasticquantizationeq}) as the perturbation of the
system of SDEs
\begin{equation}
  \partial_t z_{mn} = - A_{m \nospace n} z_{m \nospace n} + \dot{B}_t^{(m
  \nospace n)} , \label{Free}
\end{equation}
whose  solution consists of  a collection of Ornstein-Uhlenbeck processes $\{ z_{m
\nospace n} (t) \}_{m, n = 0}^{\infty}$. We choose to artificially extend the noise to negative 
times and work with the stationary solution 
$$z_{mn}(t)=\int_{-\infty}^t e^{-A_{mn}(t-s)}\dot{B}_s^{(m  \nospace n)}ds,$$
with correlation function
\[ 
\langle z_{m \nospace n} (t) z_{k \nospace l} (s) \rangle = \frac{\delta_{m
   \nospace l} \delta_{n \nospace k}}{A_{m \nospace n}} e^{- | t - s | A_{m
   \nospace n}}. \]
In particular,  for this choice   
$\{ z_{m \nospace n} (0) \}_{m, n =
0}^{\infty}$ is Gaussian with mean $0$ and covariance $\langle z_{m n} (0)
z_{k l} (0) \rangle = \frac{\delta_{m \nospace l} \delta_{n \nospace k}}{A_{m
\nospace n}}$. 
The matrix valued random process $z (t)$ takes values in  $H^{- \frac{1}{2} -
\varepsilon}$  (see Appendix \ref{Dappendix}).

We consider $\{ \phi_{m \nospace n} (t) \}_{m, n = 0}^{\infty}$ as a
perturbation of $\{ z_{m \nospace n} (t) \}_{m, n = 0}^{\infty}$, that is we
define a new variable $v_{m \nospace n} (t) \assign \phi_{m \nospace n} (t) -
z_{m \nospace n} (t)$ for all $m, n \in \mathbb{N}$. The equation for $\{ v_{m
\nospace n} (t) \}_{m, n = 0}^{\infty}$ becomes
\begin{align*}
  \partial_t v_{m \nospace n} & =  - A_{mn} v_{m \nospace n} 
  \\
   &- \frac{\theta \lambda}{4} \bigg\{ \sum_{k, l = 0}^{\infty} (v_{m
     \nospace k} v_{k \nospace l} v_{l \nospace n} + z_{m \nospace k} v_{k
     \nospace l} v_{l \nospace n} + v_{m \nospace k} z_{\nospace k \nospace l}
     v_{l \nospace n} + v_{m \nospace k} v_{k \nospace l} z_{l \nospace n} +
     z_{m \nospace k} v_{k \nospace l} z_{l \nospace n})  
   \\
   & \quad + \sum_{k = 0}^{\infty}  v_{mk}
   \Big( :\sum_{l = 0}^{\infty} z_{k l} z_{l \nospace n}: \Big)
   + \sum_{l =  0}^{\infty}:  \Big( \sum_{k = 0}^{\infty} z_{m \nospace k} z_{kl} \Big):
   v_{ln} + :\sum_{k, l = 0}^{\infty} z_{m k} z_{k l}
     z_{l n}: \bigg\}.
 \end{align*}
where the renormalised sums  $:\sum_{k = 0}^{\infty} z_{m \nospace k} z_{k \nospace l}:$ and
$:\sum_{k, l = 0}^{\infty} z_{m k} z_{k l} z_{l n}:$ are interpreted  as discussed above in Section~\ref{1section}. 
Hence the above formal equation \eqref{stochasticquantizationeq} turns into the well-defined equation
\begin{equation}
  \partial_t v_{m  n} = - A_{m n} v_{m \nospace n} - \frac{\theta
  \lambda}{4} \bigg\{ \sum_{i = 1}^7 \mathcal{N}_i (v)_{m n} + : z^3 :_{m n}
  \bigg\} \label{DPDeq} 
\end{equation}
where we use 
 the nonlinear operators
\begin{equation}
  \mathcal{N}_1 (v) = v^3, \quad \mathcal{N}_2 (v) = z \nospace v^2, \quad
  \mathcal{N}_3 (v) = v \nospace z \nospace v, \quad \mathcal{N}_4 (v) = v^2 z
  \label{Terms}
\end{equation}
and  the linear operators
\begin{align}
\mathcal{N}_5 (v) = z \nospace v \nospace z, \quad \mathcal{N}_6 (v) = v :
z^2 :, \quad \mathcal{N}_7 (v) = : z^2 : v.
  \label{Terms-2}
\end{align}

Before proceeding, we collect the various regularities of the terms we encounter.  For every sufficiently small $\varepsilon>0$ and sufficiently large $p$,
$z,:z^2:\in L^p(\Omega;C_TM^{\frac12-\varepsilon})
\hookrightarrow L^p(\Omega;C_TH^{-\frac12-\varepsilon})$; see
Lemma~\ref{lemma:M-construction-z-z2} and
Corollary~\ref{corollary:H-regularity-z-z2} in
Appendix~\ref{Dappendix}. Moreover,
$:z^3:\in L^p(\Omega;C_TH^{-\frac12-\varepsilon})$, whose construction
is also given in Appendix~\ref{Dappendix}.

 As the worst term $ z^3 :$ takes values in 
$C_T H^{-\frac{1}{2} - \varepsilon}$ and the operator $(\partial_t + A)^{- 1}$ where $A$ is the matrix $\{ A_{m
n} \}_{m, n = 0}^{\infty}$ improves ``regularity'' in the $H^\alpha$-spaces by $1$ (see Lemma~\ref{Schauder}), we expect the solution $v$ of \eqref{DPDeq}
to take values in $H^{\frac{1}{2} - \varepsilon}$,
In the analysis of \eqref{DPDeq} we will require an additional stochastic argument to show that 
the random operator $\mathcal{N}_5$
belongs to 
$L^p(\Omega\times[0,T];
\mathcal{L}(H^{\frac12-\varepsilon},H^{-\varepsilon-\varepsilon'}))$ (see Section~\ref{FPM} and Appendix~\ref{Eappendix}).

\section{Fixed Point Map}\label{FPM}

To solve the equation (\ref{DPDeq}), we study the following integral version
\begin{align}
 v_{m n} (t) = e^{- A_{m \nospace n} t} v_{m n} (0) - \frac{\theta \lambda}{4}
   \int_0^t e^{- A_{m n} (t - s)} \Big( \sum_{i = 1}^7 \mathcal{N}_i (v)_{m
       n} (s) + : z^3 :_{m n} (s) \Big) ds.
\label{DPD-fixeq}
\end{align}
In order to do a Picard iteration, we derive estimates on the
operators $\mathcal{N}_i$. Multiplicative inequalities in the matrix
spaces $H^\alpha$, $M^{\beta}$ are provided in Lemma
\ref{lem:multiplicative_inequalities}. It turns out that these
inequalities are sufficient to estimate the $\mathcal{N}_i$ for
$i = 1,2,4,6,7$, see the following Lemmas \ref{lemma-N1},
\ref{zvfunction}.  The term $\mathcal{N}_3(v) = vzv$ requires slighly
more care, as a multilinear estimate for the entire product of three
factors is needed, rather than using the multiplicative inequality
twice (Lemma \ref{lem:N3Estimate}).  The most substantial difficulty
arises when bounding $\mathcal{N}_5 (v) = z \nospace v \nospace
z$. This product will be dealt with in Lemma~\ref{lem:N5-estimate} and
Appendix~\ref{Eappendix} via stochastic estimates on the operator
$v \mapsto z v z$.

Throughout the section we work under the assumption  $v \in C_T H^{\frac{1}{2} - \varepsilon}$.
The contribution from $e^{- A_{m \nospace n} t} v_{m n} (0)$ takes values in
$C_T H^{\frac{1}{2} - \varepsilon}$ if we assume $v (0) \in H^{\frac{1}{2} -
\varepsilon}$, and $\int_0^t e^{- A_{m n} (t - s)} : z^3 :_{m n} (s) d
\nospace s$ lives in $C_T H^{\frac{1}{2} - \varepsilon}$ by the Schauder
estimate (Lemma~\ref{Schauder}) and the regularity of $: z^3 :$ (Lemma~\ref{lem:z_cube}).

\begin{lemma}
  \label{lemma-N1}
  We have the following inequalities for the $\mathcal{N}_1$ map:
  
  1. $\| \mathcal{N}_1 (v) \|_{C_T H^{\frac{1}{2} - \varepsilon}} \leqslant \|
  v \|_{C_T H^{\frac{1}{2} - \varepsilon}}^3$ for all $v \in C_T
  H^{\frac{1}{2} - \varepsilon}$;
  
  2. for all $w, v \in C_T H^{\frac{1}{2} - \varepsilon}$
  \[ \| \mathcal{N}_1 (v) -\mathcal{N}_1 (w) \|_{C_T H^{\frac{1}{2} -
     \varepsilon}} \lesssim \| v - w \|_{C_T H^{\frac{1}{2} - \varepsilon}}
     \left( \| v \|_{C_T H^{\frac{1}{2} - \varepsilon}}^2 + \| w \|_{C_T
     H^{\frac{1}{2} - \varepsilon}}^2 \right) . \]
\end{lemma}

\begin{proof}
  For the first estimate, using the inequality (\ref{multiplicativeineq}), we have
  \[
    \| \mathcal{N}_1 (v (t)) \|_{H^{\frac{1}{2} - \varepsilon}} = \| v (t)^3
     \|_{H^{\frac{1}{2} - \varepsilon}} \leqslant \| v (t) \|_{H^{\frac{1}{2}
     - \varepsilon}} \| v (t)^2 \|_{H^{\frac{1}{2} - \varepsilon}} \leqslant
     \| v (t) \|_{H^{\frac{1}{2} - \varepsilon}}^3 \]
  and taking supremum of $t$ over $t \in [0, T]$ one gets the result.
  
  For the second one, by the same inequality (\ref{multiplicativeineq}),
  \begin{align*}
    \| \mathcal{N}_1 (v (t)) -\mathcal{N}_1 (w (t)) \|_{H^{\frac{1}{2} -
    \varepsilon}}
    & =  \| v (t)^3 - w (t)^3 \|_{H^{\frac{1}{2} - \varepsilon}}\\
    & =  \| [v (t) - w (t)] v (t)^2 + w (t) [v (t) - w (t)] v (t) + w (t)^2
    [v (t) - w (t)] \|_{H^{\frac{1}{2} - \varepsilon}}\\
    & \leqslant  \| v (t) - w (t) \|_{H^{\frac{1}{2} - \varepsilon}} \left(
    \| v (t) \|_{H^{\frac{1}{2} - \varepsilon}}^2 + \| v (t)
    \|_{H^{\frac{1}{2} - \varepsilon}} \| w (t) \|_{H^{\frac{1}{2} -
    \varepsilon}} + \| w (t) \|_{H^{\frac{1}{2} - \varepsilon}}^2 \right)\\
    & \lesssim  \| v (t) - w (t) \|_{H^{\frac{1}{2} - \varepsilon}} \left(
    \| v (t) \|_{H^{\frac{1}{2} - \varepsilon}}^2 + \| w (t)
    \|_{H^{\frac{1}{2} - \varepsilon}}^2 \right) .
\end{align*}
 Taking supremum of $t$ over $t \in [0, T]$ one gets the result.
\end{proof}

The estimates for $\mathcal{N}_2$, $\mathcal{N}_4$, $\mathcal{N}_6$,
$\mathcal{N}_7$ follow from a similar argument, and we put them together.

\begin{lemma}
  \label{zvfunction}Assume $z, : z^2 : \in C_T M^{\frac{1}{2} - \varepsilon'}$
  and
  $w, v \in C_T H^{\frac{1}{2} - \varepsilon}$ (see Section~\ref{notation} for the definitions of these spaces).
  Let  $2 \alpha + 2 \beta - 2 \varepsilon' > 1$, $\beta > \epsilon'$  and $\frac{1}{2} - \varepsilon \geqslant \beta \geqslant 0$. Then 
  we have the following
  inequalities for $\mathcal{N}_2$, $\mathcal{N}_4$, $\mathcal{N}_6$,
  $\mathcal{N}_7$:
  
  1. $\| \mathcal{N}_2 (v) \|_{C_T H^{- \alpha}} \leqslant \| z \|_{C_T
  M^{\frac{1}{2} - \varepsilon'}} \| v \|_{C_T H^{\beta}}^2$ ;
  
  2. $\| \mathcal{N}_4 (v) \|_{C_T H^{- \alpha}} \leqslant \| z \|_{C_T
  M^{\frac{1}{2} - \varepsilon'}} \| v \|_{C_T H^{\beta}}^2$ ;
  
  3. $\| \mathcal{N}_6 (v) \|_{C_T H^{- \alpha}} \leqslant \| : z^2 : \|_{C_T
  M^{\frac{1}{2} - \varepsilon'}} \| v \|_{C_T H^{\beta}}$ ;
  
  4. $\| \mathcal{N}_7 (v) \|_{C_T H^{- \alpha}} \leqslant \| : z^2 : \|_{C_T
  M^{\frac{1}{2} - \varepsilon'}} \| v \|_{C_T H^{\beta}}$ ;
  
  5. $\| \mathcal{N}_2 (v) -\mathcal{N}_2 (w) \|_{C_T H^{- \alpha}} \leqslant
  \| z \|_{C_T M^{\frac{1}{2} - \varepsilon'}} \| v - w \|_{C_T H^{\beta}} (\|
  v \|_{C_T H^{\beta}} + \| w \|_{C_T H^{\beta}})$ ;
  
  6. $\| \mathcal{N}_4 (v) -\mathcal{N}_4 (w) \|_{C_T H^{- \alpha}} \leqslant
  \| z \|_{C_T M^{\frac{1}{2} - \varepsilon'}} \| v - w \|_{C_T H^{\beta}} (\|
  v \|_{C_T H^{\beta}} + \| w \|_{C_T H^{\beta}})$ ;
  
  7. $\| \mathcal{N}_6 (v) -\mathcal{N}_6 (w) \|_{C_T H^{- \alpha}} \leqslant
  \| : z^2 : \|_{C_T M^{\frac{1}{2} - \varepsilon'}} \| v - w \|_{C_T
  H^{\beta}}$ ;
  
  8. $\| \mathcal{N}_7 (v) -\mathcal{N}_7 (w) \|_{C_T H^{- \alpha}} \leqslant
  \| : z^2 : \|_{C_T M^{\frac{1}{2} - \varepsilon'}} \| v - w \|_{C_T
  H^{\beta}}$ .
\end{lemma}

\begin{proof}
The Lemma follows  by applying the  multiplicative inequalities in Lemma~\ref{lem:multiplicative_inequalities} to each of the products involved in the definitions 
of the $\mathcal{N}_i$.
\end{proof}

\begin{lemma}\label{lem:N3Estimate}
  Assume $z \in C_T M^{\frac{1}{2} - \varepsilon'}$ and $w, v \in C_T
  H^{\frac{1}{2} - \varepsilon}$. We have the following inequalities for
  $\mathcal{N}_3$ with $\frac{1}{4} + \frac{\varepsilon'}{2} < \beta \leqslant
  \frac{1}{2} - \varepsilon$ and $\alpha \leqslant \beta - \frac{1}{4} -
  \frac{\varepsilon'}{2}$:
  
  1. $\| \mathcal{N}_3 (v) \|_{C_T H^{\alpha}} \lesssim \| z \|_{C_T
  M^{\frac{1}{2} - \varepsilon'}} \| v \|_{C_T H^{\beta}}^2$;
  
  2. $\| \mathcal{N}_3 (v) -\mathcal{N}_3 (w) \|_{C_T H^{\alpha}} \lesssim \| z
  \|_{C_T M^{\frac{1}{2} - \varepsilon'}} \| v - w \|_{C_T H^{\beta}} (\| v
  \|_{C_T H^{\beta}} + \| w \|_{C_T H^{\beta}})$.
\end{lemma}

\begin{proof}
  The first one follows from
  \begin{align*}
 \| \mathcal{N}_3 (v (t)) \|_{H^{\alpha}}^2
    & =  \sum_{m, n \geqslant 0} A_{m n}^{2 \alpha} \Big| \sum_{k, l
      \geqslant 0} v_{m k} (t) z_{k l} (t) v_{l n} (t) \Big|^2
    \\
    & \leqslant  \| z (t) \|_{M^{\frac{1}{2} - \varepsilon'}}^2 \sum_{m, n
    \geqslant 0} A_{m n}^{2 \alpha} \Bigg( \sum_{k, l \geqslant 0} \frac{|
    v_{m k} (t) | | v_{l n} (t) |}{A_{k l}^{\frac{1}{2} - \varepsilon'}}
  \Bigg)^2
  \\
    & \lesssim  \| z (t) \|_{M^{\frac{1}{2} - \varepsilon'}}^2 \sum_{m, n
    \geqslant 0} A_{m m}^{2 \alpha} A_{n n}^{2 \alpha} \Bigg( \sum_{k, l
    \geqslant 0} \frac{| v_{m k} (t) | | v_{l n} (t) |}{A_{k k}^{\frac{1}{4} -
    \frac{\varepsilon'}{2}} A_{l l}^{\frac{1}{4} - \frac{\varepsilon'}{2}}}
\Bigg)^2
\\
    & =  \| z (t) \|_{M^{\frac{1}{2} - \varepsilon'}}^2 \sum_{m \geqslant 0}
    A_{m m}^{2 \alpha} \Bigg( \sum_{k \geqslant 0} \frac{A_{m k}^{\beta} |
    v_{m k} (t) |}{A_{k k}^{\frac{1}{4} - \frac{\varepsilon'}{2}} A_{m
    k}^{\beta}} \Bigg)^2 \sum_{n \geqslant 0} A_{n n}^{2 \alpha} \Bigg(
    \sum_{l \geqslant 0} \frac{A_{l n}^{\beta} | v_{l n} (t) |}{A_{l
        l}^{\frac{1}{4} - \frac{\varepsilon'}{2}} A_{l n}^{\beta}} \Bigg)^2
    \\
    & \leqslant  \| z (t) \|_{M^{\frac{1}{2} - \varepsilon'}}^2 \Bigg[
    \sum_{m \geqslant 0} A_{m m}^{2 \alpha} \Big( \sum_{k \geqslant 0}
    \frac{1}{A_{k k}^{\frac{1}{2} - \varepsilon'} A_{m k}^{2 \beta}} \Big)
    \Big( \sum_{k' \geqslant 0} A_{m k'}^{2 \beta} | v_{m k'} (t) |^2 \Big)
    \Bigg]^2
    \\
    & \leqslant  \| z (t) \|_{M^{\frac{1}{2} - \varepsilon'}}^2 \Bigg[
    \sum_{m \geqslant 0} \frac{1}{A_{m m}^{- 2 \alpha + 2 \beta - \frac{1}{2}
    - \varepsilon'}} \Big( \sum_{k' \geqslant 0} A_{m k'}^{2 \beta} | v_{m
    k'} (t) |^2 \Big) \Bigg]^2\\
    & \leqslant  \| z (t) \|_{M^{\frac{1}{2} - \varepsilon'}}^2 \Bigg[
    \sum_{m \geqslant 0} \Big( \sum_{k' \geqslant 0} A_{m k'}^{2 \beta} |
    v_{m k'} (t) |^2 \Big) \Bigg]^2 = \| z (t) \|_{M^{\frac{1}{2} -
    \varepsilon'}}^2 \| v (t) \|_{H^{\beta}}^4.
\end{align*}
We have used the simple inequality $A_{m n} \leqslant A_{m m} A_{n n} \leqslant
  A_{m n}^2$, Cauchy Schwarz and one of correlation inequalities in the
  appendix; here we require $2 \beta + \frac{1}{2} - \varepsilon' > 1$ and $-
  2 \alpha + 2 \beta - \frac{1}{2} - \varepsilon' > 0$. After taking supremum
  over $t \in [0, T]$ one gets the result. This argument also shows
  \[ \| v z w \|_{C_T H^{\alpha}} \lesssim \| z \|_{C_T M^{\frac{1}{2} -
     \varepsilon'}} \| v \|_{C_T H^{\frac{1}{2} - \varepsilon}} \| w \|_{C_T
   H^{\frac{1}{2} - \varepsilon}}
\]
  and with the same argument as before one can show the second inequality of
  the lemma.
\end{proof}

Now for $\mathcal{N}_5$, which we regard as a random linear operator
$\mathcal{N}_5 (t) : w \rightarrow z (t) w z (t)$ for any test matrix $w$. We
have the following estimation of the operator norm of $\mathcal{N}_5 (t)$.

\begin{lemma}\label{lem:N5-estimate}
  The norm of the linear operator $\mathcal{N}_5 (t) : H^{\alpha} \rightarrow
  H^{\beta}$ satisfies the following estimate
  \[ 
\| \mathcal{N}_5 (t) \|_{\mathcal{L} (H^{\alpha} ; H^{\beta})} \leqslant
     \Bigg( \sum_{k, l, \bar{k}, \bar{l}} \frac{1}{A_{k l}^{2 \alpha}
     A_{\bar{k}  \bar{l}}^{2 \alpha}} \left| \sum_{m, n} A_{m n}^{2 \beta}
     z_{m k} (t) z_{l n} (t) z_{n \bar{l}} (t) z_{\bar{k} m} (t) \right|^2
     \Bigg)^{1 / 4} ,
\]
  and we have
  \[ 
\mathbb{E} \Big[\| \mathcal{N}_5 (t) \|_{\mathcal{L} (H^{\alpha} ;
     H^{\beta})}^p\Big]^{1 / p} 
\lesssim_p \mathbb{E} \Bigg[ \sum_{k, l, \bar{k},
     \bar{l}} \frac{1}{A_{k l}^{2 \alpha} A_{\bar{k}  \bar{l}}^{2 \alpha}}
     \bigg| \sum_{m, n} A_{m n}^{2 \beta} z_{m k} (t) z_{l n} (t) z_{n
     \bar{l}} (t) z_{\bar{k} m} (t) \bigg|^2 \Bigg]^{1 / 4} \]
  for any $p \geqslant 4$.
\end{lemma}

\begin{proof}
  Assume $w \in H^{\alpha}$ is a fixed test matrix, then
  \begin{align*}
    \| \mathcal{N}_5 (t) w \|_{H^{\beta}}^2 
& =  \sum_{m, n = 0}^{+ \infty}
    A_{m n}^{2 \beta} | (\mathcal{N}_5 (t) w)_{m n} |^2
\\
    & =  \sum_{k, l, \bar{k}, \bar{l}} w_{k l} w_{\bar{k}  \bar{l}} \sum_{m,
    n} A_{m n}^{2 \beta} z_{m k} (t) z_{l n} (t) z_{n \bar{l}} (t) z_{\bar{k}
    m} (t)
  \\
    & =  \sum_{k, l, \bar{k}, \bar{l}} A_{k l}^{\alpha} w_{k l} A_{\bar{k} 
    \bar{l}}^{\alpha} w_{\bar{k}  \bar{l}} \frac{1}{A_{k l}^{\alpha}
    A_{\bar{k}  \bar{l}}^{\alpha}} \sum_{m, n} A_{m n}^{2 \beta} z_{m k} (t)
  z_{l n} (t) z_{n \bar{l}} (t) z_{\bar{k} m} (t)
  \\
    & \leqslant  \Bigg( \sum_{k, l, \bar{k}, \bar{l}} A_{k l}^{2 \alpha} |
    w_{k l} |^2 A_{\bar{k}  \bar{l}}^{2 \alpha} | w_{\bar{k}  \bar{l}} |^2
  \Bigg)^{1 / 2} 
  \Bigg( \sum_{k, l, \bar{k}, \bar{l}} \frac{1}{A_{k l}^{2 \alpha}
    A_{\bar{k}  \bar{l}}^{2 \alpha}} \bigg| \sum_{m, n} A_{m n}^{2 \beta} z_{m
    k} (t) z_{l n} (t) z_{n \bar{l}} (t) z_{\bar{k} m} (t) \bigg|^2
\Bigg)^{1 / 2}
\\
  & =  \| w \|_{H^{\alpha}}^2 \Bigg( \sum_{k, l, \bar{k}, \bar{l}}
    \frac{1}{A_{k l}^{2 \alpha} A_{\bar{k}  \bar{l}}^{2 \alpha}} \bigg|
    \sum_{m, n} A_{m n}^{2 \beta} z_{m k} (t) z_{l n} (t) z_{n \bar{l}} (t)
    z_{\bar{k} m} (t) \bigg|^2 \Bigg)^{1 / 2},
\end{align*}
where we used Cauchy Schwarz. This shows
\[
  \| \mathcal{N}_5 (t) \|_{\mathcal{L} (H^{\alpha} ; H^{\beta})} \leqslant
     \Bigg( \sum_{k, l, \bar{k}, \bar{l}} \frac{1}{A_{k l}^{2 \alpha}
     A_{\bar{k}  \bar{l}}^{2 \alpha}} \bigg| \sum_{m, n} A_{m n}^{2 \beta}
     z_{m k} (t) z_{l n} (t) z_{n \bar{l}} (t) z_{\bar{k} m} (t) \bigg|^2
     \Bigg)^{1 / 4} .
   \]
  For the second statement, we use Minkowski inequality and Gaussian
  hypercontractivity (see appendix \ref{Hyper}) to get
  \begin{align*}
   \mathbb{E} [\| \mathcal{N}_5 (t) \|_{\mathcal{L} (H^{\alpha} ;
    H^{\beta})}^p]^{1 / p}
    & \leqslant  \mathbb{E} \left[ \Bigg( \sum_{k, l, \bar{k}, \bar{l}}
    \frac{1}{A_{k l}^{2 \alpha} A_{\bar{k}  \bar{l}}^{2 \alpha}} \bigg|
    \sum_{m, n} A_{m n}^{2 \beta} z_{m k} (t) z_{l n} (t) z_{n \bar{l}} (t)
    z_{\bar{k} m} (t) \bigg|^2 \Bigg)^{p / 4} \right]^{1 / p}
\\
    & \leqslant  \left( \sum_{k, l, \bar{k}, \bar{l}} \frac{1}{A_{k l}^{2
    \alpha} A_{\bar{k}  \bar{l}}^{2 \alpha}} \mathbb{E} \Bigg[ \Bigg( \bigg|
    \sum_{m, n} A_{m n}^{2 \beta} z_{m k} (t) z_{l n} (t) z_{n \bar{l}} (t)
    z_{\bar{k} m} (t) \bigg|^2 \Bigg)^{p / 4} \Bigg]^{4 / p}
  \right)^{1/4}
\\
    & \lesssim_p  \mathbb{E} \Bigg[ \sum_{k, l, \bar{k}, \bar{l}}
    \frac{1}{A_{k l}^{2 \alpha} A_{\bar{k}  \bar{l}}^{2 \alpha}} \bigg|
    \sum_{m, n} A_{m n}^{2 \beta} z_{m k} (t) z_{l n} (t) z_{n \bar{l}} (t)
    z_{\bar{k} m} (t) \bigg|^2 \Bigg]^{1 / 4}
\end{align*}
for $p \geqslant 4$.
\end{proof}

The next lemma shows that this operator norm bound is 
almost surely finite. 
In order to carry out the involved stochastic computations a graphical algorithm is introduced. The analysis of the individual graphs is carried out in Appendix~\ref{Eappendix}.
\begin{lemma}
  For $\alpha = \frac{1}{2} - \varepsilon$ and $\beta = 0 - \varepsilon -
  \varepsilon'$, where $\varepsilon, \varepsilon'$ are positive small numbers,
  the value
  \[
    \mathbb{E} \Bigg[ \sum_{k, l, \bar{k}, \bar{l}} \frac{1}{A_{k l}^{2
     \alpha} A_{\bar{k}  \bar{l}}^{2 \alpha}} \bigg| \sum_{m, n} A_{m n}^{2
     \beta} z_{m k} (t) z_{l n} (t) z_{n \bar{l}} (t) z_{\bar{k} m} (t)
 \bigg|^2 \Bigg] \leqslant C
\]
  is bounded by some time-independent constant $C$.
\begin{proof}
First we change the form of the objects we want to estimate into the following
form:
\begin{align*}
&\mathbb{E} \Bigg[ \sum_{k, l, \bar{k}, \bar{l}} \frac{1}{A_{k l}^{2
  \alpha} A_{\bar{k}  \bar{l}}^{2 \alpha}} \bigg| \sum_{m, n} A_{m n}^{2
  \beta} z_{m k} (t) z_{l n} (t) z_{n \bar{l}} (t) z_{\bar{k} m} (t) \bigg|^2
\Bigg]
\\
  & =  \sum_{k, l, \bar{k}, \bar{l}} \frac{1}{A_{k l}^{2 \alpha} A_{\bar{k} 
  \bar{l}}^{2 \alpha}} \sum_{m, n, \overline{m}, \overline{n}} \mathbb{E}
  [A_{m n}^{2 \beta} z_{m k} z_{l n} z_{n \bar{l}} z_{\bar{k} m} A_{\bar{m} 
  \bar{n}}^{2 \beta} z_{\overline{m} \overline{k} } z_{\overline{l}
  \overline{n}} z_{\overline{n} l} z_{k \bar{m}}]
\\
  & =  \sum_{k, l, \bar{k}, \bar{l}, m, n, \overline{m}, \overline{n}}
  \frac{\mathbb{E} [z_{m k} z_{k \bar{m}} z_{\overline{m} \overline{k} }
  z_{\bar{k} m} z_{n \bar{l}} z_{\overline{l} \overline{n}} z_{\overline{n} l}
  z_{l n}]}{A_{k l}^{2 \alpha} A_{\bar{k}  \bar{l}}^{2 \alpha} A_{m n}^{- 2
  \beta} A_{\bar{m}  \bar{n}}^{- 2 \beta}}.
\end{align*}
Since $z (t)$ is Gaussian, the last expression can be expanded by
using Wick's theorem. By stationarity of $z (t)$, such contractions
are time-independent, so we could ignore the time variable $t$. In
order to show this is finite with $\alpha = \frac{1}{2} - \varepsilon$
and $\beta = 0 - \varepsilon - \varepsilon'$, we employ graphical
techniques.

We represent the matrix element $z_{m n}$ as follows
\[
  \resizebox{0.3\columnwidth}{!}{\includegraphics{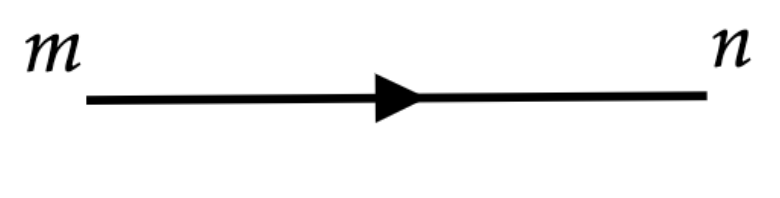}}
\]
where the arrow indicates which index is row index, which one is column index.
Putting the vertices with same indices together, the expectation $\mathbb{E}
[z_{m k} z_{k \bar{m}} z_{\overline{m} \overline{k} } z_{\bar{k} m} z_{n
\bar{l}} z_{\overline{l} \overline{n}} z_{\overline{n} l} z_{l n}]$ can be
represented as
\[
  \resizebox{0.3\columnwidth}{!}{\includegraphics{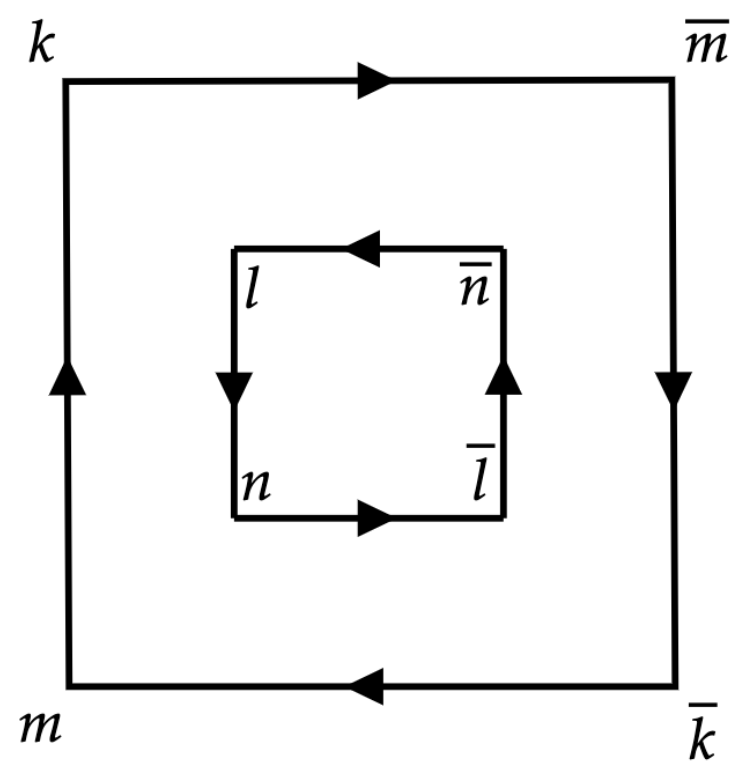}}
\]
After summing over indices, using colored lines to indicate and
distinguish weights $A_{k l}^{2 \alpha}, A_{\bar{k}  \bar{l}}^{2 \alpha}, A_{m
n}^{- 2 \beta}, A_{\bar{m}  \bar{n}}^{- 2 \beta}$, we have following basic
graph
\[
  \resizebox{0.35\columnwidth}{!}{\includegraphics{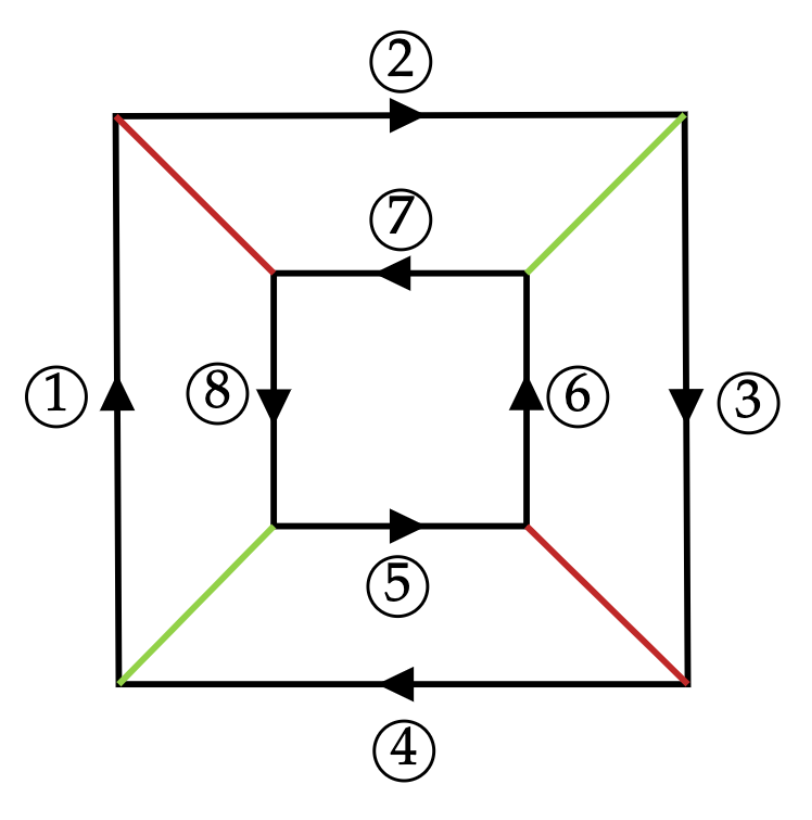}}
\]
Here red edges represents weights $2 \alpha$ and green edges represents
weights $- 2 \beta$. For future simplicity we also labeled in the same 
order as in the expectation. When we do Wick contractions, we use correlation
function $\langle z_{m \nospace n} z_{k \nospace l} \rangle = \frac{\delta_{m
\nospace l} \delta_{n \nospace k}}{A_{m \nospace n}}$, which introduce as
cancellation rule that an black directed edge should be contracted with
another one in opposite direction. There are in total 105 different ways to do
contraction and the following reduction algorithm is the way to check that all
of them are finite systematically.

The first step is to do Wick contraction as described above, and to replace the
resulting multi-connected graph as a weighted graph. The rule is the black
edge has weight $1$, the red edge has weight $2 \alpha$ and the green edge has
weight $- 2 \beta$. Here is an example:
\begin{align}
  \resizebox{1\columnwidth}{!}{\includegraphics{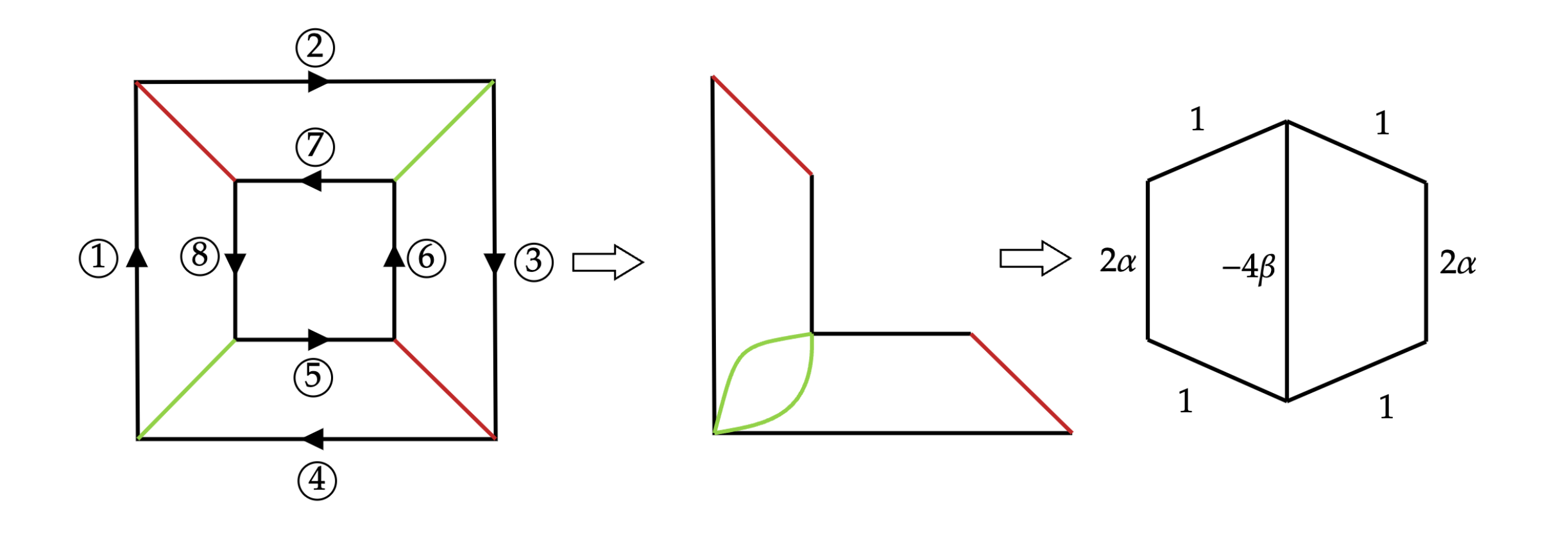}}
\label{example-105a}
\end{align}
which comes from the contraction of $(12) (34) (56) (78)$. To estimate the
result represented by the weighted graph on the left hand side, we have
the following rules:

\medskip
\paragraph{\bf Rule 0:}
When there are more than one edge connecting two vertices,
then replace them by a weighted one with weight equal to the sum of each
individual weights.

\medskip
\paragraph{\bf Rule 1:}
\[
  \resizebox{1\columnwidth}{!}{\includegraphics{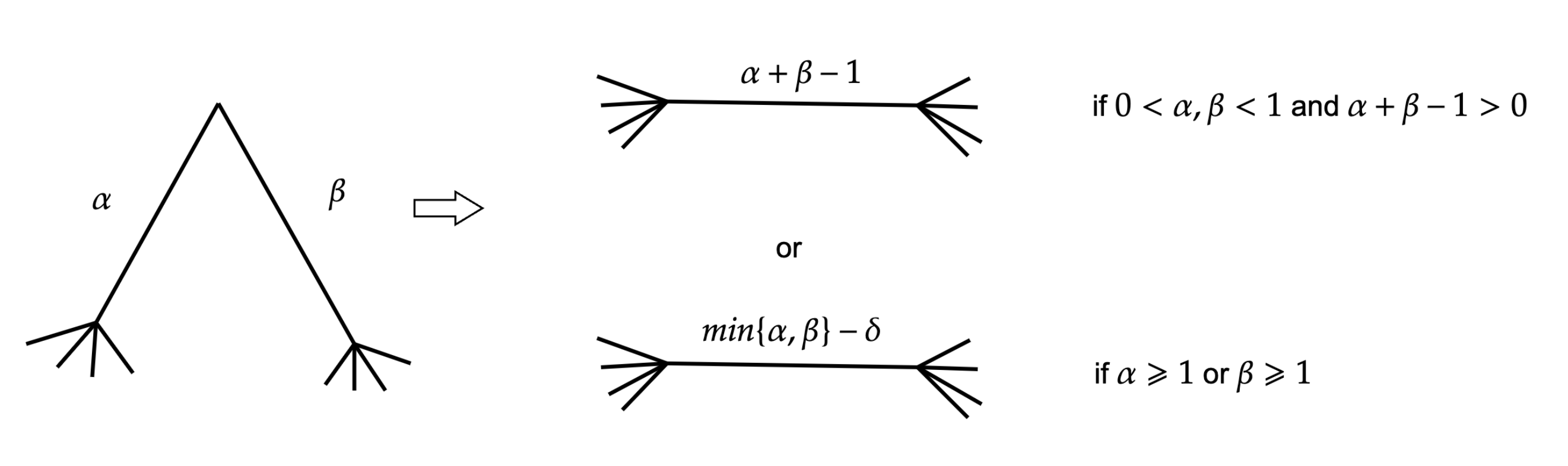}}
\]
which represents the inequalities
\begin{enumerate}

\item if $\alpha, \beta \in (0, 1)$ and $\alpha + \beta - 1 > 0$, then
\[ \sum_{k = 0}^{\infty} \frac{1}{A_{m \nospace k}^{\alpha} A_{k \nospace
   n}^{\beta}} \lesssim \frac{1}{A_{m \nospace n}^{\alpha + \beta - 1}} ; \]

\item  if $\alpha \geqslant 1$ or $\beta \geqslant 1$, then for any small
positive number $\delta$ we have
\[ \sum_{k = 0}^{\infty} \frac{1}{A_{m \nospace k}^{\alpha} A_{k \nospace
   n}^{\beta}} \lesssim \frac{1}{A^{\min \{ \alpha, \beta \} - \delta}_{m
   \nospace n}} .
\]
\end{enumerate}

\paragraph{\bf Rule 2:}
\[
  \resizebox{1\columnwidth}{!}{\includegraphics{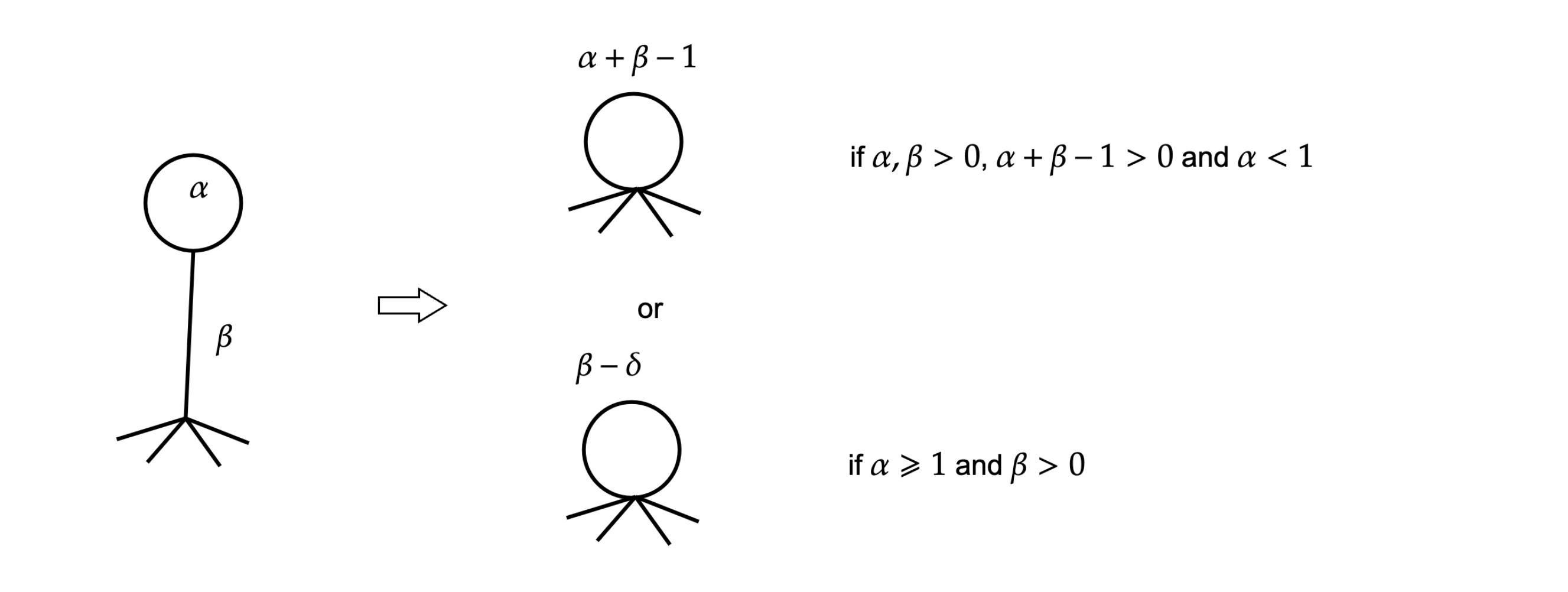}}
\]
which represents the inequalities:

\begin{enumerate}\setcounter{enumi}{2}

\item if $\alpha, \beta > 0$, $\alpha + \beta - 1 > 0$ and $\alpha < 1$, then
\[ \sum_{m = 0}^{\infty} \frac{1}{A_{m \nospace m}^{\alpha} A_{m \nospace
   n}^{\beta}} \lesssim \frac{1}{A_{n \nospace n}^{\alpha + \beta - 1}} ; \]

\item if $\beta > 0$ and $\alpha \geqslant 1$ then
\[ \sum_m \frac{1}{A^{\alpha}_{m m} A_{m n}^{\beta}} \lesssim \frac{1}{A_{n
      n}^{\beta - \delta}} .
\]
\end{enumerate}

\paragraph{\bf Rule 3:}
\[
  \resizebox{1\columnwidth}{!}{\includegraphics{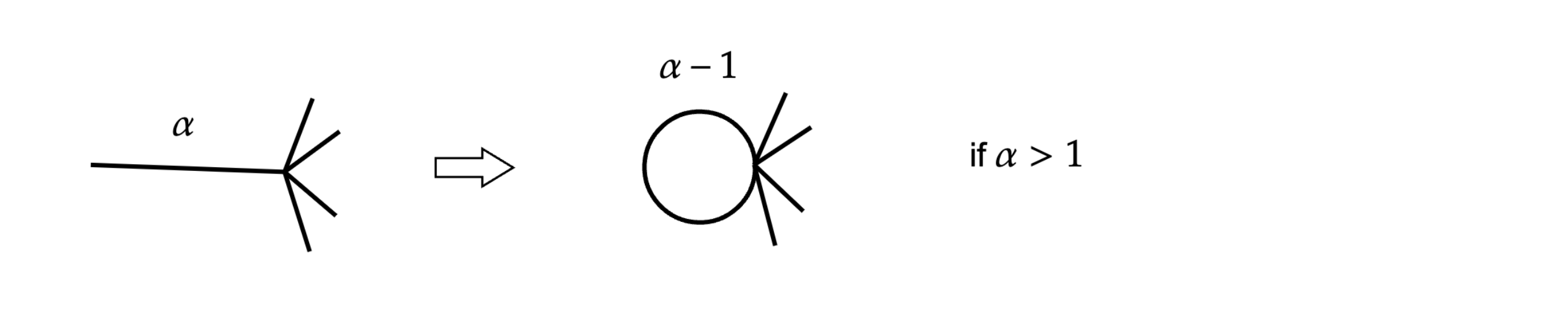}}
\]
which represents:
\begin{enumerate}\setcounter{enumi}{4}
\item if $\alpha > 1$, then $\sum_{m = 0}^{\infty} \frac{1}{A_{m \nospace
n}^{\alpha}} \sim \frac{1}{A_{n \nospace n}^{\alpha - 1}}$.
\end{enumerate}

\paragraph{\bf Rule 4:}
\[
  \resizebox{1\columnwidth}{!}{\includegraphics{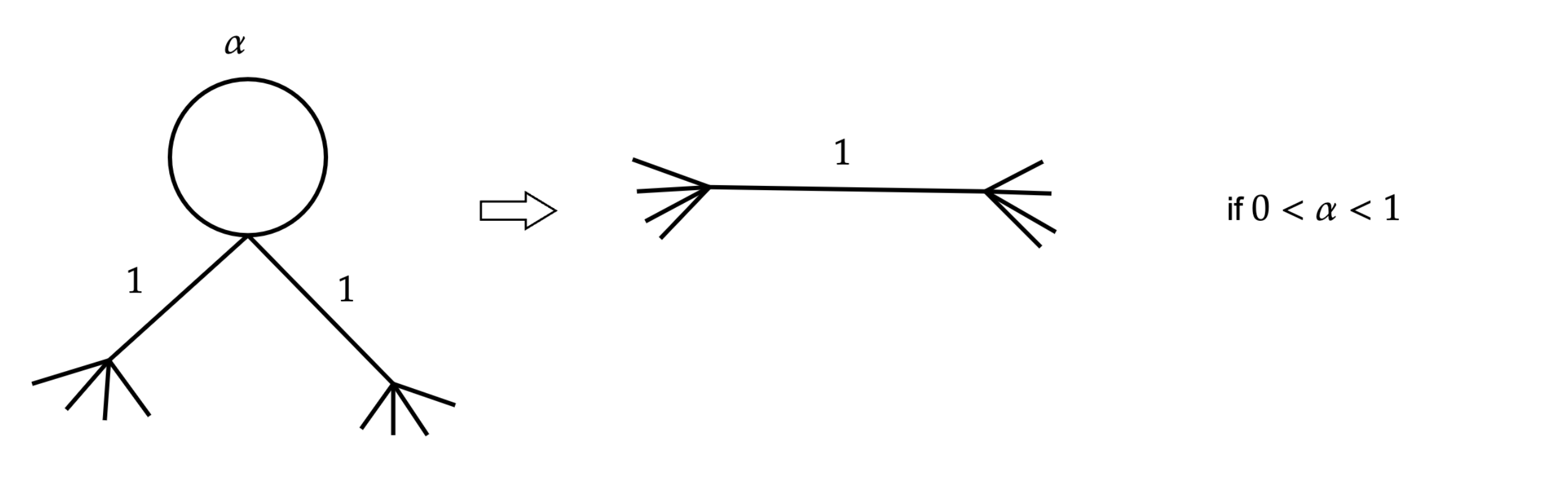}}
\]
which represents the inequality:
\begin{enumerate}\setcounter{enumi}{5}
\item if $\alpha \in (0, 1)$, then
\[ \sum_{k = 0}^{\infty} \frac{1}{A_{m \nospace k} A_{k \nospace k}^{\alpha}
    A_{k \nospace n}} \lesssim \frac{1}{A_{m \nospace n}} .
\]
\end{enumerate}

\paragraph{\bf Rule 5:}
\[
  \resizebox{1\columnwidth}{!}{\includegraphics{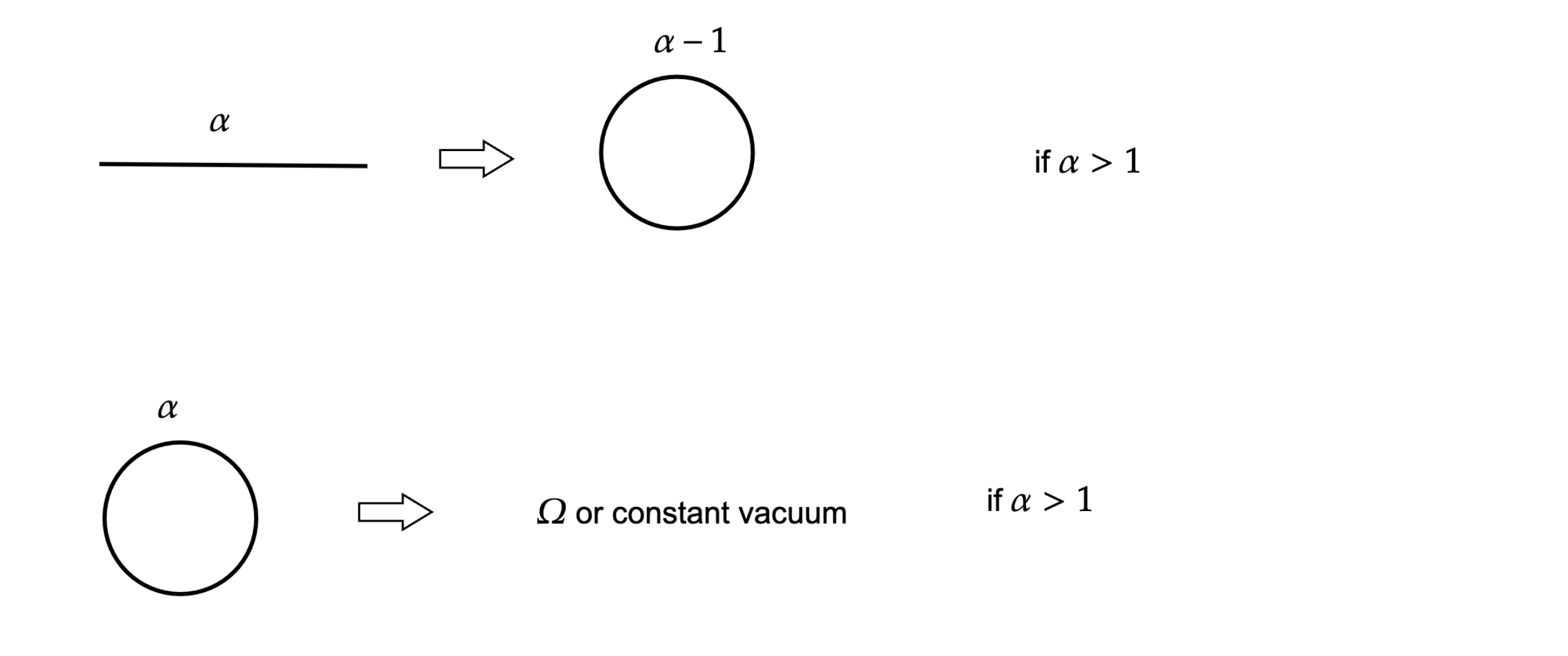}}
\]
which are simply:
\begin{enumerate}\setcounter{enumi}{6}
\item  if $\alpha > 1$, then $\sum_{m, n = 0}^{\infty} \frac{1}{A_{m \nospace
n}^{\alpha}} \sim \sum_{n = 0}^{\infty} \frac{1}{A_{n \nospace n}^{\alpha -
1}}$;

\item if $\alpha > 1$, then $\sum_{n = 0}^{\infty} \frac{1}{A_{n \nospace
n}^{\alpha}}$ is finite.
\end{enumerate}

Taking the previous example (\ref{example-105a}),
with $\alpha = \frac{1}{2} - \varepsilon$, $\beta = 0
- \varepsilon - \varepsilon'$ in mind, and choosing
$\delta < \varepsilon'$, the
reduction algorithm is done in the following way
\[
  \resizebox{1\columnwidth}{!}{\includegraphics{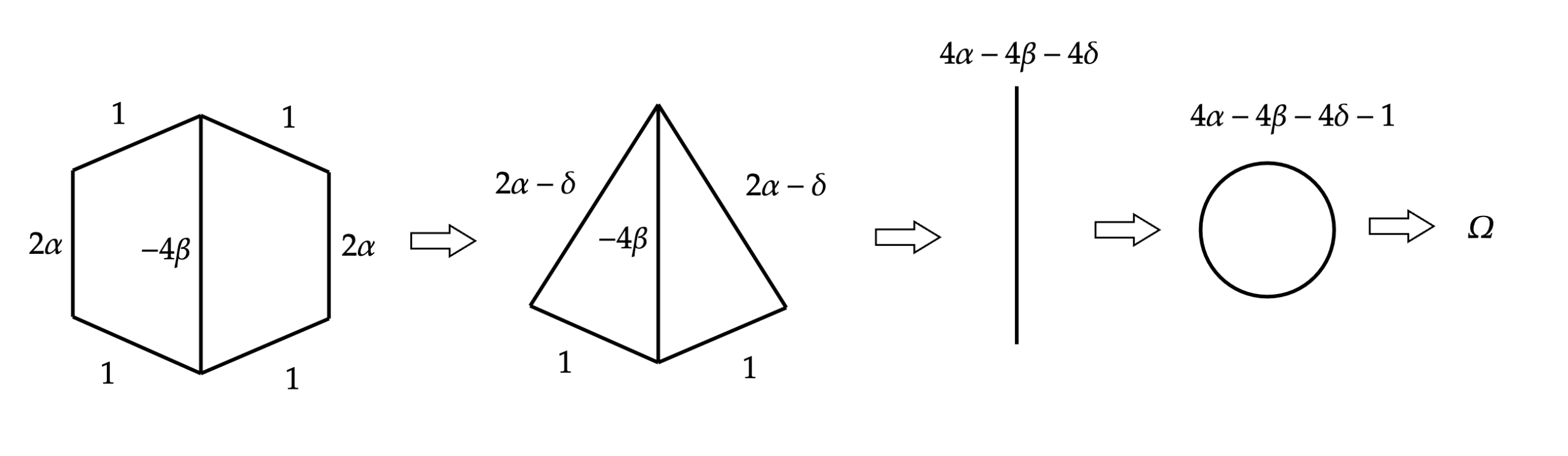}}
\]
which simply means the Wick contraction
\[ \sum_{k, l, \bar{k}, \bar{l}, m, n, \overline{m}, \overline{n}}
   \frac{\mathbb{E} [z_{m k} z_{k \bar{m}}] \mathbb{E} [z_{\overline{m}
   \overline{k} } z_{\bar{k} m}] \mathbb{E} [z_{n \bar{l}} z_{\overline{l}
   \overline{n}}] \mathbb{E} [z_{\overline{n} l} z_{l n}]}{A_{k l}^{2 \alpha}
   A_{\bar{k}  \bar{l}}^{2 \alpha} A_{m n}^{- 2 \beta} A_{\bar{m}  \bar{n}}^{-
   2 \beta}} < \infty \]
is finite. The complete verification of all 105 different contractions is done
in the in appendix \ref{Eappendix}, and this concludes the proof of the lemma.
\end{proof}
\end{lemma}

\section{Local Existence for Stochastic
  Quantization Equation}\label{LocalESQ}


We are going to solve the integral equation (\ref{DPD-fixeq}),
\[
  v_{m n} (t) = e^{- A_{m n} t} v_{m n} (0) - \frac{\theta \lambda}{4}
   \int_0^t e^{- A_{m n} (t - s)} \Big( \sum_{i = 1}^7 \mathcal{N}_i (v)_{m
       n} (s) + : z^3 :_{m n} (s) \Big)  ds ,
\]
in the space $K_T^{\frac{1}{2} - \varepsilon}$ with initial data $v (0) \in H^0$.
See Section~\ref{notation} for the definition of these spaces.
Denote the integral operator in (\ref{DPD-fixeq}) by
\begin{align}
\Phi_i (v)_{m n} \assign \int_0^t e^{- A_{m n} (t - s)} \mathcal{N}_i
   (v(s))_{m n}  d s, \quad 
\Phi (v)_{m n} \assign \sum_{i = 1}^7  \Phi_i (v)_{m n},
\end{align}
where the operators $\mathcal{N}_i$ were introduced in (\ref{Terms}) and
(\ref{Terms-2}). We have
\begin{eqnarray*}
  \| e^{- A t} v (0) \|^2_{H^{\frac{1}{2} - \varepsilon}} & = & \sum_{m, n
  \geqslant 0} A_{m n}^{1 - 2 \varepsilon} e^{- 2 A_{m \nospace n} t} | v_{m
  n} (0) |^2\\
  & = & t^{- (1 - 2 \varepsilon)} \sum_{m, n \geqslant 0} (A_{m n} t)^{1 - 2
  \varepsilon} e^{- 2 A_{m \nospace n} t} | v_{m n} (0) |^2\\
  & \lesssim & t^{- (1 - 2 \varepsilon)} \| v (0) \|_{H^0}^2,
\end{eqnarray*}
which means 
$t^{\frac{1}{2} -  \varepsilon}
\| e^{- A t} v (0) \|_{H^{\frac{1}{2} - \varepsilon}} 
 \lesssim  \| v (0) \|_{H^0}$. Taking the definition 
of  $K^{\beta}_T$ in Section~\ref{notation} into account, we conclude
$\| e^{- A t} v (0) \|_{K_T^{\frac{1}{2} - \varepsilon}} \lesssim \| v
(0) \|_{H^0}$ for any $t\in [0,T]$.

Using the Schauder type estimate and $: z^3 : \in C_T
H^{\frac{1}{2} - \varepsilon}$, we know $\int_0^{\cdot} e^{- A (\cdot - s)} :
z^3 : (s) d \nospace s \in K_T^{\frac{1}{2} - \varepsilon}$. For convenience,
denote $h_{m n} (t) \assign v_{m n} (t) - e^{- A_{m \nospace n} t} v_{m n}
(0)$ so that $h (0) = 0$. Then
\begin{align}
  h_{m n} (t) & =  - \frac{\theta \lambda}{4} \int_0^t e^{- A_{m n} (t - s)}
  \bigg( \sum_{i = 1}^7 \mathcal{N}_i (h(s) + e^{- A s} v (0))_{m n}  + : z^3
    :_{m n} (s) \bigg) d \nospace s
  \nonumber
  \\
  & =  - \frac{\theta \lambda}{4} \sum_{i = 1}^7 \Phi_i (h + e^{- A \cdot }
  v (0))_{m n}
  - \frac{\theta \lambda}{4} \int_0^t e^{- A_{m n} (t - s)} : z^3 :_{m n} (s)
  d s.
\end{align}
which defines the Picard iteration map
\begin{align}
  \Psi (h)_{m n} &= - \frac{\theta \lambda}{4} \bigg\{ \sum_{i = 1}^7
  \Phi_i (h + e^{- A \cdot } v (0))_{m n}
  + \int_0^t e^{- A_{m n} (t - s)} : z^3 :_{m n} (s) d s \bigg\} .
\end{align}
Using the Schauder estimate in Appendix \ref{spaceofmatrix}, we obtain the
following estimates. We denote $\Delta h \assign h_1 - h_2$, $\Delta \Psi (h)
\assign \Psi (h_1) - \Psi (h_2)$, $\Delta \mathcal{N}_i (h) \assign
\mathcal{N}_i (h_1 + e^{- A t} v (0)) -\mathcal{N}_i (h_2 + e^{- A t} v (0))$
and $\Delta \Phi_i (h) \assign \Phi_i (h_1 + e^{- A t} v (0)) - \Phi_i (h_2 +
e^{- A t} v (0))$. Then:

\begin{lemma}
  For $\Phi_1$ and $h \in K_T^{\frac{1}{2} - \varepsilon}$, we have
  \[ \| \Phi_1 (h + e^{- A \cdot} v (0)) \|_{K_T^{\frac{1}{2} - \varepsilon}}
     \lesssim T^{\varepsilon} \left( \| h \|_{K_T^{\frac{1}{2} - \varepsilon}}
     + \| e^{- A \cdot} v (0) \|_{K_T^{\frac{1}{2} - \varepsilon}} \right)^3
  \]
  and
  \[
    \| \Delta \Phi_1 (h) \|_{K_T^{\frac{1}{2} - \varepsilon}} \lesssim
     T^{\varepsilon} \| \Delta h \|_{K_T^{\frac{1}{2} - \varepsilon}} \sum_{k
     = 1, 2} \left( \| h_k \|_{K_T^{\frac{1}{2} - \varepsilon}} + \| e^{- A
     \cdot} v (0) \|_{K_T^{\frac{1}{2} - \varepsilon}} \right)^2 . \]
\end{lemma}

\begin{proof}
  By definition
  \begin{align*}
& \| \Phi_1 (h + e^{- A \cdot} v (0)) \|_{H^{\frac{1}{2} -
    \varepsilon}}
\\
& =  \left\| \int_0^t e^{- A (t - s)} \mathcal{N}_1 (h(s) + e^{- A s} v
    (0)) d \nospace s \right\|_{H^{- \frac{1}{2} + (1 - \varepsilon)}}
\\
&\leq 
\int_0^t
\Bigg(\sum_{m,n} A_{mn}^{2-2\epsilon}
 e^{- 2A_{mn} (t - s)} A_{mn}^{-1} \Big|\mathcal{N}_1 (h(s) + e^{- A s} v
    (0))_{mn}  \Big|^2 \Bigg)^{1/2}d  s 
\\
    & \lesssim  \int_0^t (t - s)^{- (1 - \varepsilon)} \| \mathcal{N}_1 (h +
    e^{- A s} v (0)) (s) \|_{H^{- \frac{1}{2}}} d \nospace s\\
    & \leqslant  \int_0^t (t - s)^{- (1 - \varepsilon)} \| \mathcal{N}_1 (h
    + e^{- A s} v (0)) (s) \|_{H^0} ds
    \\
    & \lesssim  \int_0^t (t - s)^{- (1 - \varepsilon)} s^{- (
    \frac{1}{2} - \varepsilon )} \| h (s) + e^{- A s} v (0) \|^2_{H^0}
    \left( s^{\frac{1}{2} - \varepsilon} \| h (s) + e^{- A s} v (0)
    \|_{H^{\frac{1}{2} - \varepsilon}} \right) d s
\\
    & \leqslant  \int_0^t (t - s)^{- (1 - \varepsilon)} s^{- \left(
    \frac{1}{2} - \varepsilon \right)} d \nospace s \left( \| h
    \|_{K_T^{\frac{1}{2} - \varepsilon}} + \| e^{- A \cdot} v (0)
    \|_{K_T^{\frac{1}{2} - \varepsilon}} \right)^3\\
    & =  t^{\varepsilon} t^{- \left( \frac{1}{2} - \varepsilon \right)}
    \int_0^1 (1 - s)^{- (1 - \varepsilon)} s^{- \left( \frac{1}{2} -
    \varepsilon \right)} d \nospace s \left( \| h \|_{K_T^{\frac{1}{2} -
    \varepsilon}} + \| e^{- A \cdot} v (0) \|_{K_T^{\frac{1}{2} -
    \varepsilon}} \right)^3\\
    & \lesssim  t^{- \left( \frac{1}{2} - \varepsilon \right)}
    T^{\varepsilon} \left( \| h \|_{K_T^{\frac{1}{2} - \varepsilon}} + \| e^{-
    A \cdot} v (0) \|_{K_T^{\frac{1}{2} - \varepsilon}} \right)^3,
\end{align*}
where Lemma~\ref{lemma-N1}.1 was used in the 6th line
and inequalities for $\|~ \|_{K_T^{\frac{1}{2} - \varepsilon}}$ in
the 7th line. 
This is rearranged to
\[
  t^{\frac{1}{2} - \varepsilon} \| \Phi_1 (h + e^{- A \cdot} v (0))
     \|_{H^{\frac{1}{2} - \varepsilon}} \lesssim T^{\varepsilon} \left( \| h
     \|_{K_T^{\frac{1}{2} - \varepsilon}} + \| e^{- A \cdot} v (0)
     \|_{K_T^{\frac{1}{2} - \varepsilon}} \right)^3 . \tag{*}
 \]
  By the same method
  \begin{align*}
\| \Phi_1 (h + e^{- A \cdot} v (0)) \|_{H^0}
    & =  \left\| \int_0^t e^{- A (t - s)} \mathcal{N}_1 (h(s) + e^{- A s} v
    (0))  ds \right\|_{H^{- (1 - \varepsilon) + (1 -
    \varepsilon)}}
\\
& \lesssim  \int_0^t (t - s)^{- (1 - \varepsilon)} \| \mathcal{N}_1
(h(s) +  e^{- A s} v (0)) \|_{H^{- (1 - \varepsilon)}} ds
\\
& \leqslant  \int_0^t (t - s)^{- (1 - \varepsilon)}
\| \mathcal{N}_1 (h(s) + e^{- A s} v (0))  \|_{H^0} d s
\\
  & \lesssim  \int_0^t (t - s)^{- (1 - \varepsilon)} \| h (s) + e^{- A s}
  v (0) \|^3_{H^0} d s
  \\
    & \leqslant  \int_0^t (t - s)^{- (1 - \varepsilon)} d \nospace s \Big(
    \| h \|_{K_T^{\frac{1}{2} - \varepsilon}} + \| e^{- A \cdot} v (0)
    \|_{K_T^{\frac{1}{2} - \varepsilon}} \Big)^3
  \\
    & =  t^{\varepsilon} \int_0^1 (1 - s)^{- (1 - \varepsilon)} d \nospace s
    \Big( \| h \|_{K_T^{\frac{1}{2} - \varepsilon}} + \| e^{- A \cdot} v (0)
    \|_{K_T^{\frac{1}{2} - \varepsilon}} \Big)^3\\
    & \lesssim  T^{\varepsilon} \left( \| h \|_{K_T^{\frac{1}{2} -
    \varepsilon}} + \| e^{- A \cdot} v (0) \|_{K_T^{\frac{1}{2} -
    \varepsilon}} \right)^3 .
\tag{**}
\end{align*}
Both estimates (*) and (**) combine to
\[
  \| \Phi_1 (h + e^{- A \cdot} v (0)) \|_{K_T^{\frac{1}{2} - \varepsilon}}
     \lesssim T^{\varepsilon} \Big( \| h \|_{K_T^{\frac{1}{2} - \varepsilon}}
     + \| e^{- A \cdot} v (0) \|_{K_T^{\frac{1}{2} - \varepsilon}} \Big)^3 .
  \]
  The proof of the estimate for
  $\Delta \Phi_1 (h) \|_{K_T^{\frac{1}{2} - \varepsilon}}$ is similar
  (see also the proof of Lemma~\ref{lemma-N1}.2).
\end{proof}

\begin{lemma}
  Suppose $z, : z^2 : \in C_T M^{\frac{1}{2} - \varepsilon'}$ and $h \in
  K_T^{\frac{1}{2} - \varepsilon}$, then
  
  1. $\| \Phi_2 (h + e^{- A \cdot} v (0)) \|_{K_T^{\frac{1}{2} - \varepsilon}}
  \lesssim T^{\varepsilon} \| z \|_{C_T M^{\frac{1}{2} - \varepsilon'}} \left(
  \| h \|_{K_T^{\frac{1}{2} - \varepsilon}} + \| e^{- A \cdot} v (0)
  \|_{K_T^{\frac{1}{2} - \varepsilon}} \right)^2$;
  
  2. $\| \Phi_4 (h + e^{- A \cdot} v (0)) \|_{K_T^{\frac{1}{2} - \varepsilon}}
  \lesssim T^{\varepsilon} \| z \|_{C_T M^{\frac{1}{2} - \varepsilon'}} \left(
  \| h \|_{K_T^{\frac{1}{2} - \varepsilon}} + \| e^{- A \cdot} v (0)
  \|_{K_T^{\frac{1}{2} - \varepsilon}} \right)^2$;
  
  3. $\| \Phi_6 (h + e^{- A \cdot} v (0)) \|_{K_T^{\frac{1}{2} - \varepsilon}}
  \lesssim T^{\varepsilon} \| : z^2 : \|_{C_T M^{\frac{1}{2} - \varepsilon'}}
  \left( \| h \|_{K_T^{\frac{1}{2} - \varepsilon}} + \| e^{- A \cdot} v (0)
  \|_{K_T^{\frac{1}{2} - \varepsilon}} \right)$;
  
  4. $\| \Phi_7 (h + e^{- A \cdot} v (0)) \|_{K_T^{\frac{1}{2} - \varepsilon}}
  \lesssim T^{\varepsilon} \| : z^2 : \|_{C_T M^{\frac{1}{2} - \varepsilon'}}
  \left( \| h \|_{K_T^{\frac{1}{2} - \varepsilon}} + \| e^{- A \cdot} v (0)
  \|_{K_T^{\frac{1}{2} - \varepsilon}} \right)$;
  
  5. $\| \Delta \Phi_2 (h) \|_{K_T^{\frac{1}{2} - \varepsilon}} \lesssim
  T^{\varepsilon} \| \Delta h \|_{K_T^{\frac{1}{2} - \varepsilon}} \sum_{k =
  1, 2} \left( \| h_k \|_{K_T^{\frac{1}{2} - \varepsilon}} + \| e^{- A \cdot}
  v (0) \|_{K_T^{\frac{1}{2} - \varepsilon}} \right)$;
  
  6. $\| \Delta \Phi_4 (h) \|_{K_T^{\frac{1}{2} - \varepsilon}} \lesssim
  T^{\varepsilon} \| \Delta h \|_{K_T^{\frac{1}{2} - \varepsilon}} \sum_{k =
  1, 2} \left( \| h_k \|_{K_T^{\frac{1}{2} - \varepsilon}} + \| e^{- A \cdot}
  v (0) \|_{K_T^{\frac{1}{2} - \varepsilon}} \right)$;
  
  7. $\| \Delta \Phi_6 (h) \|_{K_T^{\frac{1}{2} - \varepsilon}} \lesssim
  T^{\varepsilon} \| : z^2 : \|_{C_T M^{\frac{1}{2} - \varepsilon'}} \| \Delta
  h \|_{K_T^{\frac{1}{2} - \varepsilon}}$;
  
  8. $\| \Delta \Phi_7 (h) \|_{K_T^{\frac{1}{2} - \varepsilon}} \lesssim
  T^{\varepsilon} \| : z^2 : \|_{C_T M^{\frac{1}{2} - \varepsilon'}} \| \Delta
  h \|_{K_T^{\frac{1}{2} - \varepsilon}}$.
\end{lemma}

\begin{proof}
  We check $\Phi_2$ first, $\Phi_4$ follows from similar arguments. By
  definition
\begin{align*}
      & \| \Phi_2 (h + e^{- A \cdot} v (0)) \|_{H^{\frac{1}{2} -
    \varepsilon}}\\
    & =  \left\| \int_0^t e^{- A (t - s)} \mathcal{N}_2 (h + e^{- A s} v
    (0)) (s) d \nospace s \right\|_{H^{- \frac{1}{2} + (1 - \varepsilon)}}\\
    & \lesssim  \int_0^t (t - s)^{- (1 - \varepsilon)} \| \mathcal{N}_2 (h +
    e^{- A s} v (0)) (s) \|_{H^{- \frac{1}{2}}} d \nospace s\\
    & \leqslant  \int_0^t (t - s)^{- (1 - \varepsilon)} \| z (s)
    \|_{M^{\frac{1}{2} - \varepsilon'}} \| h (s) + e^{- A s} v (0) \|_{H^{2
    \varepsilon'}}^2 d \nospace s\\
    & \leqslant  \int_0^t (t - s)^{- (1 - \varepsilon)} \| z (s)
    \|_{M^{\frac{1}{2} - \varepsilon'}} \left( \| h (s) + e^{- A s} v (0)
    \|_{H^0}^{1 - \theta} \| h (s) + e^{- A s} v (0) \|_{H^{\frac{1}{2} -
    \varepsilon}}^{\theta} \right)^2 d \nospace s\\
    & \leqslant  \| z \|_{C_T M^{\frac{1}{2} - \varepsilon'}} \int_0^t (t -
    s)^{- (1 - \varepsilon)} s^{- 2 \frac{2 \varepsilon'}{\frac{1}{2} -
    \varepsilon} \left( \frac{1}{2} - \varepsilon \right)} d \nospace s \left(
    \| h \|_{K_T^{\frac{1}{2} - \varepsilon}} + \| e^{- A \cdot} v (0)
    \|_{K_T^{\frac{1}{2} - \varepsilon}} \right)^2\\
    & \sim  t^{\frac{1}{2} - 4 \varepsilon'} {t^{- \left( \frac{1}{2} -
    \varepsilon \right)}}  \| z \|_{C_T M^{\frac{1}{2} - \varepsilon'}} \left(
    \| h \|_{K_T^{\frac{1}{2} - \varepsilon}} + \| e^{- A \cdot} v (0)
    \|_{K_T^{\frac{1}{2} - \varepsilon}} \right)^2\\
    & \leqslant  T^{\frac{1}{2} - 4 \varepsilon'} {t^{- \left( \frac{1}{2} -
    \varepsilon \right)}}  \| z \|_{C_T M^{\frac{1}{2} - \varepsilon'}} \left(
    \| h \|_{K_T^{\frac{1}{2} - \varepsilon}} + \| e^{- A \cdot} v (0)
    \|_{K_T^{\frac{1}{2} - \varepsilon}} \right)^2.
\end{align*}
Here $\theta = \frac{2 \varepsilon'}{\frac{1}{2} - \varepsilon}$, and
we used the interpolation inequality. Similarly,
  \begin{align*}
  \| \Phi_2 (h + e^{- A \cdot} v (0)) \|_{H^0}
    & =  \left\| \int_0^t e^{- A (t - s)} \mathcal{N}_2 (h + e^{- A s} v
    (0)) (s) d \nospace s \right\|_{H^{- (1 - \varepsilon) + (1 -
    \varepsilon)}}\\
    & \lesssim  \int_0^t (t - s)^{- (1 - \varepsilon)} \| \mathcal{N}_2 (h +
    e^{- A s} v (0)) (s) \|_{H^{- (1 - \varepsilon)}} d \nospace s\\
    & \leqslant  \int_0^t (t - s)^{- (1 - \varepsilon)} \| z (s)
    \|_{M^{\frac{1}{2} - \varepsilon'}} \| h (s) + e^{- A s} v (0) \|_{H^0}^2
    d \nospace s\\
    & \leqslant  \int_0^t (t - s)^{- (1 - \varepsilon)} d \nospace s \| z
    \|_{C_T M^{\frac{1}{2} - \varepsilon'}} \left( \| h \|_{K_T^{\frac{1}{2} -
    \varepsilon}} + \| e^{- A \cdot} v (0) \|_{K_T^{\frac{1}{2} -
    \varepsilon}} \right)^2\\
    & \lesssim  T^{\varepsilon} \| z \|_{C_T M^{\frac{1}{2} - \varepsilon'}}
    \left( \| h \|_{K_T^{\frac{1}{2} - \varepsilon}} + \| e^{- A \cdot} v (0)
    \|_{K_T^{\frac{1}{2} - \varepsilon}} \right)^2.
\end{align*}
With assumption $T \leqslant 1$ we get 
\[
  \| \Phi_2 (h + e^{- A \cdot} v (0)) \|_{K_T^{\frac{1}{2} - \varepsilon}}
     \lesssim T^{\varepsilon} \| z \|_{C_T M^{\frac{1}{2} - \varepsilon'}}
     \left( \| h \|_{K_T^{\frac{1}{2} - \varepsilon}} + \| e^{- A \cdot} v (0)
       \|_{K_T^{\frac{1}{2} - \varepsilon}} \right)^2 .
  \]
  Next we consider $\Phi_6$:
  \begin{align*}
\| \Phi_6 (h + e^{- A \cdot} v (0)) \|_{H^{\frac{1}{2} -
    \varepsilon}}
    & =  \left\| \int_0^t e^{- A (t - s)} \mathcal{N}_6 (h + e^{- A s} v
    (0)) (s) d \nospace s \right\|_{H^{- \frac{1}{2} + (1 - \varepsilon)}}\\
    & \lesssim  \int_0^t (t - s)^{- (1 - \varepsilon)} \| \mathcal{N}_6 (h +
    e^{- A s} v (0)) (s) \|_{H^{- \frac{1}{2}}} d \nospace s\\
    & \leqslant  \int_0^t (t - s)^{- (1 - \varepsilon)} \| : z^2 :
    \|_{M^{\frac{1}{2} - \varepsilon'}} \| h (s) + e^{- A s} v (0)
    \|_{H^{\frac{1}{2} - \varepsilon}} d \nospace s\\
    & =  \int_0^t (t - s)^{- (1 - \varepsilon)} s^{- \left( \frac{1}{2} -
    \varepsilon \right)} \| : z^2 : \|_{M^{\frac{1}{2} - \varepsilon'}} \left(
    s^{\frac{1}{2} - \varepsilon} \| h (s) + e^{- A s} v (0)
    \|_{H^{\frac{1}{2} - \varepsilon}} \right) d \nospace s\\
    & \leqslant  \| : z^2 : \|_{C_T M^{\frac{1}{2} - \varepsilon'}} \int_0^t
    (t - s)^{- (1 - \varepsilon)} s^{- \left( \frac{1}{2} - \varepsilon
    \right)} d \nospace s \left( \| h \|_{K_T^{\frac{1}{2} - \varepsilon}} +
    \| e^{- A \cdot} v (0) \|_{K_T^{\frac{1}{2} - \varepsilon}} \right)\\
    & \lesssim  T^{\varepsilon} t^{- \left( \frac{1}{2} - \varepsilon
    \right)} \| : z^2 : \|_{C_T M^{\frac{1}{2} - \varepsilon'}} \left( \| h
    \|_{K_T^{\frac{1}{2} - \varepsilon}} + \| e^{- A \cdot} v (0)
    \|_{K_T^{\frac{1}{2} - \varepsilon}} \right)
\end{align*}
  and
  \begin{align*}
 \| \Phi_6 (h + e^{- A \cdot} v (0)) \|_{H^0}
    & =  \left\| \int_0^t e^{- A (t - s)} \mathcal{N}_6 (h + e^{- A s} v
    (0)) (s) d \nospace s \right\|_{H^{- (1 - \varepsilon) + (1 -
    \varepsilon)}}\\
    & \lesssim  \int_0^t (t - s)^{- (1 - \varepsilon)} \| \mathcal{N}_6 (h +
    e^{- A s} v (0)) (s) \|_{H^{- (1 - \varepsilon)}} d \nospace s\\
    & \leqslant  \int_0^t (t - s)^{- (1 - \varepsilon)} \| : z^2 :
    \|_{M^{\frac{1}{2} - \varepsilon'}} \| h (s) + e^{- A s} v (0) \|_{H^0} d
    \nospace s\\
    & \leqslant  \| : z^2 : \|_{C_T M^{\frac{1}{2} - \varepsilon'}} \int_0^t
    (t - s)^{- (1 - \varepsilon)} d \nospace s \left( \| h
    \|_{K_T^{\frac{1}{2} - \varepsilon}} + \| e^{- A \cdot} v (0)
    \|_{K_T^{\frac{1}{2} - \varepsilon}} \right)\\
    & \lesssim  T^{\varepsilon} \| : z^2 : \|_{C_T M^{\frac{1}{2} -
    \varepsilon'}} \left( \| h \|_{K_T^{\frac{1}{2} - \varepsilon}} + \| e^{-
    A \cdot} v (0) \|_{K_T^{\frac{1}{2} - \varepsilon}} \right).
\end{align*}
These combine to 
  \[
    \| \Phi_6 (h + e^{- A \cdot} v (0)) \|_{K_T^{\frac{1}{2} - \varepsilon}}
     \lesssim T^{\varepsilon} \| : z^2 : \|_{C_T M^{\frac{1}{2} -
     \varepsilon'}} \left( \| h \|_{K_T^{\frac{1}{2} - \varepsilon}} + \| e^{-
     A \cdot} v (0) \|_{K_T^{\frac{1}{2} - \varepsilon}} \right) .
\]

The estimate for $\Phi_7$ is similar, and the 
proofs for 5--8 are similar to 1--4.
\end{proof}

\begin{lemma}
  Suppose $z \in C_T M^{\frac{1}{2} - \varepsilon'}$ and $h \in
  K_T^{\frac{1}{2} - \varepsilon}$, then
  \[ \| \Phi_3 (h + e^{- A \cdot} v (0)) \|_{K_T^{\frac{1}{2} - \varepsilon}}
     \lesssim T^{\varepsilon} \| z \|_{C_T M^{\frac{1}{2} - \varepsilon'}}
     \left( \| h \|_{K_T^{\frac{1}{2} - \varepsilon}} + \| e^{- A \cdot} v (0)
     \|_{K_T^{\frac{1}{2} - \varepsilon}} \right)^2 \]
  and
  \[ \| \Delta \Phi_3 (h) \|_{K_T^{\frac{1}{2} - \varepsilon}} \lesssim
     T^{\varepsilon} \| \Delta h \|_{K_T^{\frac{1}{2} - \varepsilon}} \sum_{k
     = 1, 2} \left( \| h_k \|_{K_T^{\frac{1}{2} - \varepsilon}} + \| e^{- A
     \cdot} v (0) \|_{K_T^{\frac{1}{2} - \varepsilon}} \right) . \]
\end{lemma}

\begin{proof}
  By definition
  \begin{align*}
     & \| \Phi_3 (h + e^{- A \cdot} v (0)) \|_{H^{\frac{1}{2} -
    \varepsilon}}\\
    & =  \left\| \int_0^t e^{- A (t - s)} \mathcal{N}_3 (h + e^{- A s} v
    (0)) (s) d \nospace s \right\|_{H^{- \left( \frac{1}{2} - 2 \varepsilon' -
    \varepsilon \right) + (1 - 2 \varepsilon - 2 \varepsilon')}}\\
    & \lesssim  \int_0^t (t - s)^{- (1 - 2 \varepsilon - 2 \varepsilon')} \|
    \mathcal{N}_3 (h + e^{- A s} v (0)) (s) \|_{H^{- \left( \frac{1}{2} - 2
    \varepsilon' - \varepsilon \right)}} d \nospace s\\
    & \lesssim  \int_0^t (t - s)^{- (1 - 2 \varepsilon - 2 \varepsilon')} \|
    z (s) \|_{M^{\frac{1}{2} - \varepsilon'}} \| h (s) + e^{- A s} v (0)
    \|_{H^{\frac{1}{4} + \varepsilon'}}^2 d \nospace s\\
    & \leqslant  \int_0^t (t - s)^{- (1 - 2 \varepsilon - 2 \varepsilon')}
    \| z (s) \|_{M^{\frac{1}{2} - \varepsilon'}} \\
    &  \qquad \times  \left( \| h (s) + e^{- A s} v (0) \|_{H^0}^{\theta} \| h (s) + e^{-
    A s} v (0) \|_{H^{\frac{1}{2} - \varepsilon}}^{1 - \theta} \right)^2 d
    \nospace s\\
    & \leqslant  \| z \|_{C_T M^{\frac{1}{2} - \varepsilon'}} \int_0^t (t -
    s)^{- (1 - 2 \varepsilon - 2 \varepsilon')} s^{- 2 \frac{\frac{1}{4} +
    \varepsilon'}{\frac{1}{2} - \varepsilon} \left( \frac{1}{2} - \varepsilon
    \right)} d \nospace s \left( \| h \|_{K_T^{\frac{1}{2} - \varepsilon}} +
    \| e^{- A \cdot} v (0) \|_{K_T^{\frac{1}{2} - \varepsilon}} \right)^2\\
    & \sim  t^{\varepsilon} {t^{- \left( \frac{1}{2} - \varepsilon
    \right)}}  \| z \|_{C_T M^{\frac{1}{2} - \varepsilon'}} \left( \| h
    \|_{K_T^{\frac{1}{2} - \varepsilon}} + \| e^{- A \cdot} v (0)
    \|_{K_T^{\frac{1}{2} - \varepsilon}} \right)^2\\
    & \leqslant  T^{\varepsilon} {t^{- \left( \frac{1}{2} - \varepsilon
    \right)}}  \| z \|_{C_T M^{\frac{1}{2} - \varepsilon'}} \left( \| h
    \|_{K_T^{\frac{1}{2} - \varepsilon}} + \| e^{- A \cdot} v (0)
    \|_{K_T^{\frac{1}{2} - \varepsilon}} \right)^2,
\end{align*}
 where $\theta = 1 - \frac{\frac{1}{4} + \varepsilon'}{\frac{1}{2} -
  \varepsilon}$. Similarly,
  \begin{align*}
    &   \| \Phi_3 (h + e^{- A \cdot} v (0)) \|_{H^0}\\
    & =  \left\| \int_0^t e^{- A (t - s)} \mathcal{N}_3 (h + e^{- A s} v
    (0)) (s) d \nospace s \right\|_{H^{- (1 - x) + (1 - x)}}\\
    & \lesssim  \int_0^t (t - s)^{- (1 - x)} \| \mathcal{N}_3 (h + e^{- A s}
    v (0)) (s) \|_{H^{- (1 - x)}} d \nospace s\\
    & \lesssim  \int_0^t (t - s)^{- (1 - x)} \| z (s) \|_{M^{\frac{1}{2} -
    \varepsilon'}} \| h (s) + e^{- A s} v (0) \|_{H^{\frac{1}{4} +
    \frac{\varepsilon'}{2} + \delta}}^2 d \nospace s\\
    & \leqslant  \int_0^t (t - s)^{- (1 - x)} \| z (s) \|_{M^{\frac{1}{2} -
    \varepsilon'}} \left( \| h (s) + e^{- A s} v (0) \|_{H^0}^{\theta} \| h
    (s) + e^{- A s} v (0) \|_{H^{\frac{1}{2} - \varepsilon}}^{1 - \theta}
    \right)^2 d \nospace s\\
    & \leqslant  \| z \|_{C_T M^{\frac{1}{2} - \varepsilon'}} \int_0^t (t -
    s)^{- (1 - x)} s^{- 2 \frac{\frac{1}{4} + \frac{\varepsilon'}{2} +
    \delta}{\frac{1}{2} - \varepsilon} \left( \frac{1}{2} - \varepsilon
    \right)} d \nospace s \left( \| h \|_{K_T^{\frac{1}{2} - \varepsilon}} +
    \| e^{- A \cdot} v (0) \|_{K_T^{\frac{1}{2} - \varepsilon}} \right)^2\\
    & \sim  t^{x - \left( \frac{1}{2} + \varepsilon' + 2 \delta \right)} \|
    z \|_{C_T M^{\frac{1}{2} - \varepsilon'}} \left( \| h \|_{K_T^{\frac{1}{2}
    - \varepsilon}} + \| e^{- A \cdot} v (0) \|_{K_T^{\frac{1}{2} -
    \varepsilon}} \right)^2\\
    & \leqslant  T^{\varepsilon} \| z \|_{C_T M^{\frac{1}{2} -
    \varepsilon'}} \left( \| h \|_{K_T^{\frac{1}{2} - \varepsilon}} + \| e^{-
    A \cdot} v (0) \|_{K_T^{\frac{1}{2} - \varepsilon}} \right)^2.
\end{align*}
Here $\theta = 1 - \frac{\frac{1}{4} + \frac{\varepsilon'}{2} +
  \delta}{\frac{1}{2} - \varepsilon}$, $\delta > 0$ is a small number and
$x = \left( \frac{1}{2} + \varepsilon' + 2 \delta \right) + \varepsilon$.
The statement for $\Delta \Phi_3$ is similar.
\end{proof}

\begin{lemma}
  Suppose $z \in C_T M^{\frac{1}{2} - \varepsilon'}$ and $h \in
  K_T^{\frac{1}{2} - \varepsilon}$. Then
  \[ \| \Phi_5 (h + e^{- A \cdot} v (0)) \|_{K_T^{\frac{1}{2} - \varepsilon}}
     \lesssim 
 T^{\varepsilon} \left( \| h \|_{K_T^{\frac{1}{2} - \varepsilon}} + \|
     e^{- A \cdot} v (0) \|_{K_T^{\frac{1}{2} - \varepsilon}} \right) \left(
     \int_0^T \| \mathcal{N}_5 (s) \|^p_{\mathcal{L} \left( H^{\frac{1}{2} -
     \varepsilon} ; H^{0 - \varepsilon - \varepsilon'} \right)} d \nospace s
     \right)^{1 / p} \]
  for $p$ a large positive number, and
  \[ \| \Delta \Phi_5 (h) \|_{K_T^{\frac{1}{2} - \varepsilon}} \lesssim
     T^{\varepsilon} \| \Delta h \|_{K_T^{\frac{1}{2} - \varepsilon}} \left(
     \int_0^T \| \mathcal{N}_5 (s) \|^p_{\mathcal{L} \left( H^{\frac{1}{2} -
     \varepsilon} ; H^{0 - \varepsilon - \varepsilon'} \right)} d \nospace s
     \right)^{1 / p} . \]
\end{lemma}

\begin{proof}
  By definition and random operator estimate,
  \begin{align*}
     \| \Phi_5 (h + e^{- A \cdot} v (0)) \|_{H^{\frac{1}{2} -
    \varepsilon}}
    & =  \left\| \int_0^t e^{- A (t - s)} \mathcal{N}_5 (h + e^{- A s} v
    (0)) (s) d \nospace s \right\|_{H^{- \frac{1}{2} + (1 - \varepsilon)}}\\
    & \lesssim  \int_0^t (t - s)^{- (1 - \varepsilon)} \| \mathcal{N}_5 (h +
    e^{- A s} v (0)) (s) \|_{H^{- \frac{1}{2}}} d \nospace s\\
    & \lesssim  \int_0^t (t - s)^{- (1 - \varepsilon)} \| \mathcal{N}_5 (s)
    \|_{\mathcal{L} \left( H^{\frac{1}{2} - \varepsilon} ; H^{0 - \varepsilon
    - \varepsilon'} \right)} \| h (s) + e^{- A s} v (0) \|_{H^{\frac{1}{2} -
    \varepsilon}} d \nospace s\\
    & \leqslant  \int_0^t (t - s)^{- (1 - \varepsilon)} s^{- \left(
    \frac{1}{2} - \varepsilon \right)} {\| \mathcal{N}_5 (s) \|_{\mathcal{L}}}
    d \nospace s \left( \| h \|_{K_T^{\frac{1}{2} - \varepsilon}} + \| e^{- A
    \cdot} v (0) \|_{K_T^{\frac{1}{2} - \varepsilon}} \right)\\
    & \leqslant  \left( \| h \|_{K_T^{\frac{1}{2} - \varepsilon}} + \| e^{-
    A \cdot} v (0) \|_{K_T^{\frac{1}{2} - \varepsilon}} \right) \left(
    \int_0^t \| \mathcal{N}_5 (s) \|^p_{\mathcal{L} \left( H^{\frac{1}{2} -
    \varepsilon} ; H^{0 - \varepsilon - \varepsilon'} \right)} d \nospace s
    \right)^{1 / p} \\
    &  \qquad \times  
   \left( \int_0^t (t - s)^{- q (1 - \varepsilon)} s^{- q \left(
    \frac{1}{2} - \varepsilon \right)} d \nospace s \right)^{1 / q}\\
    & \sim  t^{\varepsilon} t^{- \left( \frac{1}{2} - \varepsilon \right)}
    \Big( \| h \|_{K_T^{\frac{1}{2} - \varepsilon}} + \| e^{- A \cdot} v (0)
    \|_{K_T^{\frac{1}{2} - \varepsilon}} \Big) \left( \int_0^t \! \|
    \mathcal{N}_5 (s) \|^p_{\mathcal{L} \left( H^{\frac{1}{2} - \varepsilon} ;
    H^{0 - \varepsilon - \varepsilon'} \right)} d \nospace s \right)^{1 / p}\\
    & \leqslant  T^{\varepsilon} t^{-(\frac{1}{2} - \varepsilon
    )} \Big( \| h \|_{K_T^{\frac{1}{2} - \varepsilon}} + \| e^{- A
    \cdot} v (0) \|_{K_T^{\frac{1}{2} - \varepsilon}} \Big) \left( \int_0^T
    \!\! \| \mathcal{N}_5 (s) \|^p_{\mathcal{L} \left( H^{\frac{1}{2} -
    \varepsilon} ; H^{0 - \varepsilon - \varepsilon'} \right)} d \nospace s
    \right)^{1 / p}
  \end{align*}
  and
  \begin{align*}
   \| \Phi_5 (h + e^{- A \cdot} v (0)) \|_{H^0}
    & =  \left\| \int_0^t e^{- A (t - s)} \mathcal{N}_5 (h + e^{- A s} v
    (0)) (s) d \nospace s \right\|_{H^{- \frac{1}{2} + \frac{1}{2}}}\\
    & \lesssim  \int_0^t (t - s)^{- \frac{1}{2}} \| \mathcal{N}_5 (h + e^{-
    A s} v (0)) (s) \|_{H^{- (1 - x)}} d \nospace s\\
    & \lesssim  \int_0^t (t - s)^{- \frac{1}{2}} \| \mathcal{N}_5 (s)
    \|_{\mathcal{L} \left( H^{\frac{1}{2} - \varepsilon} ; H^{0 - \varepsilon
    - \varepsilon'} \right)} \| h (s) + e^{- A s} v (0) \|_{H^{\frac{1}{2} -
    \varepsilon}} d \nospace s\\
    & \leqslant  \int_0^t (t - s)^{- \frac{1}{2}} s^{- \left( \frac{1}{2} -
    \varepsilon \right)} {\| \mathcal{N}_5 (s) \|_{\mathcal{L}}}  d \nospace s
    \left( \| h \|_{K_T^{\frac{1}{2} - \varepsilon}} + \| e^{- A \cdot} v (0)
    \|_{K_T^{\frac{1}{2} - \varepsilon}} \right)\\
    & \leqslant  \left( \| h \|_{K_T^{\frac{1}{2} - \varepsilon}} + \| e^{-
    A \cdot} v (0) \|_{K_T^{\frac{1}{2} - \varepsilon}} \right) \left(
    \int_0^t \| \mathcal{N}_5 (s) \|^p_{\mathcal{L} \left( H^{\frac{1}{2} -
    \varepsilon} ; H^{0 - \varepsilon - \varepsilon'} \right)} d \nospace s
    \right)^{1 / p} \\
    &  \qquad \times
    \left( \int_0^t (t - s)^{- q \frac{1}{2}} s^{- q \left( \frac{1}{2}
    - \varepsilon \right)} d \nospace s \right)^{1 / q}\\
    & \sim  t^{\varepsilon} \left( \| h \|_{K_T^{\frac{1}{2} - \varepsilon}}
    + \| e^{- A \cdot} v (0) \|_{K_T^{\frac{1}{2} - \varepsilon}} \right)
    \left( \int_0^t \| \mathcal{N}_5 (s) \|^p_{\mathcal{L} \left(
    H^{\frac{1}{2} - \varepsilon} ; H^{0 - \varepsilon - \varepsilon'}
    \right)} d \nospace s \right)^{1 / p}\\
    & \leqslant  T^{\varepsilon} \left( \| h \|_{K_T^{\frac{1}{2} -
    \varepsilon}} + \| e^{- A \cdot} v (0) \|_{K_T^{\frac{1}{2} -
    \varepsilon}} \right) \left( \int_0^T \| \mathcal{N}_5 (s)
    \|^p_{\mathcal{L} \left( H^{\frac{1}{2} - \varepsilon} ; H^{0 -
    \varepsilon - \varepsilon'} \right)} d \nospace s \right)^{1 / p} .
\end{align*}
  We used H{\"o}lder's inequality for $q$ close to 1 and $p$ large enough such
  that $\frac{1}{p} + \frac{1}{q} = 1$. So we get the result.
\end{proof}

\begin{remark}
  We could also consider the time dependent random variable $\| \mathcal{N}_5
  (s) \|_{\mathcal{L} \left( H^{\frac{1}{2} - \varepsilon} ; H^{0 -
  \varepsilon - \varepsilon'} \right)}$ with more standard methods in
  stochastic analysis. Using Kolmogorov's criterion on the time dependent
  random operator $\mathcal{N}_5 (t)$, one needs to prove an inequality like
  \[ \mathbb{E} \left[ \| \mathcal{N}_5 (t) -\mathcal{N}_5 (s)
     \|^p_{\mathcal{L} \left( H^{\frac{1}{2} - \varepsilon} ; H^{0 -
     \varepsilon - \varepsilon'} \right)} \right]^{1 / p} \leqslant M | t - s
     |^{\delta} \]
  and then conclude $\| \mathcal{N}_5 (v) \|_{C_T H^{0 - \varepsilon -
  \varepsilon'}} \lesssim \| \mathcal{N}_5 \|_{\mathcal{L} \left( C_T
  H^{\frac{1}{2} - \varepsilon} ; C_T H^{0 - \varepsilon - \varepsilon'}
  \right)} \| v \|_{C_T H^{\frac{1}{2} - \varepsilon}}$. The reason we didn't
  do this is, with time differences involved, one needs to introduce more
  features in the graph representation which increases the number of different
  graphs to study. With some calculation one can easily check and convince
  oneself this doesn't change the nature of the problem and doesn't change
  estimates too much.
\end{remark}

Now we come back to the iteration map:
\begin{align*}
  {\| \Psi (h) \|_{K_T^{\frac{1}{2} - \varepsilon}}} 
  & \lesssim  \sum_{i = 1}^7 \| \Phi_i (h + e^{- A t} v (0))
  \|_{K_T^{\frac{1}{2} - \varepsilon}} + \left\| \int_0^{\cdot} e^{- A (\cdot
  - s)} : z^3 : (s) d \nospace s \right\|_{K_T^{\frac{1}{2} - \varepsilon}}\\
  & \lesssim  T^{\varepsilon} \left( \| h \|_{K_T^{\frac{1}{2} -
  \varepsilon}} + \| e^{- A \cdot} v (0) \|_{K_T^{\frac{1}{2} - \varepsilon}}
  \right)^3 
    + 3 T^{\varepsilon} \| z \|_{C_T M^{\frac{1}{2} - \varepsilon'}} \left(
  \| h \|_{K_T^{\frac{1}{2} - \varepsilon}} + \| e^{- A \cdot} v (0)
  \|_{K_T^{\frac{1}{2} - \varepsilon}} \right)^2 \\
  &  + T^{\varepsilon} \left( \| h \|_{K_T^{\frac{1}{2} - \varepsilon}} + \|
  e^{- A \cdot} v (0) \|_{K_T^{\frac{1}{2} - \varepsilon}} \right) \left(
  \int_0^T \| \mathcal{N}_5 (s) \|^p_{\mathcal{L} \left( H^{\frac{1}{2} -
  \varepsilon} ; H^{0 - \varepsilon - \varepsilon'} \right)} d \nospace s
  \right)^{1 / p}\\
  &  + 2 T^{\varepsilon} \| : z^2 : \|_{C_T M^{\frac{1}{2} - \varepsilon'}}
  \left( \| h \|_{K_T^{\frac{1}{2} - \varepsilon}} + \| e^{- A \cdot} v (0)
  \|_{K_T^{\frac{1}{2} - \varepsilon}} \right) + \left\| \int_0^{\cdot} e^{- A
  (\cdot - s)} : z^3 : (s) d \nospace s \right\|_{K_T^{\frac{1}{2} -
  \varepsilon}}.
\end{align*}
This shows that the iteration map has an invariant closed ball for small enough
time $T \leqslant 1$, where $T$ depends on initial data $v (0)$, the random
objects, and on $\varepsilon, \varepsilon'$.

To show the iteration map is a contraction, we collect
\begin{align*}
  {\| \Delta \Psi (h) \|_{K_T^{\frac{1}{2} - \varepsilon}}}  & \leqslant 
  \sum_{i = 1}^7 \| \Delta \Phi_i (h) \|_{K_T^{\frac{1}{2} - \varepsilon}}\\
  & \lesssim  T^{\varepsilon} \| \Delta h \|_{K_T^{\frac{1}{2} -
  \varepsilon}} \sum_{k = 1, 2} \left( \| h_k \|_{K_T^{\frac{1}{2} -
  \varepsilon}} + \| e^{- A \cdot} v (0) \|_{K_T^{\frac{1}{2} - \varepsilon}}
  \right)^2 \\
  &  + 3 T^{\varepsilon} \| \Delta h \|_{K_T^{\frac{1}{2} - \varepsilon}}
  \sum_{k = 1, 2} \left( \| h_k \|_{K_T^{\frac{1}{2} - \varepsilon}} + \| e^{-
  A \cdot} v (0) \|_{K_T^{\frac{1}{2} - \varepsilon}} \right) \\
  &  + T^{\varepsilon} \| \Delta h \|_{K_T^{\frac{1}{2} - \varepsilon}}
  \left( \int_0^T \| \mathcal{N}_5 (s) \|^p_{\mathcal{L} \left( H^{\frac{1}{2}
  - \varepsilon} ; H^{0 - \varepsilon - \varepsilon'} \right)} d \nospace s
  \right)^{1 / p} 
   + 2 T^{\varepsilon} \| : z^2 : \|_{C_T M^{\frac{1}{2} - \varepsilon'}}
  \| \Delta h \|_{K_T^{\frac{1}{2} - \varepsilon}}.
\end{align*}
This means that inside the previously constructed invariant closed ball, the
Picard iteration map $\Psi$ is a contraction for small enough random time $T$.
So combined with $\| e^{- A t} v (0) \|_{K_T^{\frac{1}{2} - \varepsilon}}
\lesssim \| v (0) \|_{H^0}$, we have our following main theorem.

\begin{theorem}
  \label{Localexistence}For any initial value $v (0) \in H^0$, \ there exists
  a random time $T$, which depends on the norm of initial data $\| v (0)
  \|_{H^0}$ and the value of random objects $(z \nobracket$, $: z^2 :$ and
  $\nobracket : z^3 :)$, such that the equation
  \[ \partial_t v_{m \nospace n} = - A_{m n} v_{m \nospace n} - \frac{\theta
     \lambda}{4} \left\{ \sum_{i = 1}^7 \mathcal{N}_i (v)_{m n} + : z^3 :_{m n}
   \right\}
 \]
  has a unique solution up to time $T$ in the space $K_T^{\frac{1}{2} -
  \varepsilon}$ almost surely.
\end{theorem}

\section{A Priori Estimate}\label{APE}

Following the method in {\cite{tsatsoulis2018spectral}} and
{\cite{mourrat2017global}}, we show that the local in time solution
obtained before can be extended to a global one. This requires us to
find an a priori estimate.

We first refine one of estimations in the local solution theory. To handle the random object $\Gamma^n_{l, l'}$ from the following calculation, we
define the space
\[ G^{a, b} \assign \{ (L^i_{j k})_{i, j, k \geqslant 0} | \|L\|_{G^{a,b}}:=\sup_{i,
   j, k \geqslant 0} A_{i i}^{a} A_{j k}^{b} | L^i_{j k} | < \infty
   \} \]
and define
\[ C_T G^{a, b} \assign \{ L : [0, T] \rightarrow G^{a, b}
   \tmop{continuous} | \sup_{t \in [0, T]} \| L \|_{G^{a, b}} <
   \infty  \} \]
as usual.

\begin{lemma}
  \label{zvoperator}Suppose $\varepsilon > 0$ is small, $\beta\in(0,1/2)$, $\delta, \delta' \in (0,
  \beta)$ and $\kappa \in \left( 0, \frac{1}{2} \right)$. Let $z$ be the
  stationary solution of the free field stochastic quantization equation as
  before, and $v$ be a Hermitian matrix valued function in $C_T H^{\frac{1}{4}
  + \frac{\kappa}{2} + \varepsilon}$. Define $\Gamma^n_{l, l'} \assign \sum_{m
  \geqslant 0} \frac{: z_{l' m} z_{m l} :}{A_{m n}^{2 \beta}}$, then at any
  fixed time
  \[ \| z \nospace v \|_{H^{- \beta}}^2 \lesssim \| v \|_{H^{- (\beta -
     \delta)}}^2 + \| \Gamma \|_{G^{\beta - \delta', \frac{1}{2} - \kappa}} \|
 v \|_{H^{\frac{1}{4} + \frac{\kappa}{2} + \varepsilon}}^2.
\]
Consequently, we have
  \[ \| z \nospace v \|_{H^{- \beta}} \lesssim \left( 1 + \| \Gamma
     \|_{G^{\beta - \delta', \frac{1}{2} - \kappa}} \right)^{1 / 2} \| v
     \|_{H^{\frac{1}{4} + \frac{\kappa}{2} + \varepsilon}} . \]
 
\end{lemma}

\begin{proof}
  By definition,
  \begin{align*}
    \| z \nospace v \|_{H^{- \beta}}^2 & =
    \sum_{m, n \geqslant 0} \frac{1}{A_{m
    n}^{2 \beta}} \sum_{l \geqslant 0} z_{m l} v_{l n} \sum_{l' \geqslant 0}
    v_{n l'} z_{l' m}\\
    & =  \sum_{n, l, l' \geqslant 0} v_{l n} v_{n l'} \sum_{m \geqslant 0}
    \frac{: z_{l' m} z_{m l} :}{A_{m n}^{2 \beta}} + \sum_{n, l, l' \geqslant
    0} v_{l n} v_{n l'} \sum_{m \geqslant 0} \frac{\mathbb{E} [z_{l' m} z_{m
    l}]}{A_{m n}^{2 \beta}}\\
    & =  \sum_{n, l, l' \geqslant 0} v_{l n} v_{n l'} \Gamma^n_{l, l'} +
    \sum_{n, l, l' \geqslant 0} v_{l n} v_{n l'} \sum_{m \geqslant 0}
    \frac{\delta_{l l'}}{A_{m n}^{2 \beta} A_{m l}},
  \end{align*}
  where we denote
  \[ \Gamma^n_{l, l'} \assign \sum_{m \geqslant 0} \frac{: z_{l' m} z_{m l}
     :}{A_{m n}^{2 \beta}} . \]
  The second sum gives
  \begin{align*}
    \sum_{n, l, l' \geqslant 0} v_{l n} v_{n l'} \sum_{m \geqslant 0}
    \frac{\delta_{l l'}}{A_{m n}^{2 \beta} A_{m l}} & =  \sum_{n, l \geqslant
    0} | v_{l n} |^2 \sum_{m \geqslant 0} \frac{1}{A_{m n}^{2 \beta} A_{m
    l}}\\
    & \lesssim  \sum_{n, l \geqslant 0} | v_{l n} |^2 \frac{1}{A_{n l}^{2
    (\beta - \delta)}}\\
    & =  \| v \|_{H^{- (\beta - \delta)}}^2
  \end{align*}
  for some $\delta \in (0, \beta)$. We have used inequality  (2) in
  Appendix \ref{inequalitiescorrelation}.
  
  For the first sum, $\Gamma^n_{l, l'}$ is a collection of random Hermitian
  matrices indexed by $n$. The next lemma shows that it is almost surely in
  the space $G^{\beta - \delta', \frac{1}{2} - \kappa}$ with $\delta' \in (0,
  \beta)$ and $\kappa \in \left( 0, \frac{1}{2} \right)$. Assume $\sigma=\frac{1}{4} + \frac{\kappa}{2} + \varepsilon$ for some $\varepsilon > 0$, then (we use the simplified notation $ \| \Gamma \| $ for $ \| \Gamma \|_{G^{\beta - \delta', \frac{1}{2} - \kappa}} $)
  \begin{align*}
   \sum_{n, l, l' \geqslant 0} v_{l n} v_{n l'} \Gamma^n_{l, l'}
    & \leqslant  \| \Gamma \| \sum_{n, l, l' \geqslant 0} | v_{l n} | | v_{n
    l'} | \frac{1}{A_{n n}^{\beta - \delta'} A_{l l'}^{\frac{1}{2} -
    \kappa}}\\
    & =  \| \Gamma \| \sum_{n, l, l' \geqslant 0} A_{n l}^{\sigma} | v_{l
    n} | A_{n l'}^{\sigma} | v_{n l'} | \frac{1}{A_{n n}^{\beta - \delta'}
    A_{l l'}^{\frac{1}{2} - \kappa} A_{n l}^{\sigma} A_{n l'}^{\sigma}}\\
    & \leqslant  \| \Gamma \| \sum_{n, l \geqslant 0} A_{n l}^{\sigma} |
    v_{l n} | \left( \sum_{l' \geqslant 0} A_{n l'}^{2 \sigma} | v_{n l'}
    |^2 \right)^{1 / 2} \left( \sum_{l' \geqslant 0} \frac{1}{A_{n n}^{2 \beta
    - 2 \delta'} A_{l l'}^{1 - 2 \kappa} A_{n l}^{2 \sigma} A_{n l'}^{2
    \sigma}} \right)^{1 / 2}\\
    & \leqslant  \| \Gamma \| \sum_{n \geqslant 0} \left( \sum_{l' \geqslant
    0} A_{n l'}^{2 \sigma} | v_{n l'} |^2 \right)^{1 / 2} \left( \sum_{l
    \geqslant 0} A_{n l}^{2 \sigma} | v_{l n} |^2 \right)^{1 / 2} \left(
    \sum_{l, l' \geqslant 0} \frac{1}{A_{n n}^{2 \beta - 2 \delta'} A_{l
    l'}^{1 - 2 \kappa} A_{n l}^{2 \sigma} A_{n l'}^{2 \sigma}} \right)^{1
    / 2}\\
    & \lesssim  \| \Gamma \| \| v \|_{H^{\sigma}} \| v \|_{H^{\sigma}}
    \left( \sum_{l, l' \geqslant 0} \frac{1}{A_{l
    l}^{\frac{1}{2}-\kappa+2 \sigma} A_{l' l'}^{\frac{1}{2}-\kappa+2 \sigma}} \right)^{1 / 2}.
\end{align*}
We have used twice the Cauchy Schwarz inequality and the fact that $A_{n n}^{2 \beta - 2 \delta'}$ is bounded from below by some positive constant
since we chose $\delta' \in (0,
  \beta)$. The summation in the last
  line is clearly finite. This concludes the proof of the lemma.
\end{proof}

The following lemma gives the construction of $\Gamma$.

\begin{lemma}
	Let $0<\delta'<\beta<\frac12$ and $0<\kappa<\frac12$. For $N\geq0$, define
	\[
	\Gamma^{n,N}_{l,l'}(t)
	:=
	\sum_{m\geq0}
	\frac{
		:z^{(N)}_{l'm}(t)z^{(N)}_{ml}(t):
	}{A_{mn}^{2\beta}}.
	\]
	Then, for every sufficiently large $p$, the sequence
	$\{\Gamma^N\}_{N\geq0}$ is Cauchy in
	\[
	L^p\!\left(
	\Omega,\mathbb P;
	C_TG^{\beta-\delta',\,\frac12-\kappa}
	\right).
	\]
	We denote its limit by $\Gamma$. Entrywise,
	\[
	\Gamma^n_{l,l'}(t)
	=
	\sum_{m\geq0}
	\frac{:z_{l'm}(t)z_{ml}(t):}{A_{mn}^{2\beta}}.
	\]
	In particular,
	\[
	\Gamma\in
	L^p\!\left(
	\Omega,\mathbb P;
	C_TG^{\beta-\delta',\,\frac12-\kappa}
	\right).
	\]
\end{lemma}

\begin{proof}
	Let $r=\min\{N,M\}$ and $h=|t-s|$. The mixed covariance of the cutoff
	processes is
	\[
	\mathbb E\!\left[
	z^{(N)}_{ij}(t)z^{(M)}_{pq}(s)
	\right]
	=
	\mathbf 1_{\{i,j\leq N\}}
	\mathbf 1_{\{p,q\leq M\}}
	\frac{\delta_{iq}\delta_{jp}}{A_{ij}}
	e^{-hA_{ij}}.
	\]
	Using the Wick identity, we obtain
	\[
	\mathbb E\!\left[
	\Gamma^{n,N}_{l,l'}(t)
	\Gamma^{n,M}_{l',l}(s)
	\right]
	=
	C_r^{n,l,l'}(h),
	\]
	where
	\[
	C_r^{n,l,l'}(h)
	:=
	\mathbf 1_{\{l,l'\leq r\}}
	\left\{
	\frac{\delta_{ll'}e^{-2hA_{ll}}}
	{A_{ln}^{4\beta}A_{ll}^{2}}
	+
	\sum_{m=0}^{r}
	\frac{e^{-h(A_{ml}+A_{ml'})}}
	{A_{mn}^{4\beta}A_{ml}A_{ml'}}
	\right\}.
	\]
	
	Let $C_\infty^{n,l,l'}$ denote the same expression without cutoffs. Since
	\[
	A_{mn}^{2}\geq A_{mm}A_{nn},
	\]
	inequality (6) of Appendix D, applied with exponent $2\beta\in(0,1)$, gives
	\[
	\begin{aligned}
		C_\infty^{n,l,l'}(0)
		&\leq
		\frac{1}{A_{nn}^{2\beta}}
		\left\{
		\frac{\delta_{ll'}}{A_{ll'}^{2+2\beta}}
		+
		\sum_{m\geq0}
		\frac{1}{A_{ml'}A_{mm}^{2\beta}A_{ml}}
		\right\} \\
		&\lesssim
		\frac{1}{A_{nn}^{2\beta}}
		\left(
		\frac{\delta_{ll'}}{A_{ll'}^{2+2\beta}}
		+
		\frac{1}{A_{ll'}}
		\right).
	\end{aligned}
	\]
	
	We next estimate the time increments. Choose
	\[
	0<\eta<2\min\{\beta,\kappa\}.
	\]
	Using $1-e^{-x}\lesssim x^\eta$, together with
	\[
	A_{ml}+A_{ml'}
	=
	A_{mm}+A_{ll'}
	\lesssim
	A_{mm}A_{ll'},
	\]
	and applying inequality (6) of Appendix D with exponent
	$2\beta-\eta\in(0,1)$, we obtain, uniformly in $N$,
	\[
	0\leq
	C_N^{n,l,l'}(0)-C_N^{n,l,l'}(h)
	\lesssim
	\frac{h^\eta}{A_{nn}^{2\beta}}
	\left(
	\frac{\delta_{ll'}}
	{A_{ll'}^{2+2\beta-\eta}}
	+
	\frac{1}{A_{ll'}^{1-\eta}}
	\right).
	\]
	
	Every entry of $\Gamma^N$, as well as every cutoff or time difference, belongs
	to the second homogeneous Wiener chaos. Gaussian hypercontractivity and the
	inequality $\sup a_{nll'}\leq\sum a_{nll'}$ therefore imply that, for every
	such random tensor $L$,
	\[
	\begin{aligned}
		\|L\|_{L^p(\Omega;G^{\beta-\delta',\,\frac12-\kappa})}
		\lesssim_p
		\left(
		\sum_{n,l,l'\geq0}
		\left[
		A_{nn}^{2(\beta-\delta')}
		A_{ll'}^{1-2\kappa}
		\mathbb E|L^n_{l,l'}|^2
		\right]^{p/2}
		\right)^{1/p}.
	\end{aligned}
	\]
	
	For $N\leq M$, the mixed covariance formula at $s=t$ gives
	\[
	\mathbb E\left[
	\left|
	\Gamma^{n,M}_{l,l'}(t)
	-
	\Gamma^{n,N}_{l,l'}(t)
	\right|^2
	\right]
	=
	C_M^{n,l,l'}(0)-C_N^{n,l,l'}(0).
	\]
	Define
	\[
	b_N
	:=
	\left(
	\sum_{n,l,l'\geq0}
	\left[
	A_{nn}^{2(\beta-\delta')}
	A_{ll'}^{1-2\kappa}
	\bigl(
	C_\infty^{n,l,l'}(0)-C_N^{n,l,l'}(0)
	\bigr)
	\right]^{p/2}
	\right)^{1/p}.
	\]
	The preceding estimate shows that
	\[
	A_{nn}^{2(\beta-\delta')}
	A_{ll'}^{1-2\kappa}
	C_\infty^{n,l,l'}(0)
	\lesssim
	\frac{1}{A_{nn}^{2\delta'}}
	\left(
	\frac{\delta_{ll'}}
	{A_{ll'}^{1+2\kappa+2\beta}}
	+
	\frac{1}{A_{ll'}^{2\kappa}}
	\right).
	\]
	The $p/2$-power of the right-hand side is summable for sufficiently large
	$p$. Since
	\[
	C_N^{n,l,l'}(0)
	\uparrow
	C_\infty^{n,l,l'}(0),
	\]
	the dominated convergence theorem gives $b_N\to0$. Consequently,
	\[
	\left\|
	\Gamma^M(t)-\Gamma^N(t)
	\right\|_
	{L^p(\Omega;G^{\beta-\delta',\,\frac12-\kappa})}
	\lesssim_p b_N.
	\]
	
	Similarly,
	\[
	\mathbb E\left[
	\left|
	\Gamma^{n,N}_{l,l'}(t)
	-
	\Gamma^{n,N}_{l,l'}(s)
	\right|^2
	\right]
	=
	2\bigl(
	C_N^{n,l,l'}(0)-C_N^{n,l,l'}(h)
	\bigr).
	\]
	Hence, uniformly in $N$,
	\[
	\left\|
	\Gamma^N(t)-\Gamma^N(s)
	\right\|_
	{L^p(\Omega;G^{\beta-\delta',\,\frac12-\kappa})}
	\lesssim_p
	|t-s|^{\eta/2},
	\]
	provided $p$ is sufficiently large; it is enough to require
	\[
	p\delta'>1,
	\qquad
	p(2\kappa-\eta)>4.
	\]
	
	Finally, set $Y^{N,M}:=\Gamma^M-\Gamma^N$. Interpolating the fixed-time and
	time-increment estimates, for every $\lambda\in(0,1)$,
	\[
	\left\|
	Y^{N,M}(t)-Y^{N,M}(s)
	\right\|_
	{L^p(\Omega;G^{\beta-\delta',\,\frac12-\kappa})}
	\lesssim
	b_N^\lambda
	|t-s|^{(1-\lambda)\eta/2}.
	\]
	Taking $p$ still larger if necessary and applying \cite[Theorem A.10]{friz2010multidimensional}  as in
	Appendix E, we conclude that
	\[
	\left\|
	\Gamma^M-\Gamma^N
	\right\|_
	{L^p(\Omega;C_TG^{\beta-\delta',\,\frac12-\kappa})}
	\lesssim
	b_N^\lambda
	\longrightarrow0.
	\]
	Thus $\{\Gamma^N\}_{N\geq0}$ is Cauchy in the asserted space. For every fixed
	$n,l,l'$, its limit agrees with the infinite series in the statement. This
	completes the proof.
\end{proof}

To get an a priori estimate, the easiest approach to simply multiply the remainder equation (\ref{DPDeq})
with the matrix $v_{n m}$ and sum over the indices $m$ and $n$ does not work,
because  one cannot estimate the term $\tmop{tr} (: z^3 : v)$ which would arise on the right hand side by the $H^{\frac{1}{2}}$ norm
of $v$ and $H^0$ norm of $v^2$ available on the left hand side.
The solution of this problem is to expand the solution one step further. Let  
\[
y_{mn}(t)=- \frac{\theta \lambda}{4}\int_{-\infty}^t e^{-A_{mn}(t-s)} : z^3 :_{m n}(s)ds
\]
be the stationary solution of the equation
\[ \partial_t y_{m n} = - A_{m n} y_{m n} - \frac{\theta \lambda}{4} : z^3 :_{m n}
\]
starting from $t=-\infty$. Since $: z^3 : \in C_T H^{- \frac{1}{2} -}$,
we have $y \in C_T H^{\frac{1}{2} -}$ by the Schauder estimate.
Denote $w$ to be the second order
remainder, which means $w \assign v - y$. Then the equation for $w$ reads
\[ \partial_t w_{m \nospace n} = - A_{m n} w_{m \nospace n} - \frac{\theta
   \lambda}{4} [w^3_{m n} + S_2 (w, y, z)_{m n} + S_1 (w, y, z)_{m n} + S_0 (y,
 z)_{m n}]
\]
where 
\begin{align*}
 S_2 (w, y, z) &\assign w^2 z + w \nospace z \nospace w + z \nospace w^2 +
   w^2 y + w \nospace y \nospace w + y \nospace w^2, 
   \\
 S_1 (w, y, z) &\assign w (: z^2 : + y^2 + \nospace y \nospace z + \nospace z
   \nospace y) + (: z^2 : + y^2 + \nospace y \nospace z + \nospace z \nospace
   y) w + z \nospace w \nospace z + y \nospace w \nospace y + z \nospace w
   \nospace y + y \nospace w \nospace z ,
\\   
 S_0 (y, z) &\assign y^3 + y^2 z + y \nospace z \nospace y + z \nospace y^2 +
   y : z^2 : + z \nospace y \nospace z + : z^2 : y.
 \end{align*}
Notice that since $y$ has positive regularity, we do not need further
renormalization in the second order expansion equation; all multiplications of
matrices are well defined. With the help of the bounds in section \ref{FPM},
the second order remainder $w$ is in the space $C_T H^{1 -}$. Now we test the
second order remainder equation by $w$, which is
\[ w_{n m} \partial_t w_{m \nospace n} + A_{m n} w_{m \nospace n} w_{n m} + 
  \frac{\theta \lambda}{4} w^3_{m n} w_{n m}
  = - \frac{\theta \lambda}{4} (S_2 + S_1 +
  S_0)_{m n} w_{n m}.
\]
We take the sum over $m$ and $n$, which means to take the  trace of matrices,
to get
\[ \frac{1}{2} \partial_t \| w \|_{H^0}^2 + \| w \|_{H^{\frac{1}{2}}}^2 +
  \frac{\theta \lambda}{4} \| w^2 \|_{H^0}^2 = - \frac{\theta \lambda}{4} \tmop{tr}
   [(S_2 + S_1 + S_0) w] . \]
We have the following a priori estimate.

\begin{theorem}[a priori estimate]
  We have
  \[ \partial_t \| w \|_{H^0}^2 + \| w \|_{H^{\frac{1}{2}}}^2 +  \frac{\theta
     \lambda}{4} \| w^2 \|_{H^0}^2 \leqslant C \nospace F [y, z] \]
  where $C$ is a positive number and the positive function $F [y, z]$ (see
  formula (\ref{Function})) only depends on $y$ and $z$, and has time
  independent stochastic moments of all orders. Moreover, we have
  \[ \| w \|_{H^0}^2 (t) \leqslant e^{- t} \| w \|_{H^0}^2 (0) + C \int_0^t
     e^{- (t - s)} F [y, z] (s) d \nospace s. \]
\end{theorem}

\begin{proof}
  First we deal with terms with only one $w$, which is $\tmop{tr} (S_0 w)$.
  The bounds in section \ref{FPM} show that $S_0$ has negative regularity. So
  using the duality inequality in lemma \ref{duality}, we have
  \[ | \tmop{tr} (S_0 w) | \leqslant \| S_0 \|_{H^{- \varepsilon}} \| w
     \|_{H^{\varepsilon}} \leqslant \| S_0 \|_{H^{- \varepsilon}} \| w
     \|_{H^{\frac{1}{2}}} \leqslant \frac{\theta \lambda}{16 \sigma} \|
     S_0 \|_{H^{- \varepsilon}}^2 + \frac{4\sigma}{\theta \lambda} \| w
     \|_{H^{\frac{1}{2}}}^2 ,
   \]
   where $\varepsilon>0$. We have introduced a parameter $\sigma$
   whose value will be determined later.
  
  Next we estimate  the terms with three $w$'s. By cyclic symmetry of the trace
  there are two possibilities, namely $\tmop{tr} (z \nospace w^3)$
  and $\tmop{tr} (y \nospace w^3)$. For \ $\tmop{tr} (z \nospace w^3)$, using
  lemma \ref{zvoperator} at the beginning of this section, we have for small enough $\beta>0$
  \begin{align*}
    | \tmop{tr} (z \nospace w^3) | & \leqslant  \| z \nospace w \|_{H^{-
    \beta}} \| w^2 \|_{H^{\beta}}\\
    & \leqslant  C \left( 1 + \| \Gamma \|_{G^{\beta - \delta', \frac{1}{2}
    - \kappa}} \right)^{1 / 2} \| w \|_{H^{\frac{1}{4} + \frac{\kappa}{2} +
    \varepsilon}} \| w^2 \|_{H^{\beta}}\\
    & \leqslant  C (1 + \| \Gamma \|)^{1 / 2} \| w \|_{H^{- \frac{1}{2} -
    \varepsilon'}}^{\mu} \| w \|_{H^{\frac{1}{2}}}^{1 - \mu} \| w^2
    \|_{H^0}^{1 - 2 \beta} \| w^2 \|_{H^{\frac{1}{2}}}^{2 \beta}\\
    & \leqslant  C (1 + \| \Gamma \|)^{1 / 2} \| w^2
    \|_{H^0}^{\frac{\mu}{2}} \| w \|_{H^{\frac{1}{2}}}^{1 - \mu} \| w^2
    \|_{H^0}^{1 - 2 \beta} \| w \|_{H^{\frac{1}{2}}}^{4 \beta}\\
    & =  C (1 + \| \Gamma \|)^{1 / 2} \| w \|_{H^{\frac{1}{2}}}^{1 - \mu + 4
    \beta} \| w^2 \|_{H^0}^{1 - 2 \beta + \frac{\mu}{2}}\\
                                   & \leqslant  C \frac{\frac{\mu}{2}
                                                 - 2 \beta}{2} \cdot \frac{
            (\theta \lambda/4)^{\frac{1 - \mu + 4 \beta}{\frac{\mu}{2} - 2
    \beta}}}{\sigma^{\frac{2 + 2 \beta - \frac{\mu}{2}}{\frac{\mu}{2} - 2
    \beta}}} (1 + \| \Gamma \|)^{\frac{1}{\frac{\mu}{2} - 2 \beta}} + \frac{1
    - \mu + 4 \beta}{2} \cdot \frac{4\sigma}{\theta \lambda} \| w
    \|_{H^{\frac{1}{2}}}^2
     + \frac{1 - 2 \beta + \frac{\mu}{2}}{2} \sigma \| w^2 \|_{H^0}^2 ,
  \end{align*}
  where $\mu = \frac{\frac{1}{4} - \frac{\kappa}{2} - \varepsilon}{1 +
  \varepsilon'}$. See previous lemmas for conditions on exponents and
  the definition of $\Gamma$. Here $C$ is some constant which only depends on
  parameters in the exponents and may differ from line to line, we also used
  lemma \ref{specialineq} at fourth inequality. For $\tmop{tr} (y \nospace
  w^3)$, we have
  \begin{align*}
    | \tmop{tr} (y \nospace w^3) | & \leqslant  \| y \nospace w \|_{H^{-
    \beta}} \| w^2 \|_{H^{\beta}}\\
    & \leqslant  \| y \nospace w \|_{H^0} \| w^2 \|_{H^{\beta}}\\
    & \leqslant  \| y \nospace \|_{H^0} \| \nospace w \|_{H^0} \| w^2
    \|_{H^{\beta}}\\
    & \leqslant  \| y \nospace \|_{H^0} \| \nospace w \|_{H^{- \frac{1}{2} -
    \varepsilon'}}^{\pi} \| \nospace w \|_{H^{\frac{1}{2}}}^{1 - \pi} \| w^2
    \|_{H^0}^{1 - 2 \beta} \| w^2 \|_{H^{\frac{1}{2}}}^{2 \beta}\\
    & \leqslant  \| y \nospace \|_{H^0} \| \nospace w^2
    \|_{H^0}^{\frac{\pi}{2}} \| \nospace w \|_{H^{\frac{1}{2}}}^{1 - \pi} \|
    w^2 \|_{H^0}^{1 - 2 \beta} \| w \|_{H^{\frac{1}{2}}}^{4 \beta}\\
    & =  \| y \nospace \|_{H^0} \| \nospace w \|_{H^{\frac{1}{2}}}^{1 - \pi
    + 4 \beta} \| w^2 \|_{H^0}^{1 - 2 \beta + \frac{\pi}{2}}\\
                                   & \leqslant  \frac{\frac{\pi}{2} - 2 \beta}{2} \cdot
       \frac{(\theta
    \lambda/4)^{\frac{1 - \pi + 4 \beta}{\frac{\pi}{2} - 2
    \beta}}}{\sigma^{\frac{2 + 2 \beta - \frac{\pi}{2}}{\frac{\pi}{2} - 2
    \beta}}} \| y \nospace \|_{H^0}^{\frac{2}{\frac{\pi}{2} - 2 \beta}} +
    \frac{1 - \pi + 4 \beta}{2} \cdot \frac{4\sigma}{\theta \lambda} \| w
    \|_{H^{\frac{1}{2}}}^2
     + \frac{1 - 2 \beta + \frac{\pi}{2}}{2} \sigma \| w^2 \|_{H^0}^2,
    \end{align*}
    where $\pi = \frac{1}{2 (1 + \varepsilon')}$.
  
  The last case are the terms with two $w$'s. By cyclic symmetry 
  there are four possibilities, namely $\tmop{tr} (z \nospace w
  \nospace z \nospace w)$, $\tmop{tr} (y \nospace w \nospace y \nospace w)$,
  $\tmop{tr} (z \nospace w \nospace y \nospace w)$ and $\tmop{tr} ((: z^2 : +
  y^2 + \nospace y \nospace z + \nospace z \nospace y) w^2)$. For $\tmop{tr}
  (z \nospace w \nospace z \nospace w)$, we have
  \begin{align*}
    | \tmop{tr} (z \nospace w \nospace z \nospace w) | & \leqslant  \| z
    \nospace w \nospace z \|_{H^{- \beta}} \| w \|_{H^{\beta}}\\
    & \leqslant  \| \mathcal{N}_5 \|_{\mathcal{L} \left( H^{\frac{1 -
    \beta}{2}} ; H^{- \beta} \right)} \| w \|_{H^{\frac{1 - \beta}{2}}} \| w
    \|_{H^{\beta}}\\
    & \leqslant  \| \mathcal{N}_5 \| \| w \|_{H^{- \frac{1}{2} -
    \varepsilon'}}^{\kappa_1} \| w \|_{H^{\frac{1}{2}}}^{1 - \kappa_1} \| w
    \|_{H^{- \frac{1}{2} - \varepsilon'}}^{\kappa_2} \| w
    \|_{H^{\frac{1}{2}}}^{1 - \kappa_2}\\
    & \leqslant  C \| \mathcal{N}_5 \| \| w \|_{H^{\frac{1}{2}}}^{2 -
    \kappa_1 - \kappa_2} \| w^2 \|_{H^0}^{\frac{\kappa_1 + \kappa_2}{2}}\\
    & \leqslant  C \frac{\kappa_1 + \kappa_2}{4} \cdot \frac{(\theta
    \lambda/4)^{\frac{2 (2 - \kappa_1 - \kappa_2)}{\kappa_1 +
    \kappa_2}}}{\sigma^{\frac{4 - \kappa_1 - \kappa_2}{\kappa_1 + \kappa_2}}}
    \| \mathcal{N}_5 \|^{\frac{4}{\kappa_1 + \kappa_2}} + \frac{2 - \kappa_1 -
    \kappa_2}{2} \cdot \frac{4\sigma}{\theta \lambda} \| w
    \|_{H^{\frac{1}{2}}}^2
    + \frac{\kappa_1 + \kappa_2}{4} \sigma \| w^2 \|_{H^0}^2,
  \end{align*}
  where $\kappa_1 = \frac{\beta}{2 (1 + \varepsilon')}$ and $\kappa_2 =
  \frac{\frac{1}{2} - \beta}{1 + \varepsilon'}$. For $\tmop{tr} (y \nospace w
  \nospace y \nospace w)$, we have
  \begin{align*}
    | \tmop{tr} (y \nospace w \nospace y \nospace w) | & \leqslant  \| y
    \nospace w \|_{H^{- \beta}} \| y \nospace w \|_{H^{\beta}}\\
    & \leqslant  \| y \nospace w \|_{H^{\beta}}^2\\
    & \leqslant  \| y \|_{H^{\beta}}^2 \| w \|_{H^{\frac{1 - \beta}{2}}} \|
    w \|_{H^{\beta}}\\
    & \leqslant  C \frac{\kappa_1 + \kappa_2}{4} \cdot \frac{(\theta
    \lambda/4)^{\frac{2 (2 - \kappa_1 - \kappa_2)}{\kappa_1 +
    \kappa_2}}}{\sigma^{\frac{4 - \kappa_1 - \kappa_2}{\kappa_1 + \kappa_2}}}
    \| y \|_{H^{\beta}}^{\frac{8}{\kappa_1 + \kappa_2}} + \frac{2 - \kappa_1 -
    \kappa_2}{2} \cdot \frac{4\sigma}{\theta \lambda} \| w
    \|_{H^{\frac{1}{2}}}^2
    + \frac{\kappa_1 + \kappa_2}{4} \sigma \| w^2 \|_{H^0}^2
  \end{align*}
  where the choice of constants is the same as in the case of $\tmop{tr} (z
  \nospace w \nospace z \nospace w)$. For $\tmop{tr} (z \nospace w \nospace y
  \nospace w)$, we have
  \begin{align*}
    | \tmop{tr} (z \nospace w \nospace y \nospace w) | & \leqslant  \| z
    \nospace w \|_{H^{- \beta}} \| y \nospace w \|_{H^{\beta}}\\
    & \leqslant  C \| z \|_{M^{\frac{1}{2} - \frac{\beta}{3}}} \| y
    \|_{H^{\beta}} \| \nospace w \|_{H^{\frac{1 - \beta}{2}}} \| w
    \|_{H^{\beta}}\\
    & \leqslant  C \frac{\kappa_1 + \kappa_2}{4} \cdot \frac{(\theta
    \lambda/4)^{\frac{2 (2 - \kappa_1 - \kappa_2)}{\kappa_1 +
    \kappa_2}}}{\sigma^{\frac{4 - \kappa_1 - \kappa_2}{\kappa_1 + \kappa_2}}}
    \left( \| z \|_{M^{\frac{1}{2} - \frac{\beta}{3}}} \| y \|_{H^{\beta}}
    \right)^{\frac{4}{\kappa_1 + \kappa_2}} \\
    &+ \frac{2 - \kappa_1 - \kappa_2}{2} \cdot \frac{4\sigma}{\theta
    \lambda} \| w \|_{H^{\frac{1}{2}}}^2 + \frac{\kappa_1 + \kappa_2}{4}
    \sigma \| w^2 \|_{H^0}^2,
  \end{align*}
  where the choice of constants is the same as before.
  For $\tmop{tr} ((: z^2 : + y^2 +
  \nospace y \nospace z + \nospace z \nospace y) w^2)$ we have
  \begin{align*}
    | \tmop{tr} ((: z^2 : + y^2 + \nospace y \nospace z + \nospace z
    \nospace y) w^2) |
    & \leqslant  \| (: z^2 : + y^2 + \nospace y \nospace z + \nospace z
    \nospace y) w \|_{H^{- \beta}} \| w \|_{H^{\beta}}\\
    & \leqslant  C \| (: z^2 : + y^2 + \nospace y \nospace z + \nospace z
    \nospace y) \|_{M^{\frac{1}{2} - \frac{\beta}{3}}} \| \nospace w
    \|_{H^{\frac{1 - \beta}{2}}} \| w \|_{H^{\beta}}\\
    & \leqslant  C \frac{\kappa_1 + \kappa_2}{4} \cdot \frac{(\theta
    \lambda/4)^{\frac{2 (2 - \kappa_1 - \kappa_2)}{\kappa_1 +
    \kappa_2}}}{\sigma^{\frac{4 - \kappa_1 - \kappa_2}{\kappa_1 + \kappa_2}}}
    \| (: z^2 : + y^2 + \nospace y \nospace z + \nospace z \nospace y)
    \|_{M^{\frac{1}{2} - \frac{\beta}{3}}}^{\frac{4}{\kappa_1 + \kappa_2}} \\
    &  \qquad + \frac{2 - \kappa_1 - \kappa_2}{2} \cdot \frac{4\sigma}{\theta
    \lambda} \| w \|_{H^{\frac{1}{2}}}^2 + \frac{\kappa_1 + \kappa_2}{4}
    \sigma \| w^2 \|_{H^0}^2,
  \end{align*}
  where the choice of constants is the same as before.
Notice that with simple
  arguments, $: z^2 : + y^2 + \nospace y \nospace z + \nospace z \nospace y \in
  C_T M^{\frac{1}{2} -}$.
  
  Finally, putting everything together, we have
  \begin{align*}
    &   | \frac{\theta \lambda}{4} \tmop{tr} [(S_2 + S_1 + S_0) w] |\\
    & \leqslant  C \nospace F [y, z] + \sigma \left( 1 + \frac{3 (1 - \mu +
    4 \beta)}{2} + \frac{3 (1 - \pi + 4 \beta)}{2} + 3 (2 - \kappa_1 -
    \kappa_2) \right) \| w \|_{H^{\frac{1}{2}}}^2 \\
    &  \qquad + \frac{\theta \lambda \sigma}{4} \left( 1 + \frac{3 \left( 1 - 2 \beta +
    \frac{\mu}{2} \right)}{2} + \frac{3 \left( 1 - 2 \beta + \frac{\pi}{2}
    \right)}{2} + \frac{3 (\kappa_1 + \kappa_2)}{2} \right) \| w^2 \|_{H^0}^2,
\end{align*}
  where
 \begin{align}
     F [y, z] & \assign  \frac{(\theta \lambda)^2}{64 \sigma} \| S_0
     \|_{H^{- \varepsilon}}^2 + 3 \frac{\frac{\mu}{2} - 2 \beta}{2} \cdot
     \frac{(\frac{\theta \lambda}{4})^{\frac{1 - \mu + 4 \beta}{\frac{\mu}{2} - 2
     \beta} + 1}}{\sigma^{\frac{2 + 2 \beta - \frac{\mu}{2}}{\frac{\mu}{2} - 2
     \beta}}} (1 + \| \Gamma \|)^{\frac{1}{\frac{\mu}{2} - 2 \beta}} +
 3 \frac{\frac{\pi}{2} - 2 \beta}{2} \cdot \frac{(\frac{\theta
   \lambda}{4})^{\frac{1 - \pi + 4 \beta}{\frac{\pi}{2} - 2 \beta}
        +
     1}}{\sigma^{\frac{2 + 2 \beta - \frac{\pi}{2}}{\frac{\pi}{2} - 2 \beta}}}
     \| y \nospace \|_{H^0}^{\frac{2}{\frac{\pi}{2} - 2 \beta}} 
     \nonumber\\
      & 
    + \frac{\kappa_1 + \kappa_2}{4} \cdot \frac{(\theta \lambda/4)^{\frac{2
     (2 - \kappa_1 - \kappa_2)}{\kappa_1 + \kappa_2} + 1}}{\sigma^{\frac{4 -
     \kappa_1 - \kappa_2}{\kappa_1 + \kappa_2}}} 
     \left( \| \mathcal{N}_5 \|^{\frac{4}{\kappa_1 + \kappa_2}} + \| y
     \|_{H^{\beta}}^{\frac{8}{\kappa_1 + \kappa_2}} + 2 \left( \| z
     \|_{M^{\frac{1}{2} - \frac{\beta}{3}}} \| y \|_{H^{\beta}}
          \right)^{\frac{4}{\kappa_1 + \kappa_2}} \right.
          \nonumber
     \\
     &\qquad \left.+ 2 \| (: z^2 : + y^2 + \nospace y
     \nospace z + \nospace z \nospace y) \|_{M^{\frac{1}{2} -
         \frac{\beta}{3}}}^{\frac{4}{\kappa_1 + \kappa_2}} \right).
   \label{Function}    
 \end{align}
We choose $\sigma > 0$ small enough so that
\begin{align*}
 \sigma \Big( 1 + \frac{3 (1 - \mu + 4 \beta)}{2} + \frac{3 (1 - \pi + 4
     \beta)}{2} + 3 (2 - \kappa_1 - \kappa_2) \Big) &< \frac{1}{2} ,
 \\
 \sigma \Big( 1 + \frac{3 \left( 1 - 2 \beta + \frac{\mu}{2} \right)}{2}
     + \frac{3 \left( 1 - 2 \beta + \frac{\pi}{2} \right)}{2} + \frac{3
     (\kappa_1 + \kappa_2)}{2} \Big) &< \frac{1}{2} .
\end{align*}
This finishes the proof of the first part of the theorem.
The second inequality follows
  from
  \[ \partial_t \| w \|_{H^0}^2 + \| w \|_{H^{\frac{1}{2}}}^2 + \frac{\theta
     \lambda}{4} \| w^2 \|_{H^0}^2 \geqslant \partial_t \| w \|_{H^0}^2 + \| w
     \|_{H^{\frac{1}{2}}}^2 \geqslant \partial_t \| w \|_{H^0}^2 + \| w
     \|_{H^0}^2 . \]
\end{proof}

Going back to $v$, we have the following corollary.
\begin{corollary}
  \label{cor-v-H0}
  The following inequality holds for $v$ and $t \in [0, T^{\ast}]$
  \[ \| v \|^2_{H^0} (t) \leqslant 2 \| y \|^2_{C_{T^{\ast}} H^0} + 2 e^{- t}
     \| v \|_{H^0}^2 (0) + 2 e^{- t} \| y \|_{H^0}^2 (0) + 2 C \int_0^t e^{-
     (t - s)} F [y, z] (s) d \nospace s. \]
\end{corollary}

\begin{proof}
  Since $w = v - y$ and
  \[ \| w \|_{H^0}^2 (t) \leqslant e^{- t} \| w \|_{H^0}^2 (0) + C \int_0^t
    e^{- (t - s)} F [y, z] (s) d \nospace s ,
  \]
we have
  \begin{eqnarray*}
    \| v \|^2_{H^0} (t) & \leqslant & \left( \sqrt{e^{- t} \| w \|_{H^0}^2 (0)
    + C \int_0^t e^{- (t - s)} F [y, z] (s) d \nospace s} + \| y \|_{H^0} (t)
    \right)^2\\
    & \leqslant & 2 \| y \|^2_{H^0} (t) + 2 e^{- t} \| w \|_{H^0}^2 (0) + 2 C
    \int_0^t e^{- (t - s)} F [y, z] (s) d \nospace s\\
    & \leqslant & 2 \| y \|^2_{H^0} (t) + 2 e^{- t} \| v \|_{H^0}^2 (0) + 2
    e^{- t} \| y \|_{H^0}^2 (0) + 2 C \int_0^t e^{- (t - s)} F [y, z] (s) d
    \nospace s\\
    & \leqslant & 2 \| y \|^2_{C_{T^{\ast}} H^0} + 2 e^{- t} \| v \|_{H^0}^2
    (0) + 2 e^{- t} \| y \|_{H^0}^2 (0) + 2 C \int_0^t e^{- (t - s)} F [y, z]
    (s) d \nospace s
  \end{eqnarray*}
  which concludes the result.
\end{proof}

With the same argument, we have the estimate for $\phi$.
\begin{corollary}
  \label{Gdef}The following inequality holds for $v$ and $t \in [0, T^{\ast}]$
  \[ \| \phi \|^2_{H^{- \frac{1}{2} - \varepsilon}} (t) \leqslant 4 e^{- t} \|
     v \|_{H^0}^2 (0) + G [y, z] \]
  where
  \[ G [y, z] \assign 2 \| z \|^2_{H^{- \frac{1}{2} - \varepsilon}} (t) + 4 \|
     y \|^2_{C_{T^{\ast}} H^0} + 4 e^{- t} \| y \|_{H^0}^2 (0) + 4 C \int_0^t
     e^{- (t - s)} F [y, z] (s) d \nospace s. \]
\end{corollary}

We arrive at our main theorem for global existence.
\begin{theorem}
  The remainder equation (\ref{DPDeq}) can be solved on $[0, \infty)$ almost
  surely.
\end{theorem}

\begin{proof}
  For any positive time $T^{\ast}$, from  Corollary \ref{cor-v-H0}
  we have the
  estimate
  \begin{align*}
    \| v \|^2_{H^0} (t) & \leqslant  2 \| y \|^2_{C_{T^{\ast}} H^0} + 2 e^{-
    t} \| v \|_{H^0}^2 (0) + 2 e^{- t} \| y \|_{H^0}^2 (0) + 2 C \int_0^t e^{-
    (t - s)} F [y, z] (s) d \nospace s\\
    & \leqslant  2 \| y \|^2_{C_{T^{\ast}} H^0} + 2 \| v \|_{H^0}^2 (0) + 2
    \| y \|_{H^0}^2 (0) + 2 C \int_0^t e^s F [y, z] (s) d \nospace s\\
    & \leqslant  2 \| y \|^2_{C_{T^{\ast}} H^0} + 2 \| v \|_{H^0}^2 (0) + 2
    \| y \|_{H^0}^2 (0) + 2 C \int_0^{T^{\ast}} e^s F [y, z] (s) d \nospace s.
  \end{align*}
  In the local existence result theorem \ref{Localexistence} for $v$, the
  solution can be established up to a time $T$ starting from initial value $v
  (0)$; $T$ depends on the norm of initial value and random objects (including
  $z$, $: z^2 :$, $: z^3 :$). The first step is to apply the local existence
  result for $T$ corresponding to norm of length $\sqrt{2 \| y
  \|^2_{C_{T^{\ast}} H^0} + 2 \| v \|_{H^0}^2 (0) + 2 \| y \|_{H^0}^2 (0) + 2
  C \int_0^{T^{\ast}} e^s F [y, z] (s) d \nospace s}$. Then solve the
  equation on the interval $[0, T \exterior T^{\ast}]$ (where $a \exterior b
  \assign \min \{ a, b \} \tmop{for} a, b \in \mathbb{R}$). Since the solution
  is in $K_T^{\frac{1}{2} - \varepsilon}$, $\| v \|_{H^0} (T)$ is finite and
  by the a priori estimate the equation starting with initial value $v (T)$ 
  can be solved on the interval $[T \exterior T^{\ast}, 2 T \exterior T^{\ast}]$.
  Using the a priori estimate on $[0, 2 T \exterior T^{\ast}]$ we can continue to
  solve the equation starting at time $2 T$, and we get solution on $[2 T
  \exterior T^{\ast}, 3 T \exterior T^{\ast}]$. Continuing this procedure,
   we get the
  solution on the whole interval $[0, T^{\ast}]$. Since $T^{\ast}$ is
  arbitrary, we get the solution on $[0, \infty)$.
\end{proof}

\section{Existence of an Invariant Measure}\label{EIM}

In this section, we construct an invariant measure for the renormalized stochastic quantization
equation
\[ \partial_t \phi_{mn} = - A_{m \nospace n} \phi_{m \nospace n} - \frac{
   \theta \lambda}{4} \sum_{k, l} : \phi_{m \nospace k} \phi_{k \nospace l}
   \phi_{l \nospace n} : + \dot{B}_t^{(m \nospace n)} \]
by the Krylov--Bogoliubov method 
{\cite{da1996ergodicity,tsatsoulis2018spectral}}. We define the  probability
measures
\[ 
R_t^{\ast} \mu \assign \frac{1}{t} \int_0^t \mu_s d \nospace s 
\]
where $\mu$ is the law of the initial value, that is to say we run the stochastic quantization equation with initial value $\phi (0) = z (0)+v(0)$, and $\mu_t$ is the distribution of the solution at time $t$.

\begin{theorem}
  Suppose $\phi (0) = z (0)+v(0)  \in H^{- \frac{1}{2} - \varepsilon}$ with the random field $v(0)  \in H^{0}$ a.s., then there exists a
  sequence of time variables $t_k \rightarrow \infty$, such that the sequence
  of probability measures
  \[ \frac{1}{t_k} \int_0^{t_k} \mu_s d \nospace s \]
  has a weak limit in $\mathcal{M}_1 \left( H^{- \frac{1}{2} - \varepsilon}
  \right)$.
\end{theorem}

\begin{proof}
  According to Markov's and Jensen's inequalities
  \[ \mathbb{P} \left[ \| \phi \|_{H^{- \frac{1}{2} -
     \frac{\varepsilon}{2}}} > a  \right] \leqslant \frac{\mathbb{E} \left[ \|
     \phi \|_{H^{- \frac{1}{2} - \frac{\varepsilon}{2}}}
     \right]}{a} \leqslant \frac{\mathbb{E} \left[ \| \phi \|^{2
     p}_{H^{- \frac{1}{2} - \frac{\varepsilon}{2}}} \right]^{\frac{1}{2
     p}}}{a} \]
  for any $a > 0$ and $2 p \geqslant 1$. Then
  \begin{align*}
R_t^{\ast} \mu \left( \left\{ \| \phi 
    \|_{H^{- \frac{1}{2} - \frac{\varepsilon}{2}}} > a  \right\} \right)
    & =  \frac{1}{t} \int_0^t \mathbb{P} \left[ \| \phi (s)
    \|_{H^{- \frac{1}{2} - \frac{\varepsilon}{2}}} > a  \right] d \nospace s\\
    & \leqslant  \frac{1}{a \nospace t} \int_0^t \mathbb{E} \left[ \| \phi
    (s) \|^{2 p}_{H^{- \frac{1}{2} - \frac{\varepsilon}{2}}}
    \right]^{\frac{1}{2 p}} d \nospace s\\
    & \leqslant  \frac{1}{a \nospace t} \int_0^t \mathbb{E} [(4 e^{- s} \| v
    \|_{H^0}^2 (0) + G [y, z] (s))^p]^{\frac{1}{2 p}} d \nospace s\\
    & \leqslant  \frac{1}{a \nospace t} \int_0^t \left( 4 e^{- s} \mathbb{E}
    [\| v \|_{H^0}^{2 p} (0)]^{\frac{1}{p}} +\mathbb{E} [G [y, z]^p
    (s)]^{\frac{1}{p}} \right)^{\frac{1}{2}} d \nospace s.
\end{align*}
  Here $G$ is defined in Corollary \ref{Gdef}. The stationarity of $y$ and $z$ implies that $\mathbb{E} [G [y, z]^p
  (s)]^{\frac{1}{p}}$ is a time independent constant and the function
  $\frac{1}{\nospace t} \int_0^t (A \nospace e^{- s} + B)^{\frac{1}{2}} d
  \nospace s$ is bounded and continuous function on $[0, \infty)$. Therefore,
  there is a constant $C$ such that 
  \[ R_t^{\ast} \mu \left( \left\{ \| \phi 
     \|_{H^{- \frac{1}{2} - \frac{\varepsilon}{2}}} > a  \right\} \right)
     \leqslant \frac{C}{a} . \]
 For a fixed $\delta >0$  we take $a = \frac{C}{\delta}$ and define $K_\delta =\left\{ \| \phi  \|_{H^{- \frac{1}{2} - \frac{\varepsilon}{2}}} \leqslant
  \frac{C}{\delta}  \right\}$. According to Lemma   \ref{Compactembedding} $K_\delta$ is compact in
  $H^{- \frac{1}{2} - \varepsilon}$ and therefore 
the collection of probability measure $\{ R_t^{\ast} \mu, t
  \geqslant 0 \}$ is tight in $H^{- \frac{1}{2} - \varepsilon}$. We conclude by invoking Corollary 3.1.2 in {\cite{da1996ergodicity}}. 
\end{proof}

\section{Outlook}

The construction of (\ref{GW-action}) for $\Omega = 1$ in dimension $d = 2$ is
interesting and important on its own right. But of course our dream is to
tackle the critical case $d = 4$ in near future. Our hope
that this should be possible rests on the remarkable result of
{\cite{Grosse:2019qps}} that the planar sector of the $d = 4$-dimensional
model (\ref{GW-action}) (at $\Omega = 1$) lives effectively in spectral
dimension $4 - \frac{2}{\pi} \arcsin (\lambda \pi)$ for $0 \leq \lambda \leq
\frac{1}{\pi}$ and in spectral dimension $3$ for $\lambda \geq \frac{1}{\pi}$.

We propose to take as reference distribution $z$ (see formula (\ref{Free}))
not the linear Gaussian theory, whose irregularity of dimension $4$ would be
intractable, {\tmem{but the stochastic process (to construct!) which
corresponds to the restriction to the planar theory}}, which effectively lives
in subcritical dimension. Both the planar and the Gaussian theory are exactly
solvable. If $z$ is available, the task is to control the remainder $v = \phi
- z$ in a similar way as we did for $d = 2$ in this paper. For that one should
first generalize this paper to fractional subcritical dimension $4 -
\epsilon$, which probably needs some of the refined methods developed for standard
$\lambda \phi^4_3$. The solution $z$ for the planar 4d model has a concrete
$\epsilon = \frac{2}{\pi} \arcsin (\lambda \pi)$, but also a modified
non-linearity. From $(v + z) \star (v + z) \star (v + z)$ discussed in this
paper a certain planar part (still to understand) of $z \star z \star z$ (and
some $z$-linear term) is subtracted. This only affects the constant $: z^3$:
in (3) (the difference will have improved regularity!) but not the decisive
operators $\mathcal{N}_1 (v), ..., \mathcal{N}_7 (v)$ (see formula
(\ref{Terms})) and its bounds. Since the non-planarity is captured by the
operators $\mathcal{N}_i (v)$, which we now control for $d = 2$ and with
reasonable hope soon for $d = 4 - \epsilon$, we are confident that along this
strategy the full construction of the $\lambda \phi^4$ Euclidean QFT on
4-dimensional Moyal space can succeed.

\appendix\section{Gaussian Hypercontractivity}\label{Hyper}

This appendix reviews the well-known Gaussian hypercontractivity bounds. An introduction can be found in  
{\cite[Appendix D.4]{friz2010multidimensional}} or Chapter 1.4.3 in
{\cite{nualart2006malliavin}}.

Recall that the Hermite polynomials $(H_k)_{k \geq 0}$ can be defined recursively via
\[
\begin{aligned}
	&\textnormal{(1)} \quad H_0(x)= 1;\\
	&\textnormal{(2)} \quad
	(H_k)'(x) = k H_{k-1}(x)
	\quad \text{for } k\geq 1;\\
	&\textnormal{(3)} \quad
	\mathbb{E}\bigl[H_k(X)]=0
	\quad \text{for } k\geq 1,
\end{aligned}
\]
where the last identity holds for a standard Gaussian random variable
$X \sim \mathcal{N}(0,1)$.

Let $\mathcal{G} \subset L^2(\Omega,\mathcal{F},\mathbb{P})$ be a
separable Gaussian Hilbert space, and let $(X_i)_{i\in I}$ be an
orthonormal basis of $\mathcal{G}$, where $I$ is finite or countable.
Thus, the $X_i$ are independent standard Gaussian random variables.

Let $\mathbb{N}_0^{(I)}$ denote the set of multi-indices
$\alpha=(\alpha_i)_{i\in I}$ with finite support. For
$\alpha\in\mathbb{N}_0^{(I)}$, set
\[
|\alpha|
:=
\sum_{i\in I}\alpha_i,
\qquad
\alpha!
:=
\prod_{i\in I}\alpha_i!,
\qquad
H_\alpha
:=
\prod_{i\in I}H_{\alpha_i}(X_i).
\]
All products above are finite. The family
\[
\left\{
\frac{1}{\sqrt{\alpha!}}H_\alpha
\,:\,
\alpha\in\mathbb{N}_0^{(I)}
\right\}
\]
forms an orthonormal basis of
$L^2(\Omega,\sigma(\mathcal{G}),\mathbb{P})$.

The homogeneous Wiener chaos of degree $n$, denoted by
$\mathcal{W}^{(n)}$, is the $L^2(\mathbb{P})$-closure of the linear
subspace generated by all Wick monomials of total degree $n$:
\[
\mathcal{W}^{(n)}
:=
\overline{
	\operatorname{span}
	\left\{
	H_\alpha 
	\,:\,
	|\alpha|=n
	\right\}
}^{\,L^2(\mathbb{P})}.
\]
The in-homogeneous Wiener chaos of degree $n$, denoted by
$\mathcal{C}^{(n)}$, is given by
\[
\mathcal{C}^{(n)}
:=
\bigoplus_{k=0}^{n}\mathcal{W}^{(k)}.
\]

\begin{theorem}
  The following bounds are true:
  
  (1) If $\psi \in \mathcal{W}^{(n)}$ and $1 < p < q < \infty$, then
  \[ \| \psi \|_{L^p (\mathbb{P})} \leqslant \| \psi \|_{L^q (\mathbb{P})}
     \leqslant \left( \frac{q - 1}{p - 1} \right)^{\frac{n}{2}} \| \psi
     \|_{L^p (\mathbb{P})} ; \]

  (2) If $\psi \in \mathcal{C}^{(n)}$ and $1 < p < q < \infty$, then
  \[ \| \psi \|_{L^p (\mathbb{P})} \leqslant \| \psi \|_{L^q (\mathbb{P})}
     \leqslant (n + 1) (q - 1)^{\frac{n}{2}} \max \{ 1, (p - 1)^{- n} \} \|
     \psi \|_{L^p (\mathbb{P})} . \]
%
%
\end{theorem}
%

\section{Moyal Product and Matrix Basis}\label{Aappendix}

The main reference in this appendix is {\cite{gracia1988algebras}} and
{\cite{Grosse:2003nw}}. Given the definition of Moyal product in Section
\ref{1section}, we list a few of its properties.

\begin{lemma}
Let  $f, g \in \mathcal{S}(\mathbb{R}^d)$ be complex valued Schwartz functions. Then
  
  1. $f \star g = \overline{\bar{g} \star \bar{f}}$;
  
  2. If $f$ is real valued, then so is $f \star f$;
  
  3. $\int_{\mathbb{R}^d} (f \star g) (x) d \nospace x = \int_{\mathbb{R}^d} f
  (x) g (x) d \nospace x = \int_{\mathbb{R}^d} (g \star f) (x) d \nospace x$.
\end{lemma}
The Moyal product can be extended to a large class of tempered distributions,
see {\cite{gracia1988algebras}}.

The matrix basis in the two-dimensional case $d=2$ is defined as
follows: starting from
\[ 
b_{00} (x) \assign 2 e^{- \frac{\| x \|^2}{\theta}}, \quad a = \frac{x_1 +
   i \nospace x_2}{\sqrt{2}}, \quad \bar{a} = \frac{x_1 - i \nospace
   x_2}{\sqrt{2}}  \qquad x=(x_1,x_2) \in \mathbb{R}^2
   \] 
one gets a two parameter family of functions
\[ 
b_{m \nospace n} (x) \assign \frac{\bar{a}^{\star m} \star b_{00} \star
   a^{\star n}}{\sqrt{m!n! \theta^{m + n}}} (x) 
   \]
where $a^{\star n}$ denotes the $n$-fold Moyal product of $a$ with itself. A
few properties of the matrix basis are listed below.

\begin{lemma}
  For the matrix basis $\{ b_{m \nospace n} \}_{m, n = 0}^{+ \infty}$ defined
  above, we have:
  
  1. $b_{00} \star b_{00} = b_{00}, \quad a \star b_{00} = 0 = b_{00} \star
  \bar{a}, \quad [\bar{a}, a]_{\star} \assign \bar{a} \star a - a \star
  \bar{a} = \theta$;
  
  2. $\{ b_{m \nospace n} \}_{m, n = 0}^{+ \infty}$ forms an orthonormal basis
  for $L^2 (\mathbb{R}^2)$;
  
  3. $b_{k \nospace l} \star b_{m \nospace n} = \delta_{l \nospace m} b_{k
  \nospace n}$, hence if two Schwartz functions $f, g \in \mathcal{S}
  (\mathbb{R}^d)$ are expanded in this basis as
  \[ f (x) = \sum_{m, n = 0}^{\infty} f_{m \nospace n} b_{m \nospace n} (x),
     \quad g (x) = \sum_{m, n = 0}^{\infty} g_{m \nospace n} b_{m \nospace n}
     (x) \]
  then the coefficients of Moyal product are given by  a matrix product of
 the  corresponding coefficients
  \[ (f \star g) (x) = \sum_{m, n = 0}^{\infty} \left( \sum_{k = 0}^{\infty}
     f_{m \nospace k} g_{k \nospace n} \right) b_{m \nospace n} (x) ; \]

  4. $b_{m \nospace n} (x) = 2 (- 1)^m \sqrt{\frac{m!}{n!}} \left(
  \sqrt{\frac{2}{\theta}} (x_1 + i \nospace x_2) \right)^{n - m} L^{n - m}_m
  \left( \frac{2}{\theta} \| x \|^2 \right) e^{- \frac{\| x \|^2}{\theta}}$
  where $L_m^{\alpha}$ are associate Laguerre polynomials.
\end{lemma}

\section{Spaces of Matrices}\label{spaceofmatrix}

Recall the definitions of the following Banach spaces of matrices
\[ H^{\alpha} = \left\{ (c_{m n}) : \| c \|_{H^{\alpha}} \assign \left(
   \sum_{m, n = 0}^{+ \infty} A_{m \nospace n}^{2 \alpha} | c_{m \nospace n}
   |^2 \right)^{\frac{1}{2}} < + \infty \right\} \]
and
\[
C_T H^\alpha
:=
C\bigl([0,T];H^\alpha\bigr)
=
\left\{
c:[0,T]\longrightarrow H^\alpha
\ \middle|\
c \text{ is continuous}
\right\},
\qquad
\|c\|_{C_T H^\alpha}
:=
\sup_{t\in[0,T]}\|c(t)\|_{H^\alpha}.
\]

One clearly has $\| c \|_{H^{\beta}} \leqslant \|
c \|_{H^{\alpha}}$ if $\alpha \geqslant \beta$, and hence $H^{\alpha} \subset
H^{\beta}$ for $\alpha \geqslant \beta$. Moreover, this embedding is compact.

\begin{lemma}[Compact embedding]
  \label{Compactembedding}If $\alpha > \beta$, then the embedding $i :
  H^{\alpha} \hookrightarrow H^{\beta}$ is compact.
\end{lemma}

\begin{proof}
  The embedding $i$ can be approximated in operator norm by the finite-rank operators
   $i_N : H^{\alpha} \rightarrow H^{\beta}$ as $N \rightarrow \infty$ 
  defined through the formula
  \[ i_N (c)_{m n} \assign c_{m n} \tmop{if} m, n \leqslant N \infixand i_N
     (c)_{m n} \assign 0 \tmop{if} m > N \tmop{or} n > N. 
     \]
\end{proof}

We have the following multiplicative inequality.

\begin{lemma}\label{lem:multiplicative_inequalities}
\begin{enumerate}
\item	If $v,w\in H^\alpha$ for $\alpha\geq 0$, then
	\begin{equation}\label{multiplicativeineq}
		\|wv\|_{H^{\alpha}}
		\lesssim_{\alpha}
		\|w\|_{H^\alpha}\|v\|_{H^0}
		+
		\|w\|_{H^0}\|v\|_{H^\alpha}.
	\end{equation}
\item Let $\alpha, \beta, \gamma$ satisify $\alpha \in \mathbb{R}$, $\gamma, \beta  \in (0, \frac12)$, $\beta+ \gamma > \frac12$  and $\alpha + \beta + \gamma > 1$. 
If $v \in H^{\beta}$ and $z \in M^{\gamma}$, then 
\[
\| z v\|_{H^{-\alpha}}   +  \| vz \|_{H^{-\alpha}}   \lesssim \| v \|_{H^{\beta}} \| z \|_{M^{\gamma}}
\]

\end{enumerate}
\end{lemma}

\begin{proof}
(1) The Cauchy Schwartz inequality yields
  \begin{align}
\notag
    \| w \nospace v \|_{H^{\alpha}}^2  = & \sum_{m, n \geqslant 0} A_{m n}^{2
    \alpha} | (w \nospace v)_{m n} |^2
     =  \sum_{m, n \geqslant 0} A_{m n}^{2 \alpha} \Big| \sum_{k \geqslant
    0} w_{m k} v_{k n} \Big|^2\\
      \label{eq:proof-multiplicative-inequality}
    & \leqslant  \sum_{m, n \geqslant 0} A_{m n}^{2 \alpha} \sum_{k
    \geqslant 0} | w_{m k} |^2 \sum_{k' \geqslant 0} | v_{k' n} |^2.
   \end{align}
   Now, using the simple identity 
   \[
   A_{m n} = m+n+1+ \frac{\theta M^2}{4} = \frac12 ( A_{m m} + A_{n n} ),
   \]
   which yields $A_{m n}^{2 \alpha}  \leq C(\alpha) ( A_{m m}^{2 \alpha} + A_{nn}^{2 \alpha}) $ this estimate turns into  
     \begin{align*}
  \text{R.H.S. of \eqref{eq:proof-multiplicative-inequality}} &\lesssim
 \sum_{m,k    \geqslant 0}  A_{m m}^{2 \alpha}  | w_{m k} |^2 \sum_{n,k' \geqslant 0} | v_{k' n} |^2 +  \sum_{m,k    \geqslant 0}  | w_{m k} |^2 \sum_{n,k' \geqslant 0} A_{n n}^{2 \alpha}  | v_{k' n} |^2 
 \\
 &\lesssim   \sum_{m,k    \geqslant 0}  A_{m k}^{2 \alpha}  | w_{m k} |^2 \sum_{n,k' \geqslant 0} | v_{k' n} |^2 +  \sum_{m,k    \geqslant 0}  | w_{m k} |^2 \sum_{n,k' \geqslant 0} A_{n k'}^{2 \alpha}  | v_{k' n} |^2 ,
   \end{align*}
   which is the right hand side of \eqref{multiplicativeineq}.
   
(2) We give the argument for $ z v$, the argument for $v z$ follows analogously: 
  \begin{align*}
  \| z  v  \|_{H^{- \alpha}}^2
    & =  \sum_{m, n \geqslant 0} \frac{1}{A_{m n}^{2 \alpha}} \Big| \sum_{k
      \geqslant 0} z_{m k}  v_{k n}  \Big|^2
    \\
    & \leqslant  \sum_{m, n \geqslant 0} \frac{1}{A_{m n}^{2 \alpha}} \Big(
    \sum_{k \geqslant 0} \frac{\| z  \|_{M^{\gamma}}}{A_{m k}^{\gamma}} | v_{k n}  |
\Big)^2
\\
    & =  \| z  \|_{M^{\gamma}}^2 \sum_{m, n \geqslant 0} \frac{1}{A_{m n}^{2 \alpha}} \Big( \sum_{k
    \geqslant 0} \frac{1}{A_{m
    k}^{\gamma} A_{k n}^{\beta}} A_{k n}^{\beta} | v_{k n}
 | \Big)^2
\\
    & \leqslant  \| z  \|_{M^{\gamma}}^2 \sum_{m, n
    \geqslant 0} \frac{1}{A_{m n}^{2 \alpha}} \Big( \sum_{k \geqslant 0}
    \frac{1}{A_{m k}^{2 \gamma} A_{k n}^{2 \beta}} \Big) \Big(
    \sum_{k' \geqslant 0} A_{k' n}^{2 \beta} | v_{k' n}  |^2 \Big)\\
    & \leqslant  \| z  \|_{M^{\gamma}}^2 \sum_{m, n
    \geqslant 0} \frac{1}{A_{m n}^{2 \alpha}}  \frac{1}{A_{m n}^{2 \gamma + 2 \beta -1
 }} \Big( \sum_{k' \geqslant 0} A_{k' n}^{2 \beta} | v_{k'
  n}  |^2 \Big)
\\
    & \lesssim  \| z  \|_{M^{\gamma}}^2 \sum_{m, n
    \geqslant 0} \frac{1}{A_{m m}^{2 \alpha + 2 \beta + 2 \gamma-1}}
    \Big( \sum_{k' \geqslant 0} A_{k' n}^{2 \beta} | v_{k' n}  |^2
    \Big)
    \\
    & =  \| z  \|_{M^{\gamma}}^2 \sum_{m \geqslant 0}
    \frac{1}{A_{m m}^{2 \alpha + 2 \beta + 2 \gamma-1}} \Big( \sum_{k', n
      \geqslant 0} A_{k' n}^{2 \beta} | v_{k' n}  |^2 \Big)
    \\
    & =  \| z  \|_{M^{\gamma}}^2 \| v 
    \|_{H^{\beta}}^2 \sum_{m \geqslant 0} \frac{1}{A_{m m}^{2 \alpha + 2 \beta + 2 \gamma-1}}.
\end{align*}
Here, in the fifth line we have made use of Lemma \ref{lem:AppendixDEstimates} (1) for $2 \beta$, $2 \gamma \in (0,1)$ with $2 \beta + 2 \gamma >1$, in the sixth line we have used $A_{mm} \geq \frac12 A_{nm}$
and in the last line we have used $2\alpha + 2 \beta + 2 \gamma -1 > 1$. 
\end{proof}

\begin{lemma}[Interpolation Inequality]
  Suppose $\alpha, \alpha_1, \alpha_2 \in \mathbb{R}$ such that $\alpha =
  \theta \alpha_1 + (1 - \theta) \alpha_2$ for $\theta \in (0, 1)$, then
  \[ \| \phi \|_{H^{\alpha}} \leqslant \| \phi \|_{H^{\alpha_1}}^{\theta} \|
     \phi \|_{H^{\alpha_2}}^{1 - \theta} . \]
\end{lemma}

\begin{proof}
  By definition and H\"older's inequality
  \begin{align*}
    \| \phi \|_{H^{\alpha}}^2 & =  \sum_{m, n} A_{m n}^{2 \alpha} | \phi_{m
    n} |^2\\
    & =  \sum_{m, n} A_{m n}^{2 \theta \alpha_1} | \phi_{m n} |^{2 \theta}
    A_{m n}^{2 (1 - \theta) \alpha_2} | \phi_{m n} |^{2(1 - \theta)}\\
    & \leqslant  \left[ \sum_{m, n} (A_{m n}^{2 \theta \alpha_1} | \phi_{m
    n} |^{2 \theta})^{\frac{1}{\theta}} \right]^{\theta} \left[ \sum_{m, n}
    (A_{m n}^{2 (1 - \theta) \alpha_2} | \phi_{m n} |^{2(1 -
    \theta)})^{\frac{1}{1 - \theta}} \right]^{1 - \theta}\\
    & =  \left[ \sum_{m, n} A_{m n}^{2 \alpha_1} | \phi_{m n} |^2
    \right]^{\theta} \left[ \sum_{m, n} A_{m n}^{2 \alpha_2} | \phi_{m n} |^2 
    \right]^{1 - \theta}\\
    & =  \| \phi \|_{H^{\alpha_1}}^{2 \theta} \| \phi \|_{H^{\alpha_2}}^{2
    (1 - \theta)}.
\end{align*}
 \end{proof}

\begin{lemma}[Duality]
  \label{duality}Suppose we have two matrices $a = (a)_{m, n \in \mathbb{N}}$
  and $b = (b)_{m, n \in \mathbb{N}}$, then
  \[ | \tmop{tr} (a \nospace b) | \leqslant \| a \|_{H^{- \alpha}} \| b
     \|_{H^{\alpha}} \]
  where the trace is defined as
  \[ \tmop{tr} (a \nospace b) \assign \sum_{m, n \geqslant 0} a_{m n} b_{n m}
     . \]
\end{lemma}

\begin{proof}
  By definition and the Cauchy Schwarz inequality we get
  \begin{align*}
    | \tmop{tr} (a \nospace b) | & =  \left| \sum_{m, n} a_{m n} b_{n m}
    \right|\\
    & =  \left| \sum_{m, n} A_{m n}^{- \alpha} a_{m n} A_{m n}^{\alpha} b_{n
    m} \right|\\
    & \leqslant  \left( \sum_{m, n = 0}^{\infty} | A_{m n}^{- \alpha} a_{m
    n} |^2 \right)^{1 / 2} \left( \sum_{m, n = 0}^{\infty} | A_{m n}^{\alpha}
    b_{m n} |^2 \right)^{1 / 2} \\
    & =  \| a \|_{H^{- \alpha}} \| b \|_{H^{\alpha}}.
  \end{align*}
\end{proof}

The following lemma is useful in the proof of the a priori estimate.

\begin{lemma}
	\label{specialineq}Given a Hermitian matrix $v\in H^0$, for any $\varepsilon>0$ we have
	\[
	\|v\|_{H^{-\frac14-\varepsilon}}^2
	\lesssim
	\|v^2\|_{H^0}.
	\]
\end{lemma}

\begin{proof}
	By Lemma~C.2, $v^2\in H^0$. Since $A_{mm}\leq 2A_{mn}$ and $v$ is
	Hermitian, we have
	\[
	\begin{aligned}
		\|v\|_{H^{-\frac14-\varepsilon}}^2
		&=
		\sum_{m,n\geq0}
		\frac{|v_{mn}|^2}{A_{mn}^{\frac12+2\varepsilon}}
		\lesssim
		\sum_{m\geq0}
		\frac{1}{A_{mm}^{\frac12+2\varepsilon}}
		\sum_{n\geq0}|v_{mn}|^2                                      \\
		&=
		\sum_{m\geq0}
		\frac{|(v^2)_{mm}|}{A_{mm}^{\frac12+2\varepsilon}}
		\leq
		\left(
		\sum_{m\geq0}\frac{1}{A_{mm}^{1+4\varepsilon}}
		\right)^{\frac12}
		\left(
		\sum_{m\geq0}|(v^2)_{mm}|^2
		\right)^{\frac12}                                             \\
		&\lesssim
		\|v^2\|_{H^0},
	\end{aligned}
	\]
	where we used the Cauchy--Schwarz inequality and the fact that
	$\sum_{m\geq0}A_{mm}^{-1-4\varepsilon}<\infty$.
\end{proof}

We have the following Schauder estimates.

\begin{lemma}[Schauder estimates]
  \label{Schauder}Suppose we have a system of equations for a matrix $\phi$ of
  the form
  \[ \partial_t \phi_{m n} = - A_{m n} \phi_{m n} + \psi_{m n} \qquad
    \tmop{for} m, n
    \in \mathbb{N} ,
  \]
   where $A_{m n} = m+n+1+\frac{\theta M^2}{4}$ and
   $\psi \in C_T H^{\alpha}$. If the initial value is $\phi (0) =
  0$, then for any $\varepsilon \in (0,1)$
  \[ \| \phi (t) \|_{H^{\alpha + (1 - \varepsilon)}} \lesssim \int_0^t (t -
     s)^{- (1 - \varepsilon)} \| \psi (s) \|_{H^{\alpha}} d \nospace s \]
  and, moreover, $\| \phi \|_{C_T H^{\alpha + (1 - \varepsilon)}} \lesssim
  T^{\varepsilon} \| \psi \|_{C_T H^{\alpha}}$.
\end{lemma}

\begin{proof}
  From the equation we get
  \[ \phi_{m n} (t) = e^{- A_{m n} t} \int_0^t e^{A_{m n} s} \psi_{m n} (s) d
    \nospace s \qquad \infixor \qquad
    \phi (t) = \int_0^t e^{A (s - t)} \psi (s) d \nospace
     s .\]
  Then
  \begin{align*}
 \| e^{A (s - t)} \psi (s) \|_{H^{\alpha + (1 - \varepsilon)}}^2
    & =  \sum_{m, n \geqslant 0} A_{m n}^{2 \alpha + 2 (1 - \varepsilon)}
    e^{- 2 A_{m \nospace n} (t - s)} | \psi_{m n} (s) |^2\\
    & =  (t - s)^{- 2 (1 - \varepsilon)} \sum_{m, n \geqslant 0} A_{m n}^{2
    \alpha} ((t - s) A_{m n})^{2 (1 - \varepsilon)} e^{- 2 A_{m \nospace n} (t
    - s)} | \psi_{m n} (s) |^2\\
    & \lesssim  (t - s)^{- 2 (1 - \varepsilon)} \sum_{m, n \geqslant 0} A_{m
    n}^{2 \alpha} | \psi_{m n} (s) |^2\\
    & =  (t - s)^{- 2 (1 - \varepsilon)} \| \psi (s) \|_{H^{\alpha}}^2,
  \end{align*}
  so
  \begin{align*}
    \| \phi (t) \|_{H^{\alpha + (1 - \varepsilon)}} & =  \left\| \int_0^t
    e^{A (s - t)} \psi (s) d \nospace s \right\|_{H^{\alpha + (1 -
    \varepsilon)}}\\
    & \leqslant  \int_0^t \| e^{A (s - t)} \psi (s) \|_{H^{\alpha + (1 -
    \varepsilon)}} d \nospace s\\
    & \lesssim  \int_0^t (t - s)^{- (1 - \varepsilon)} \| \psi (s)
    \|_{H^{\alpha}} d \nospace s\\
    & \leqslant  \int_0^t (t - s)^{- (1 - \varepsilon)} d \nospace s \| \psi
    \|_{C_t H^{\alpha}}\\
    & \sim  t^{\varepsilon} \| \psi \|_{C_t H^{\alpha}}
  \end{align*}
  for $\varepsilon \in (0, 1)$. Hence the result follows.
\end{proof}

Recall that, for $p\in\mathbb{R}$, we define
\[
M^p
\assign
\left\{
c=(c_{mn})_{m,n\geqslant 0}
\ \middle|\
\|c\|_{M^p}
\assign
\sup_{m,n\geqslant 0}A_{mn}^p|c_{mn}|<\infty
\right\}.
\]
and that its time-dependent counterpart is defined by
\[
C_T M^p
\assign
C([0,T];M^p),
\qquad
\|c\|_{C_T M^p}
\assign
\sup_{t\in[0,T]}\|c(t)\|_{M^p}.
\]

We have the following embedding.

\begin{lemma}
	\label{MtoHembedding}
	Suppose that $p,\alpha\in\mathbb{R}$ satisfy $p>\alpha+1$. Then the
	embeddings
	\[
	M^p\hookrightarrow H^\alpha
	\qquad\text{and}\qquad
	C_T M^p\hookrightarrow C_T H^\alpha
	\]
	are continuous. In particular, for every $\varepsilon>0$,
	\[
	M^p\hookrightarrow H^{p-1-\varepsilon},
	\qquad
	C_T M^p\hookrightarrow C_T H^{p-1-\varepsilon}.
	\]
\end{lemma}

\begin{proof}
	For $c\in M^p$, by the definition of $M^p$, we have
	\[
	|c_{mn}|
	\leqslant
	\frac{\|c\|_{M^p}}{A_{mn}^p}.
	\]
	Therefore,
	\[
	\begin{aligned}
		\|c\|_{H^\alpha}^2
		&=
		\sum_{m,n\geqslant 0}
		A_{mn}^{2\alpha}|c_{mn}|^2                                      \\
		&\leqslant
		\|c\|_{M^p}^2
		\sum_{m,n\geqslant 0}
		\frac{1}{A_{mn}^{2(p-\alpha)}}
		\lesssim
		\|c\|_{M^p}^2,
	\end{aligned}
	\]
	where the last double sum is finite since
	$2(p-\alpha)>2$. Hence
	\[
	\|c\|_{H^\alpha}\lesssim\|c\|_{M^p},
	\]
	which proves the first embedding.
	
	For $c\in C_T M^p$, applying the previous estimate at every
	$t\in[0,T]$ gives
	\[
	\|c\|_{C_T H^\alpha}
	\lesssim
	\|c\|_{C_T M^p}.
	\]
	Moreover,
	\[
	\|c(t)-c(s)\|_{H^\alpha}
	\lesssim
	\|c(t)-c(s)\|_{M^p},
	\]
	so the continuity of $c$ with values in $M^p$ implies its continuity
	with values in $H^\alpha$. This concludes the proof.
\end{proof}

\section{Inequalities Related to Correlation
Functions}\label{inequalitiescorrelation}

We have following inequalities involving correlation functions.

\begin{lemma}\label{lem:AppendixDEstimates}
  Given $A_{m \nospace n} \assign m+n+1+\frac{\theta M^2}{4}
  $ for $m, n \in \mathbb{N}$, we have following
  inequalities:
  
  (1) if $\alpha, \beta \in (0, 1)$ and $\alpha + \beta - 1 > 0$, then
  \[ \sum_{k = 0}^{\infty} \frac{1}{A_{m \nospace k}^{\alpha} A_{k \nospace
     n}^{\beta}} \lesssim \frac{1}{A_{m \nospace n}^{\alpha + \beta - 1}} ; \]

  (2) if $\alpha \geqslant 1$ or $\beta \geqslant 1$, and both $\alpha$ and $\beta$ are positive, then for any $\delta > 0$ we have
  \[ \sum_{k = 0}^{\infty} \frac{1}{A_{m \nospace k}^{\alpha} A_{k \nospace
     n}^{\beta}} \lesssim \frac{1}{A^{\min \{ \alpha, \beta \} - \delta}_{m
     \nospace n}} ; \]

  (3) if $\alpha, \beta > 0$, $\alpha + \beta - 1 > 0$ and $\alpha < 1$, then
  \[ \sum_{m = 0}^{\infty} \frac{1}{A_{m \nospace m}^{\alpha} A_{m \nospace
     n}^{\beta}} \lesssim \frac{1}{A_{n \nospace n}^{\alpha + \beta - 1}} ; \]

  (4) if $\beta > 0$ and $\alpha \geqslant 1$ then
  \[ \sum_m \frac{1}{A^{\alpha}_{m m} A_{m n}^{\beta}} \lesssim \frac{1}{A_{n
     n}^{\beta - \delta}} \]

  (5) if $\alpha > 1$, then $\sum_{m = 0}^{\infty} \frac{1}{A_{m \nospace
  n}^{\alpha}} \sim \frac{1}{A_{n \nospace n}^{\alpha - 1}}$;
  
  (6) if $\alpha \in (0, 1)$, then
  \[ \sum_{k = 0}^{\infty} \frac{1}{A_{m \nospace k} A_{k \nospace k}^{\alpha}
     A_{k \nospace n}} \lesssim \frac{1}{A_{m \nospace n}} . \]
\end{lemma}

\begin{proof}

For (1): Without loss of generality, assume $n \leq m$. We write 
  \begin{align*}
   \sum_{k = 0}^{\infty} \frac{1}{A_{m \nospace k}^{\alpha} A_{k \nospace
     n}^{\beta}} =     \sum_{k \leq n} \cdots +  \sum_{k = n+1}^m \cdots + \sum_{k \geq m+1} \cdots  = I + II + III   ,
     \end{align*}
where we use $\dots$ as short-hand for $\frac{1}{A_{m \nospace k}^{\alpha} A_{k \nospace
     n}^{\beta}} $. We bound the terms individually and get 
     \begin{align*}
     I \lesssim \sum_{k < n} \frac{1}{(1+m)^\alpha}   \frac{1}{(1+n)^\beta}  \leq \frac{1}{(1+m)^\alpha}   \frac{n}{(1+n)^\beta}   \leq  \frac{1}{(1+m)^{\alpha+\beta-1}}     \frac{(1+n)^{1-\beta} }{ (1+m)^{1-\beta}}  \lesssim  \frac{1}{A_{m \nospace n}^{\alpha + \beta - 1}} ,
     \end{align*}
where in the last inequality we have used that $m \geq n$ and $(1-\beta)>0$. For the second term we get
     \begin{align*}
     II \lesssim \sum_{k =n}^m \frac{1}{(1+m)^\alpha}   \frac{1}{(1+k)^\beta}      \leq  \frac{1}{(1+m)^\alpha}   \sum_{k =n}^m \frac{1}{(1+k)^\beta} \lesssim  \frac{1}{(1+m)^{\alpha + \beta - 1}} \lesssim        \frac{1}{A_{m \nospace n}^{\alpha + \beta - 1}} ,
     \end{align*}
where again we have used $\beta<1$ to bound the sum. Finally, we write 
     \begin{align*}
     III \lesssim \sum_{k \geq m+1} \frac{1}{(1+k)^\alpha}   \frac{1}{(1+k)^\beta}    \lesssim \frac{1}{(1+m)^{\alpha + \beta - 1}} \lesssim        \frac{1}{A_{m \nospace n}^{\alpha + \beta - 1}} ,
     \end{align*}
using the fact that $\alpha + \beta >1$ to estimate the sum, thus concluding the argument in case (1).

For (2), set $r:=\min\{\alpha,\beta\}$ and fix $0<\delta<r$.
By symmetry, interchanging $(m,\alpha)$ with $(n,\beta)$ if necessary,
we may assume that $n\leq m$. As in (1), we write
\[
\sum_{k=0}^{\infty}\frac{1}{A_{mk}^{\alpha}A_{kn}^{\beta}}
=
\sum_{k=0}^{n}\cdots
+\sum_{k=n+1}^{m}\cdots
+\sum_{k=m+1}^{\infty}\cdots
=:I+II+III.
\]

For the first term, we have
\[
I
\lesssim
\sum_{k=0}^{n}
\frac{1}{(1+m)^\alpha(1+n)^\beta}
=
\frac{(1+n)^{1-\beta}}{(1+m)^\alpha}.
\]
If $\beta\geq1$, then $(1+n)^{1-\beta}\leq1$, and since
$r\leq\alpha$,
\[
I
\lesssim
\frac{1}{(1+m)^\alpha}
\leq
\frac{1}{(1+m)^r}.
\]
If $\beta<1$, then the assumption in (2) implies that $\alpha\geq1$,
and hence $r=\beta$. Using $n\leq m$, we obtain
\[
I
\lesssim
\frac{(1+m)^{1-\beta}}{(1+m)^\alpha}
=
\frac{1}{(1+m)^{\alpha+\beta-1}}
\leq
\frac{1}{(1+m)^\beta}
=
\frac{1}{(1+m)^r}.
\]

For the second term,
\[
II
\lesssim
\frac{1}{(1+m)^\alpha}
\sum_{k=n+1}^{m}\frac{1}{(1+k)^\beta}.
\]
We distinguish three cases. If $\beta<1$, then
\[
\sum_{k=n+1}^{m}\frac{1}{(1+k)^\beta}
\lesssim
(1+m)^{1-\beta}.
\]
As above, $\alpha\geq1$ and $r=\beta$, so
\[
II
\lesssim
\frac{1}{(1+m)^{\alpha+\beta-1}}
\leq
\frac{1}{(1+m)^r}.
\]
If $\beta=1$, then
\[
\sum_{k=n+1}^{m}\frac{1}{1+k}
\lesssim
\log(1+m)
\lesssim
(1+m)^\delta.
\]
Consequently,
\[
II
\lesssim
\frac{1}{(1+m)^{\alpha-\delta}}
\leq
\frac{1}{(1+m)^{r-\delta}},
\]
where we used $r=\min\{\alpha,1\}\leq\alpha$. Finally, if
$\beta>1$, then
\[
\sum_{k=n+1}^{m}\frac{1}{(1+k)^\beta}
\leq
\sum_{k=0}^{\infty}\frac{1}{(1+k)^\beta}
<\infty,
\]
and therefore
\[
II
\lesssim
\frac{1}{(1+m)^\alpha}
\leq
\frac{1}{(1+m)^r}.
\]

For the last term, since $\alpha+\beta>1$, we have
\[
III
\lesssim
\sum_{k=m+1}^{\infty}
\frac{1}{(1+k)^{\alpha+\beta}}
\lesssim
\frac{1}{(1+m)^{\alpha+\beta-1}}.
\]
Moreover, $\max\{\alpha,\beta\}\geq1$ implies
\[
\alpha+\beta-1
\geq
\min\{\alpha,\beta\}
=
r,
\]
and hence
\[
III
\lesssim
\frac{1}{(1+m)^r}.
\]

Combining the estimates for $I$, $II$, and $III$, and using
$A_{mn}\sim1+m$ for $n\leq m$, we conclude that
\[
\sum_{k=0}^{\infty}
\frac{1}{A_{mk}^{\alpha}A_{kn}^{\beta}}
\lesssim
\frac{1}{(1+m)^{r-\delta}}
\sim
\frac{1}{A_{mn}^{\min\{\alpha,\beta\}-\delta}},
\]
which proves (2).

For (3), we split the sum into two parts:
\[
\sum_{m=0}^{\infty}
\frac{1}{A_{mm}^{\alpha}A_{mn}^{\beta}}
=
\sum_{m=0}^{n}
\frac{1}{A_{mm}^{\alpha}A_{mn}^{\beta}}
+
\sum_{m=n+1}^{\infty}
\frac{1}{A_{mm}^{\alpha}A_{mn}^{\beta}}
=:I+II.
\]
For the first term, since $m\leq n$ and $\alpha<1$, we have
\[
I \lesssim \frac{1}{(1+n)^\beta}
\sum_{m=0}^{n}\frac{1}{(1+m)^\alpha}
\lesssim \frac{1}{(1+n)^\beta}(1+n)^{1-\alpha}
= \frac{1}{(1+n)^{\alpha+\beta-1}}.
\]

For the second term, since $m>n$, both $A_{mm}$ and $A_{mn}$ are bounded from below by a constant multiple of $1+m$, and  $\alpha+\beta>1$, we have
\[
II \lesssim \sum_{m=n+1}^{\infty}\frac{1}{(1+m)^{\alpha+\beta}}
\lesssim \frac{1}{(1+n)^{\alpha+\beta-1}}.
\]

Combining the two estimates and using $A_{nn}\sim1+n$, we conclude that
\[
\sum_{m=0}^{\infty}
\frac{1}{A_{mm}^{\alpha}A_{mn}^{\beta}}
\lesssim
\frac{1}{(1+n)^{\alpha+\beta-1}}
\sim
\frac{1}{A_{nn}^{\alpha+\beta-1}},
\]
which proves (3).

For (4), let $0<\delta<\beta$. Since $A_{mn}\gtrsim A_{mm}$ and
$A_{mn}\gtrsim A_{nn}$, we have
\begin{align*}
	\sum_{m=0}^{\infty}\frac{1}{A_{mm}^{\alpha}A_{mn}^{\beta}}
	&=
	\sum_{m=0}^{\infty}
	\frac{1}{A_{mm}^{\alpha}A_{mn}^{\delta}
		A_{mn}^{\beta-\delta}} \\
	&\lesssim
	\frac{1}{A_{nn}^{\beta-\delta}}
	\sum_{m=0}^{\infty}\frac{1}{A_{mm}^{\alpha+\delta}}
	\lesssim
	\frac{1}{A_{nn}^{\beta-\delta}},
\end{align*}
where the last series converges because $\alpha+\delta>1$.

For (5), since $A_{mn}\sim 1+m+n$ and $A_{nn}\sim1+n$, a change
of summation index gives
\[
\sum_{m=0}^{\infty}\frac{1}{A_{mn}^{\alpha}}
\sim
\sum_{m=0}^{\infty}\frac{1}{(1+m+n)^\alpha}
=
\sum_{j=n}^{\infty}\frac{1}{(1+j)^\alpha}
\sim
\frac{1}{(1+n)^{\alpha-1}}
\sim
\frac{1}{A_{nn}^{\alpha-1}},
\]
where we used the standard tail estimate for a convergent $p$-series,
valid for $\alpha>1$.

For (6), the identity
$A_{mk}+A_{kn}=A_{mn}+A_{kk}$ implies
$A_{mk}+A_{kn}\geq A_{mn}$. Hence
\begin{align*}
	\sum_{k=0}^{\infty}
	\frac{1}{A_{mk}A_{kk}^{\alpha}A_{kn}}
	&\leq
	\frac{1}{A_{mn}}
	\sum_{k=0}^{\infty}
	\left(
	\frac{1}{A_{mk}A_{kk}^{\alpha}}
	+
	\frac{1}{A_{kn}A_{kk}^{\alpha}}
	\right) \\
	&\lesssim
	\frac{1}{A_{mn}}
	\sum_{k=0}^{\infty}\frac{1}{A_{kk}^{1+\alpha}}
	\lesssim
	\frac{1}{A_{mn}},
\end{align*}
where we used $A_{mk},A_{kn}\gtrsim A_{kk}$.
  
\end{proof}

\section{Construction of $: z^2 :$ and $: z^3 :$}\label{Dappendix}

Let $z=(z_{mn})_{m,n\geq0}$ be the solution of the system of SDEs
\[
\partial_t z_{mn}
=
-A_{mn}z_{mn}+\dot B_t^{(mn)},
\qquad
\mathbb{E}\big[\dot B_t^{(mn)}\dot B_s^{(kl)}\big]
=
2\delta(t-s)\delta_{ml}\delta_{nk},
\]
where
\[
B_t^{(mn)}=\overline{B_t^{(nm)}},
\]
and the initial condition
\[
z(0)=\bigl(z_{mn}(0)\bigr)_{m,n\geq0}
\]
is a centered jointly Gaussian Hermitian matrix, independent of the driving
Brownian motions, with covariance
\[
\mathbb{E}\big[z_{mn}(0)z_{kl}(0)\big]
=
\frac{\delta_{ml}\delta_{nk}}{A_{mn}}.
\]
Then $z$ is a centered stationary Hermitian Gaussian process with correlation
function
\[
\mathbb{E}\big[z_{mn}(t)z_{kl}(s)\big]
=
\frac{\delta_{ml}\delta_{nk}}{A_{mn}}
e^{-|t-s|A_{mn}},
\qquad s,t\geq0.
\]

For $N\in\mathbb{N}$, define the cutoff matrix $z^{(N)}$ by
\[
z_{mn}^{(N)}
=
\begin{cases}
	z_{mn},&0\leq m,n\leq N,\\
	0,&\text{otherwise},
\end{cases}
\]
and define its renormalized square by
\[
:z^2:_{mn}^{(N)}
:=
\sum_{k=0}^{N}:z_{mk}^{(N)}z_{kn}^{(N)}:
=
\sum_{k=0}^{N}
\left(
z_{mk}^{(N)}z_{kn}^{(N)}
-
\mathbb{E}\big[z_{mk}^{(N)}z_{kn}^{(N)}\big]
\right).
\]
Both $z^{(N)}$ and $:z^2:^{(N)}$ are Hermitian matrices.

\begin{lemma}\label{lemma:M-construction-z-z2}
	For every sufficiently small $\varepsilon>0$ and sufficiently large $p$,
	the sequences $\{z^{(N)}\}_{N\geq0}$ and
	$\{:z^2:^{(N)}\}_{N\geq0}$ are Cauchy sequences in
	\[
	L^p\bigl(\Omega,\mathbb{P};C_TM^{\frac12-\varepsilon}\bigr).
	\]
	We denote their respective limits by $z$ and $:z^2:$. In particular,
	\[
	z,\ :z^2:
	\in
	L^p\bigl(\Omega,\mathbb{P};C_TM^{\frac12-\varepsilon}\bigr).
	\]
\end{lemma}

\begin{proof}
	Fix $0<\varepsilon<\frac12$ and take $p\geq2$ sufficiently large. We first
	record an estimate that will be used repeatedly. If
	$X=(X_{mn})_{m,n\geq0}$ is a random matrix whose entries belong to a fixed
	homogeneous Wiener chaos, then Gaussian hypercontractivity from
	Appendix~\ref{Hyper} and $\sup a_{mn}\leq\sum a_{mn}$ give
	\begin{align}
		\mathbb{E}\left[\|X\|_{M^\rho}^p\right]^{1/p}
		&=
		\mathbb{E}\left[
		\sup_{m,n\geq0}A_{mn}^{p\rho}|X_{mn}|^p
		\right]^{1/p} \notag\\
		&\leq
		\left(
		\sum_{m,n\geq0}
		A_{mn}^{p\rho}\mathbb{E}\left[|X_{mn}|^p\right]
		\right)^{1/p} \notag\\
		&\lesssim_p
		\left(
		\sum_{m,n\geq0}
		\left(
		A_{mn}^{2\rho}\mathbb{E}\left[|X_{mn}|^2\right]
		\right)^{p/2}
		\right)^{1/p}.
		\label{eq:M-chaos-estimate}
	\end{align}
	
	Set
	\[
	\chi_N(m,n):=\mathbf{1}_{\{m\leq N,\ n\leq N\}}.
	\]
	We first consider $z^{(N)}$. For $0\leq N\leq M$, the definition of the
	cutoff and the covariance of $z$ imply
	\[
	\mathbb{E}\left[
	|z_{mn}^{(M)}(t)-z_{mn}^{(N)}(t)|^2
	\right]
	=
	\frac{\chi_M(m,n)-\chi_N(m,n)}{A_{mn}}.
	\]
	Indeed, the cutoff indicators are nested and take only the values zero and
	one. Applying \eqref{eq:M-chaos-estimate} with
	$\rho=\frac12-\varepsilon$, we obtain
	\begin{align*}
		\mathbb{E}\left[
		\|z^{(M)}(t)-z^{(N)}(t)\|_{M^{\frac12-\varepsilon}}^p
		\right]^{1/p} 
                &\lesssim_p
		\left(
		\sum_{m,n\geq0}
		\left[
		A_{mn}^{1-2\varepsilon}
		\frac{\chi_M(m,n)-\chi_N(m,n)}{A_{mn}}
		\right]^{p/2}
		\right)^{1/p} \\
		&=
		\left(
		\sum_{m,n\geq0}
		\frac{\chi_M(m,n)-\chi_N(m,n)}
		{A_{mn}^{p\varepsilon}}
		\right)^{1/p}
		\\ &\leq \left(
		\sum_{\max\{m,n\}>N}\frac{1}{A_{mn}^{p\varepsilon}}
		\right)^{1/p}.
	\end{align*}

	Since the double series $\sum_{m,n\geq0}A_{mn}^{-p\varepsilon}$ is finite
	whenever $p\varepsilon>2$, we have the $RHS\to0$ as $N\to\infty$.
	
	For the time increments, writing $h:=|t-s|$, we have
	\[
	\mathbb{E}\left[
	|z_{mn}^{(N)}(t)-z_{mn}^{(N)}(s)|^2
	\right]
	=
	\frac{2\chi_N(m,n)}{A_{mn}}
	\left(1-e^{-hA_{mn}}\right).
	\]
	Using $1-e^{-x}\lesssim x^\varepsilon$ in
	\eqref{eq:M-chaos-estimate}, we get
	\begin{align*}
		\mathbb{E}\left[
		\|z^{(N)}(t)-z^{(N)}(s)\|_{M^{\frac12-\varepsilon}}^p
		\right]^{1/p} 
		&\lesssim_p
		\left(
		\sum_{m,n\geq0}
		\left[
		A_{mn}^{1-2\varepsilon}
		\frac{\chi_N(m,n)}{A_{mn}}
		\left(1-e^{-hA_{mn}}\right)
		\right]^{p/2}
		\right)^{1/p} \\
		&\lesssim
		h^{\varepsilon/2}
		\left(
		\sum_{m,n\geq0}
		\frac{1}{A_{mn}^{p\varepsilon/2}}
		\right)^{1/p}
		\lesssim
		h^{\varepsilon/2},
	\end{align*}
	uniformly in $N$, provided $p$ is sufficiently large.
	
	We next consider the Wick square and write
	\[
	X^{(N)}:=:z^2:^{(N)}.
	\]
	If $r=\min\{N,M\}$, Wick's rule and the correlation function of $z$ give
	\begin{align*}
		&\mathbb{E}\left[
		X_{mn}^{(N)}(t)X_{nm}^{(M)}(s)
		\right] \\
		&=
		\chi_r(m,n)
		\sum_{k=0}^{N}\sum_{l=0}^{M}
		\Big\{
		\mathbb{E}[z_{mk}(t)z_{nl}(s)]
		\mathbb{E}[z_{kn}(t)z_{lm}(s)]
		+
		\mathbb{E}[z_{mk}(t)z_{lm}(s)]
		\mathbb{E}[z_{kn}(t)z_{nl}(s)]
		\Big\} \\
		&=
		\chi_r(m,n)
		\left\{
		\frac{\delta_{mn}}{A_{mm}^2}e^{-2hA_{mm}}
		+
		\sum_{k=0}^{r}
		\frac{e^{-h(A_{mk}+A_{kn})}}{A_{mk}A_{kn}}
		\right\}.
	\end{align*}
	Taking $M=N$ and expanding the square of the time increment, we obtain
	\begin{align*}
		\mathbb{E}\left[
		|X_{mn}^{(N)}(t)-X_{mn}^{(N)}(s)|^2
		\right] 
		&=
		2\chi_N(m,n)
		\left\{
		\frac{\delta_{mn}}{A_{mm}^2}
		\left(1-e^{-2hA_{mm}}\right)
		+
		\sum_{k=0}^{N}
		\frac{1-e^{-h(A_{mk}+A_{kn})}}
		{A_{mk}A_{kn}}
		\right\}.
	\end{align*}
	Using $1-e^{-x}\lesssim x^\varepsilon$ and
	$(a+b)^\varepsilon\lesssim a^\varepsilon+b^\varepsilon$, and applying Lemma~D.1(2) with loss $\varepsilon/2$ to the two sums on the right-hand side yields
	\begin{align*}
		\mathbb{E}\left[
		|X_{mn}^{(N)}(t)-X_{mn}^{(N)}(s)|^2
		\right] 
		&\lesssim
		h^\varepsilon
		\left\{
		\frac{\delta_{mn}}{A_{mm}^{2-\varepsilon}}
		+
		\sum_{k\geq0}
		\left(
		\frac{1}{A_{mk}^{1-\varepsilon}A_{kn}}
		+
		\frac{1}{A_{mk}A_{kn}^{1-\varepsilon}}
		\right)
		\right\}
		\\ & \lesssim
		h^\varepsilon
		\left(
		\frac{\delta_{mn}}{A_{mm}^{2-\varepsilon}}
		+
		\frac{1}{A_{mn}^{1-\frac32\varepsilon}}
		\right).
	\end{align*}
	
	Every entry of $X^{(N)}$ belongs to the second homogeneous Wiener chaos.
	Thus, applying \eqref{eq:M-chaos-estimate} once more,
	\begin{align*}
		\mathbb{E}\left[
		\|X^{(N)}(t)-X^{(N)}(s)\|_{M^{\frac12-\varepsilon}}^p
		\right]^{1/p} 
		&\lesssim_p
		h^{\varepsilon/2}
		\left(
		\sum_{m,n\geq0}
		\left[
		A_{mn}^{1-2\varepsilon}
		\left(
		\frac{\delta_{mn}}{A_{mm}^{2-\varepsilon}}
		+
		\frac{1}{A_{mn}^{1-\frac32\varepsilon}}
		\right)
		\right]^{p/2}
		\right)^{1/p} \\
		&=
		h^{\varepsilon/2}
		\left(
		\sum_{m,n\geq0}
		\left(
		\frac{\delta_{mn}}{A_{mm}^{1+\varepsilon}}
		+
		\frac{1}{A_{mn}^{\varepsilon/2}}
		\right)^{p/2}
		\right)^{1/p}
		\lesssim
		h^{\varepsilon/2}.
	\end{align*}
	The last series is finite for sufficiently large $p$, uniformly in $N$.
	
	It remains to estimate the cutoff difference. Define
	\[
	q_N(m,n,k)
	:=
	\chi_N(m,n)\mathbf{1}_{\{k\leq N\}}.
	\]
	For $N\leq M$, the preceding mixed covariance formula at $s=t$ gives
	\begin{align*}
		\mathbb{E}\left[
		|X_{mn}^{(M)}(t)-X_{mn}^{(N)}(t)|^2
		\right] 
		&=
		\frac{\delta_{mn}\bigl(\chi_M(m,n)-\chi_N(m,n)\bigr)}
		{A_{mm}^2}
		+
		\sum_{k\geq0}
		\frac{q_M(m,n,k)-q_N(m,n,k)}
		{A_{mk}A_{kn}}.
	\end{align*}
	Here we used the fact that, for $N\leq M$, the mixed covariance between
	$X^{(M)}$ and $X^{(N)}$ is equal to the variance of $X^{(N)}$. Since
	\[
	q_M(m,n,k)-q_N(m,n,k)
	\leq
	\mathbf{1}_{\{\max\{m,n,k\}>N\}},
	\]
	we obtain
	\[
	\mathbb{E}\left[
	|X_{mn}^{(M)}(t)-X_{mn}^{(N)}(t)|^2
	\right]
	\leq
	\frac{\delta_{mn}\mathbf{1}_{\{m>N\}}}{A_{mm}^2}
	+
	\sum_{k\geq0}
	\frac{\mathbf{1}_{\{\max\{m,n,k\}>N\}}}
	{A_{mk}A_{kn}}.
	\]
	
	Therefore, by \eqref{eq:M-chaos-estimate},
	\[
	\mathbb{E}\left[
	\|X^{(M)}(t)-X^{(N)}(t)\|_{M^{\frac12-\varepsilon}}^p
	\right]^{1/p}
	\lesssim b_N,
	\]
	where
	\[
	b_N
	:=
	\left(
	\sum_{m,n\geq0}
	\left(
	A_{mn}^{1-2\varepsilon}(\frac{\delta_{mn}\mathbf{1}_{\{m>N\}}}{A_{mm}^2}
	+
	\sum_{k\geq0}
	\frac{\mathbf{1}_{\{\max\{m,n,k\}>N\}}}
	{A_{mk}A_{kn}})
	\right)^{p/2}
	\right)^{1/p}.
	\]
	For every fixed $m,n$, we have $\frac{\delta_{mn}\mathbf{1}_{\{m>N\}}}{A_{mm}^2}
	+
	\sum_{k\geq0}
	\frac{\mathbf{1}_{\{\max\{m,n,k\}>N\}}}
	{A_{mk}A_{kn}}\to0$. On the other hand,
	Lemma~D.1(2), applied with exponents $1,1$ and loss $\varepsilon$, gives
	\[
	\sum_{k\geq0}\frac{1}{A_{mk}A_{kn}}
	\lesssim
	\frac{1}{A_{mn}^{1-\varepsilon}},
	\]
	and hence
	\[
	A_{mn}^{1-2\varepsilon}\Big(\frac{\delta_{mn}\mathbf{1}_{\{m>N\}}}{A_{mm}^2}
	+
	\sum_{k\geq0}
	\frac{\mathbf{1}_{\{\max\{m,n,k\}>N\}}}
	{A_{mk}A_{kn}}\Big)
	\lesssim
	\frac{\delta_{mn}}{A_{mm}^{1+2\varepsilon}}
	+
	\frac{1}{A_{mn}^{\varepsilon}}.
	\]
	The right-hand side, raised to the power $p/2$, is summable for
	sufficiently large $p$. The dominated convergence theorem therefore implies
	that $b_N\to0$.
	
	Finally, let $Y^{(N,M)}$ denote either
	$z^{(M)}-z^{(N)}$ or $X^{(M)}-X^{(N)}$. The fixed-time estimates above give,
	uniformly in $t\in[0,T]$,
	\[
	\|Y^{(N,M)}(t)\|_{L^p(\Omega;M^{\frac12-\varepsilon})}
	\lesssim c_N,
	\]
        where $c_N=a_N$ or $c_N=b_N$, respectively, and $c_N\to0$. By
        the triangle inequality and the uniform time-increment
        estimates,
	\[
	\begin{aligned}
		\|Y^{(N,M)}(t)-Y^{(N,M)}(s)\|
		_{L^p(\Omega;M^{\frac12-\varepsilon})} 
		&\lesssim
		\min\left\{c_N,\ |t-s|^{\varepsilon/2}\right\}.
	\end{aligned}
	\]
	Consequently, for every $\lambda\in(0,1)$,
	\[
	\|Y^{(N,M)}(t)-Y^{(N,M)}(s)\|
	_{L^p(\Omega;M^{\frac12-\varepsilon})}
	\lesssim
	c_N^\lambda
	|t-s|^{(1-\lambda)\varepsilon/2}.
	\]
	Choose $p$ sufficiently large so that
	$p(1-\lambda)\varepsilon/2>1$. Applying
	\cite[Theorem A.10]{friz2010multidimensional}, together with the fixed-time
	estimate, we conclude that
	\[
	\mathbb{E}\left[
	\|Y^{(N,M)}\|_{C_TM^{\frac12-\varepsilon}}^p
	\right]^{1/p}
	\lesssim
	c_N^\lambda
	\longrightarrow0.
	\]
	Thus both sequences are Cauchy in
	$L^p(\Omega,\mathbb{P};C_TM^{\frac12-\varepsilon})$. The limit of
	$z^{(N)}$ agrees entrywise with $z$, and we denote the limit of
	$:z^2:^{(N)}$ by $:z^2:$.
\end{proof}

\begin{corollary}\label{corollary:H-regularity-z-z2}
	For every sufficiently small $\varepsilon>0$ and sufficiently large $p$,
	\[
	z^{(N)}\longrightarrow z,
	\qquad
	:z^2:^{(N)}\longrightarrow:z^2:
	\qquad \text{in} \quad
	L^p\bigl(\Omega,\mathbb{P};C_TH^{-\frac12-\varepsilon}\bigr).
	\]
	In particular,
	\[
	z,\ :z^2:
	\in
	L^p\bigl(\Omega,\mathbb{P};C_TH^{-\frac12-\varepsilon}\bigr).
	\]
\end{corollary}

\begin{proof}
	Apply Lemma~\ref{lemma:M-construction-z-z2} with $\varepsilon/2$. Since
	$\frac12-\frac{\varepsilon}{2}>
	(-\frac12-\varepsilon)+1$, Lemma~\ref{MtoHembedding} gives the continuous
	embedding $C_TM^{\frac12-\frac{\varepsilon}{2}}
	\hookrightarrow C_TH^{-\frac12-\varepsilon}$. The convergence and the
	assertion on the limits therefore follow directly.
\end{proof}

The renormalized third power is defined as
\begin{equation}
    : z^3 :_{m \nospace n}^{(N)} \assign \sum_{k, l = 0}^N (z_{m \nospace k}^{(N)} z_{k \nospace l}^{(N)} z_{l
   n}^{(N)} -\mathbb{E} [z_{m \nospace k}^{(N)} z_{k \nospace l}^{(N)}] z_{l
   n}^{(N)} - z_{m \nospace k}^{(N)} \mathbb{E} [z_{k \nospace l}^{(N)} z_{l
   n}^{(N)}]) 
\end{equation} 
   and we show it has a well-defined limit in the next lemma.
   
   \begin{remark}
   	The renormalization above is tailored to the ordered matrix product in our
   	model and does not coincide with the standard Wick ordering recalled in
   	Appendix~\ref{Hyper}. In particular, the contraction between the first and
   	third factors, $z_{mk}^{(N)}$ and $z_{ln}^{(N)}$, is not subtracted.
   \end{remark}
   
\begin{lemma}\label{lem:z_cube}
	For every sufficiently small $\varepsilon > 0$ and sufficiently large $p$, the sequence of renormalized cubic powers $\left\{ :z^3:^{(N)} \right\}_{N \geq 0}$ is Cauchy in $L^p\left(\Omega, \mathbb{P}; \mathcal{C}_T H^{-\frac{1}{2}-\varepsilon}\right)$. We denote its limit by $:z^3:$.
\end{lemma}

\begin{proof}
	Let $W^{(N)}$ denote the fully Wick-ordered cutoff
	\[
	W_{mn}^{(N)}(t)
	:=
	: z^3 :_{mn}^{(N)}-R_{mn}^{(N)}(t)
	\]
	
	where
	
	\[
	R_{mn}^{(N)}(t)
	=
	\chi_N(m,n)\delta_{mn}
	\sum_{k=0}^{N}
	\frac{z_{kk}^{(N)}(t)}{A_{mk}}.
	\]
	
	To control $R^{(N)}$, fix $0<\varepsilon<\frac34$ and set
	$\alpha=\frac12-\frac{\varepsilon}{3}$. Define
	$(\mathcal Rc)_{mn}:=\delta_{mn}\sum_{k\geq0}\frac{c_{kk}}{A_{mk}}$.
	Since $|c_{kk}|\leq\|c\|_{M^\alpha}A_{kk}^{-\alpha}$ and
	$A_{mk}\gtrsim A_{mm},A_{kk}$, we have
	\begin{align*}
		\sum_{k\geq0}\frac{|c_{kk}|}{A_{mk}}
		&\leq \|c\|_{M^\alpha}
		\sum_{k\geq0}\frac{1}{A_{mk}A_{kk}^{\alpha}} \\
		&\lesssim
		\frac{\|c\|_{M^\alpha}}
		{A_{mm}^{\frac12-\frac{2\varepsilon}{3}}}
		\sum_{k\geq0}
		\frac{1}{A_{kk}^{\frac12+\frac{2\varepsilon}{3}+\alpha}}
		\lesssim
		\frac{\|c\|_{M^\alpha}}
		{A_{mm}^{\frac12-\frac{2\varepsilon}{3}}},
	\end{align*}
	where we used
	$\frac12+\frac{2\varepsilon}{3}+\alpha
	=1+\frac{\varepsilon}{3}>1$. Consequently,
	\[
	\|\mathcal Rc\|_{H^{-\varepsilon}}^2
	\lesssim
	\|c\|_{M^\alpha}^2
	\sum_{m\geq0}\frac{1}{A_{mm}^{1+\frac{2\varepsilon}{3}}}
	\lesssim \|c\|_{M^\alpha}^2.
	\]
	Thus $\mathcal R:M^\alpha\to H^{-\varepsilon}$ is continuous.
	
	Since
	$R_{mn}^{(N)}
	=\chi_N(m,n)(\mathcal Rz^{(N)})_{mn}$,
	set $R:=\mathcal Rz$. Using the convergence
	$z^{(N)}\to z$ in $L^p(\Omega;C_TM^\alpha)$ and the fact that multiplication
	by $\chi_N$ is contractive on $H^{-\varepsilon}$, we obtain
	\begin{align*}
		\|R^{(N)}-R\|_{L^p(\Omega;C_TH^{-\varepsilon})}
		&\leq
		\|\mathcal R(z^{(N)}-z)\|_{L^p(\Omega;C_TH^{-\varepsilon})} 
		+
		\|(\chi_N-1)R\|_{L^p(\Omega;C_TH^{-\varepsilon})}
		\longrightarrow0.
	\end{align*}
	Here the second term converges to zero because the matrix cutoffs
	$\chi_NR$ converge to $R$ in $L^p(\Omega;C_TH^{-\varepsilon})$.
	Hence $R^{(N)}$ is Cauchy in this space and therefore also in
	$L^p(\Omega;C_TH^{-\frac12-\varepsilon})$. It remains to consider
	$W^{(N)}$.
	
	Let $r=\min\{N,M\}$, $h=|t-s|$, and
	$\chi_r(m,n):=\mathbf{1}_{\{m\leq r,\ n\leq r\}}$.
	Expanding the two cutoff cubic powers, we obtain
	\begin{align*}
		\mathbb{E}\left[
		W_{mn}^{(N)}(t)W_{nm}^{(M)}(s)
		\right] 
		&=
		\sum_{k,l=0}^{N}\sum_{a,b=0}^{M}
		\mathbb{E}\left[
		:z_{mk}^{(N)}(t)z_{kl}^{(N)}(t)z_{ln}^{(N)}(t):
		:z_{na}^{(M)}(s)z_{ab}^{(M)}(s)z_{bm}^{(M)}(s):
		\right].
	\end{align*}
	For notational convenience, set
	\[
	\begin{aligned}
		(X_1^{(N)},X_2^{(N)},X_3^{(N)})
		&=
		\bigl(
		z_{mk}^{(N)}(t),
		z_{kl}^{(N)}(t),
		z_{ln}^{(N)}(t)
		\bigr),\\
		(Y_1^{(M)},Y_2^{(M)},Y_3^{(M)})
		&=
		\bigl(
		z_{na}^{(M)}(s),
		z_{ab}^{(M)}(s),
		z_{bm}^{(M)}(s)
		\bigr).
	\end{aligned}
	\]
	Wick's rule then gives
	\begin{align*}
		\mathbb{E}\left[
		W_{mn}^{(N)}(t)W_{nm}^{(M)}(s)
		\right] 
		&=
		\sum_{k,l=0}^{N}\sum_{a,b=0}^{M}
		\sum_{\sigma\in S_3}
		\prod_{j=1}^{3}
		\mathbb{E}\left[
		X_j^{(N)}Y_{\sigma(j)}^{(M)}
		\right].
	\end{align*}
	The mixed covariance of the cutoff processes is
	\[
	\mathbb{E}\left[
	z_{ij}^{(N)}(t)z_{pq}^{(M)}(s)
	\right]
	=
	\chi_N(i,j)\chi_M(p,q)
	\frac{\delta_{iq}\delta_{jp}}{A_{ij}}
	e^{-hA_{ij}}.
	\]
	Substituting this covariance into the six Wick contractions and eliminating
	the indices $a,b$ using the Kronecker deltas yields
	
	\[
	\mathbb{E}\left[
	W_{mn}^{(N)}(t)W_{nm}^{(M)}(s)
	\right]
	=
	C_r^{mn}(h),
	\]
	where
	\[
	\begin{aligned}
		C_r^{mn}(h)
		=\chi_r(m,n)\Bigg\{&
		\frac{e^{-3hA_{mn}}}{A_{mn}^3}
		+2\delta_{mn}\sum_{l=0}^{r}
		\frac{e^{-h(A_{mm}+2A_{ml})}}
		{A_{mm}A_{ml}^2} \\
		&+\frac{e^{-h(2A_{mm}+A_{mn})}}
		{A_{mm}^2A_{mn}}
		+\frac{e^{-h(A_{mn}+2A_{nn})}}
		{A_{mn}A_{nn}^2} 
		+\sum_{k,l=0}^{r}
		\frac{e^{-h(A_{mk}+A_{kl}+A_{ln})}}
		{A_{mk}A_{kl}A_{ln}}
		\Bigg\}.
	\end{aligned}
	\]

	Let $C_\infty^{mn}$ denote the same expression with all cutoffs removed.
	The correlation estimates of Appendix D, applied twice to the final double
	sum, give
	\[
	C_\infty^{mn}(0)
	\lesssim
	\frac{1}{A_{mn}^{1-\varepsilon}},
	\qquad
	0
	\leq
	C_r^{mn}(0)-C_r^{mn}(h)
	\lesssim
	\frac{h^{\varepsilon/3}}{A_{mn}^{1-\varepsilon}},
	\]
	uniformly in $r$. Indeed, the only non-immediate term is controlled by
	\[
	\sum_{k,l\geq 0}
	\frac{(A_{mk}+A_{kl}+A_{ln})^{\varepsilon/3}}
	{A_{mk}A_{kl}A_{ln}}
	\lesssim
	\frac{1}{A_{mn}^{1-\varepsilon}},
	\]
	using
	$(a+b+c)^{\varepsilon/3}
	\lesssim
	a^{\varepsilon/3}+b^{\varepsilon/3}+c^{\varepsilon/3}$;
	the remaining four terms are simpler consequences of the same estimates.
	
	Since every entry of $W^{(N)}$ belongs to the third homogeneous Wiener chaos,
	Minkowski's inequality and Gaussian hypercontractivity imply, for any such
	random matrix $Y$,
	\[
	\mathbb{E}
	\left[
	\|Y\|_{H^{-\frac12-\varepsilon}}^p
	\right]^{1/p}
	\lesssim_p
	\left(
	\sum_{m,n\geq 0}
	A_{mn}^{-1-2\varepsilon}
	\mathbb{E}[|Y_{mn}|^2]
	\right)^{1/2}.
	\]
	
	For $N\leq M$, the mixed covariance formula at $s=t$ yields
	\[
	\mathbb{E}
	\left[
	\left|
	W_{mn}^{(M)}(t)-W_{mn}^{(N)}(t)
	\right|^2
	\right]
	=
	C_M^{mn}(0)-C_N^{mn}(0).
	\]
	
	Define
	\[
	b_N^2
	:=
	\sum_{m,n\geq 0}
	A_{mn}^{-1-2\varepsilon}
	\left(
	C_\infty^{mn}(0)-C_N^{mn}(0)
	\right).
	\]
	
	Because $C_N^{mn}(0)\uparrow C_\infty^{mn}(0)$ and
	\[
	A_{mn}^{-1-2\varepsilon}
	C_\infty^{mn}(0)
	\lesssim
	A_{mn}^{-2-\varepsilon},
	\]
	the dominated convergence theorem gives $b_N\to 0$. Consequently,
	\[
	\left\|
	W^{(M)}(t)-W^{(N)}(t)
	\right\|_{L^p(\Omega;H^{-\frac12-\varepsilon})}
	\lesssim_p
	b_N.
	\]
	
	Similarly,
	\[
	\mathbb{E}
	\left[
	\left|
	W_{mn}^{(N)}(t)-W_{mn}^{(N)}(s)
	\right|^2
	\right]
	=
	2\left(
	C_N^{mn}(0)-C_N^{mn}(h)
	\right),
	\]
	and hence, uniformly in $N$,
	\[
	\left\|
	W^{(N)}(t)-W^{(N)}(s)
	\right\|_{L^p(\Omega;H^{-\frac12-\varepsilon})}
	\lesssim_p
	h^{\varepsilon/6}.
	\]
	
	Thus, for $Y^{(N,M)}=W^{(M)}-W^{(N)}$ and every $\lambda\in(0,1)$,
	\[
	\left\|
	Y^{(N,M)}(t)-Y^{(N,M)}(s)
	\right\|_{L^p(\Omega;H^{-\frac12-\varepsilon})}
	\lesssim
	b_N^\lambda
	|t-s|^{(1-\lambda)\varepsilon/6}.
	\]
	
	Choosing $p$ sufficiently large and applying \cite[Theorem A.10]{friz2010multidimensional} as in the preceding lemma gives
	\[
	\mathbb{E}
	\left[
	\left\|
	W^{(M)}-W^{(N)}
	\right\|_{C_T H^{-\frac12-\varepsilon}}^p
	\right]^{1/p}
	\lesssim
	b_N^\lambda
	\longrightarrow
	0.
	\]
	
	Therefore both $W^{(N)}$, and hence
	$: z^3 :^{(N)}=W^{(N)}+R^{(N)}$, are Cauchy in the asserted space.
\end{proof}

\section{105 Terms Verification}\label{Eappendix}

This appendix is devoted to use the graph reduction algorithm to check all 105
Wick contraction terms are finite. As before, we label the fundamental graph
as following
\[
  \resizebox{0.35\columnwidth}{!}{\includegraphics{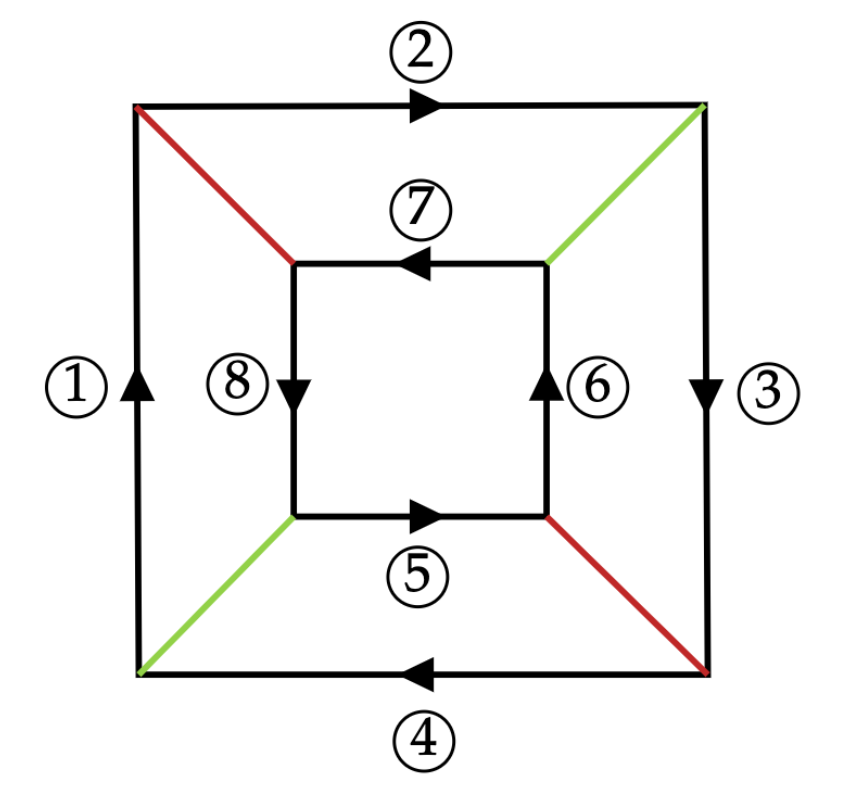}}
\]
and we generate all the pairings of first 8 numbers, we put the graph with
same structure after pairing identification together. Set $\alpha =
\frac{1}{2} - \varepsilon$, $\beta = 0 - \varepsilon - \varepsilon'$ and
$\delta > 0$ small enough.

1. $(12) (34) (56) (78)$
\[
  \resizebox{1\columnwidth}{!}{\includegraphics{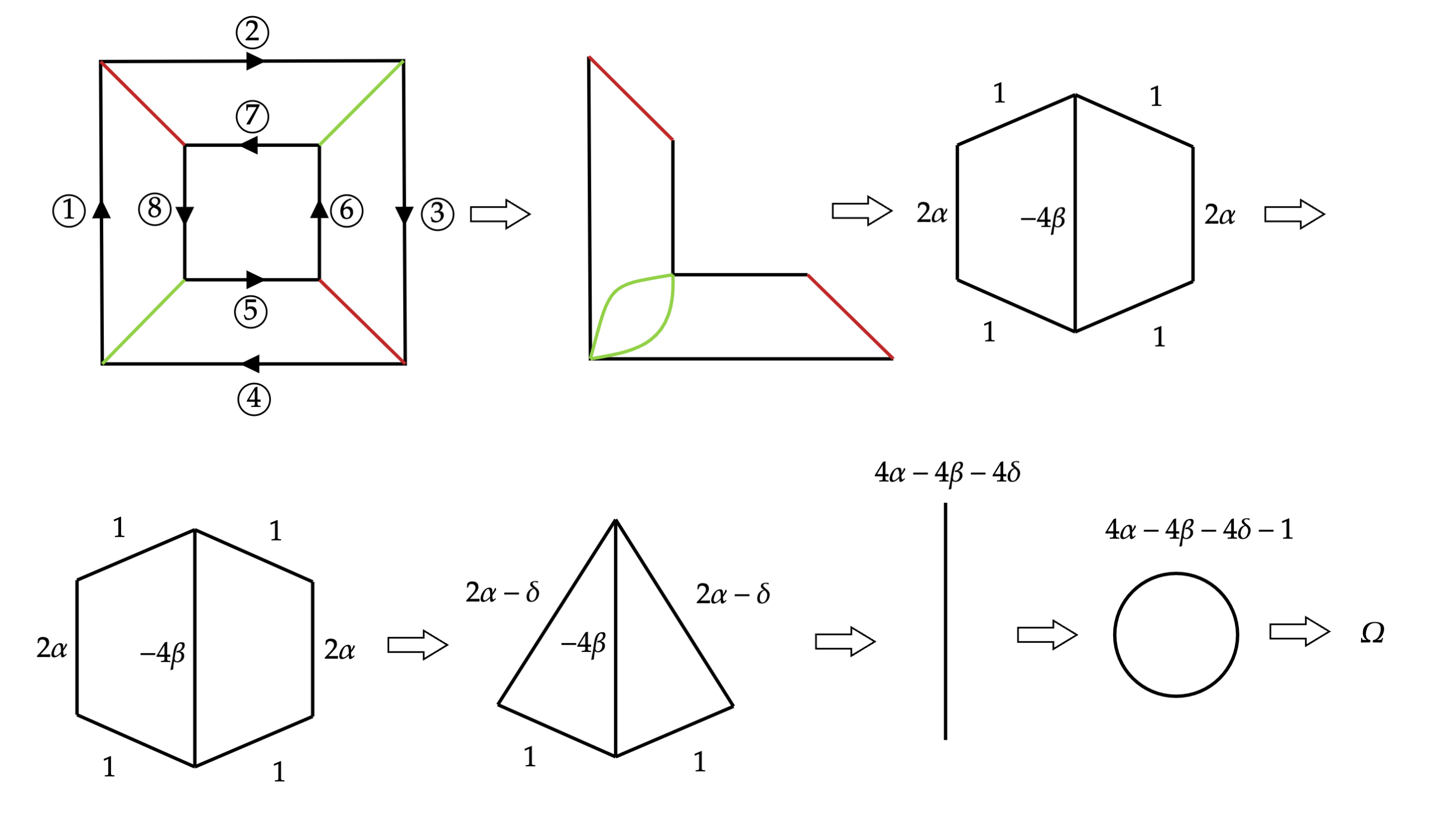}}
\]
2. (12)(34)(57)(68), (13)(24)(56)(78)
\[
  \resizebox{1\columnwidth}{!}{\includegraphics{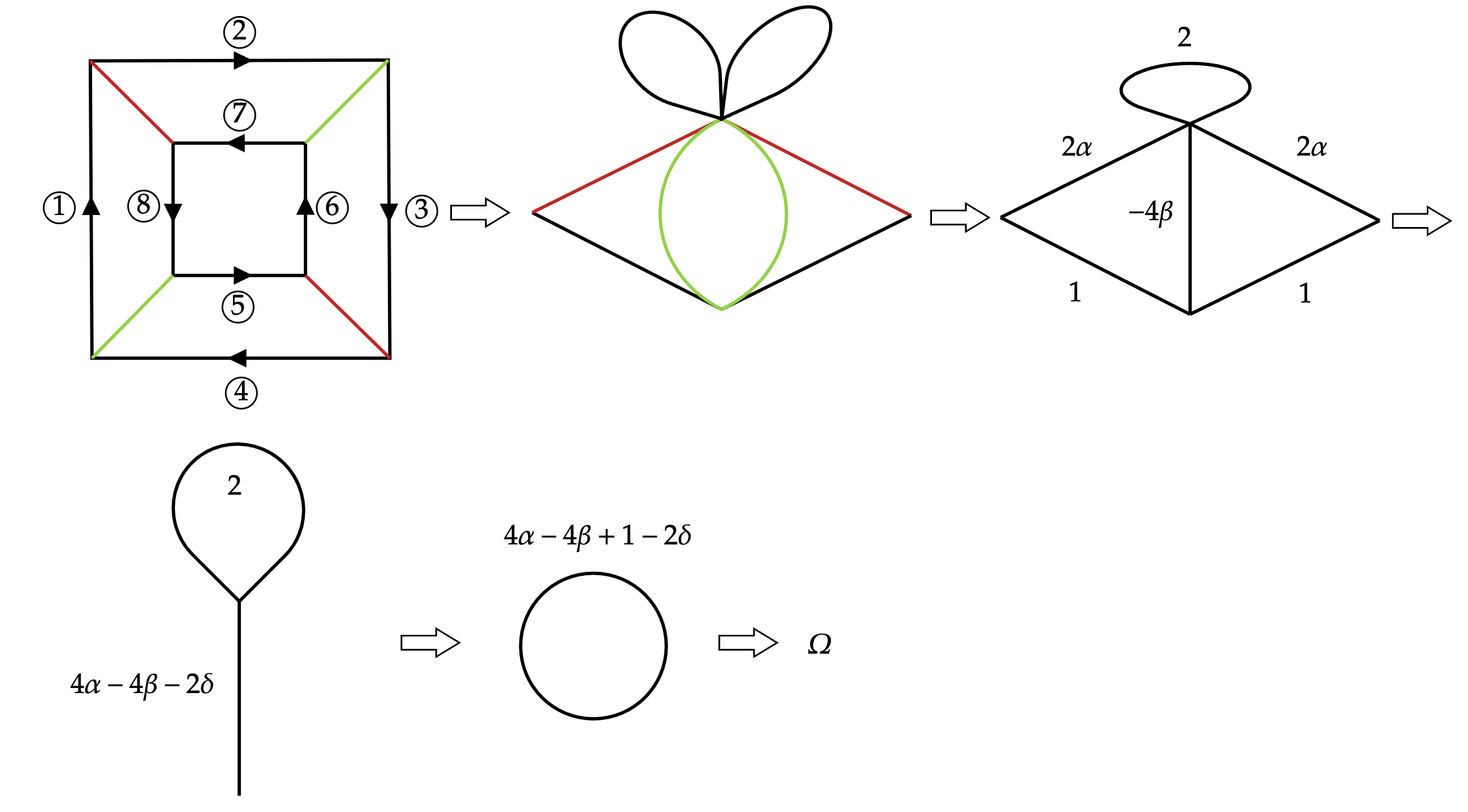}}
\]
3. (12)(34)(58)(67)
\[
  \resizebox{1\columnwidth}{!}{\includegraphics{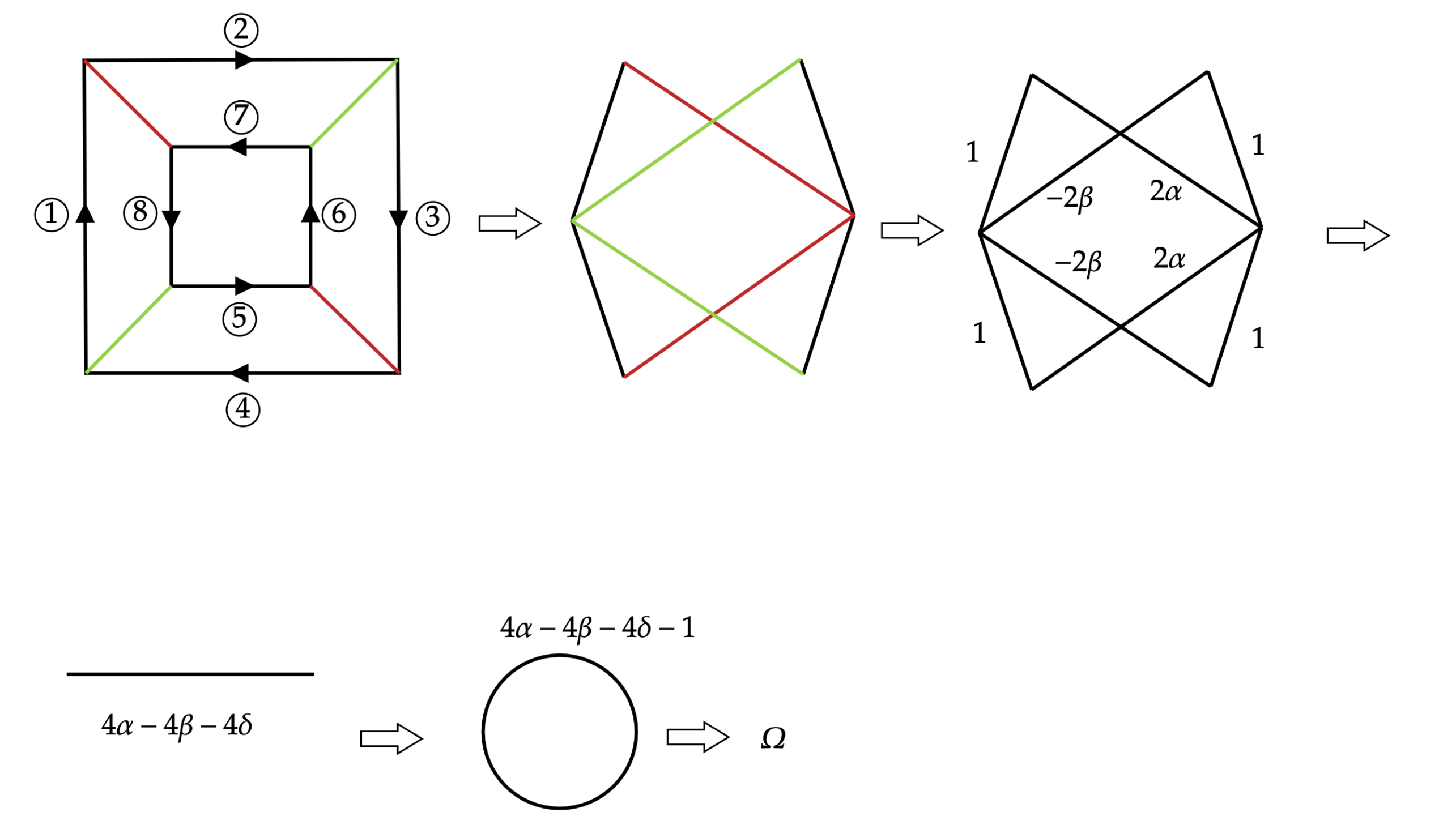}}
\]
4. (12)(35)(46)(78), (17)(28)(34)(56)
\[
  \resizebox{1\columnwidth}{!}{\includegraphics{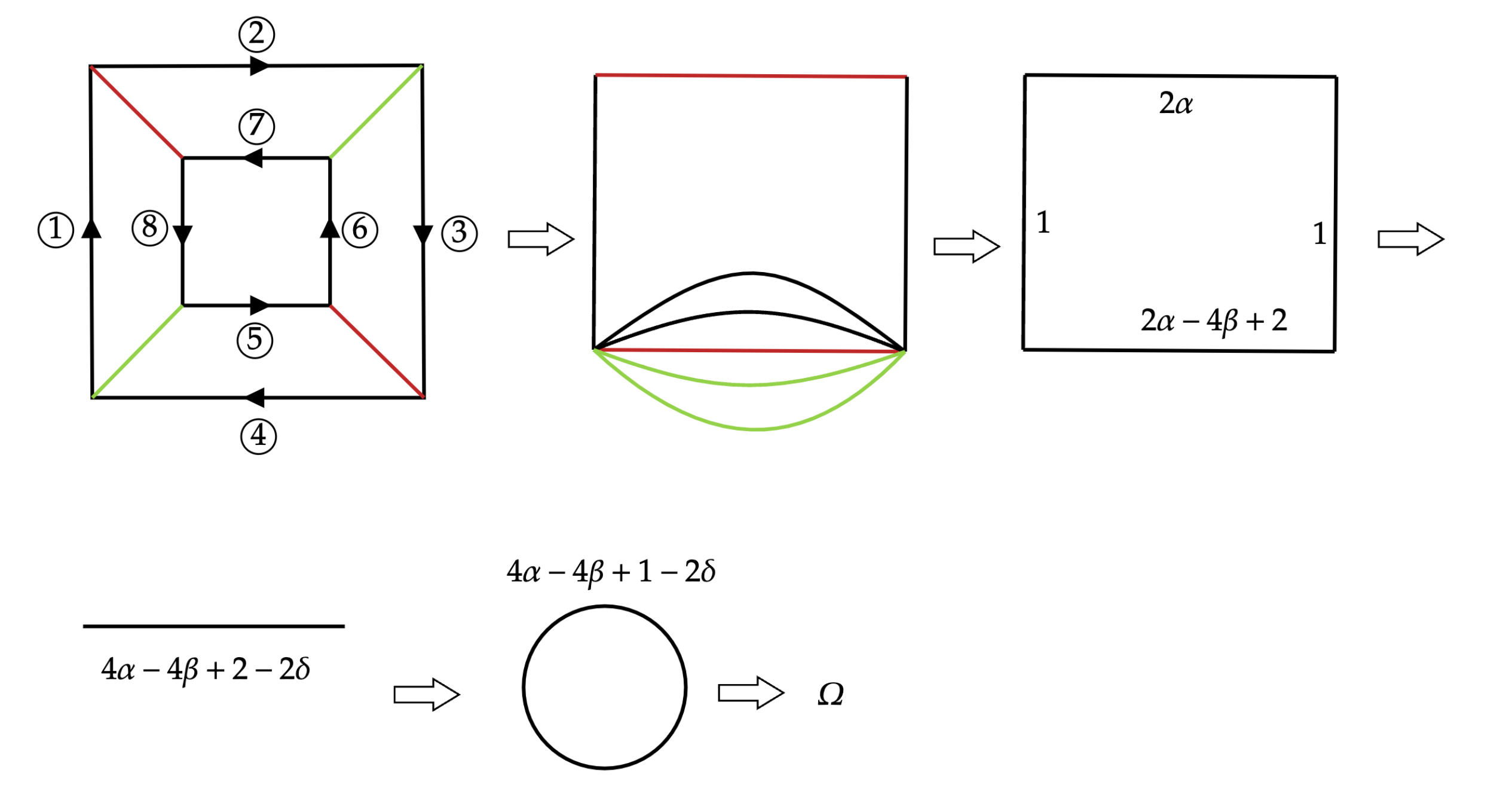}}
\]
5. (12)(35)(47)(68), (12)(36)(48)(57), (12)(37)(45)(68), (12)(38)(46)(57),
(13)(25)(46)(78), (13)(26)(45)(78), (13)(27)(48)(56), (13)(28)(47)(56),
(15)(24)(36)(78), (15)(27)(34)(68), (16)(24)(35)(78), (16)(28)(34)(57),
(17)(24)(38)(56), (17)(25)(34)(68), (18)(24)(37)(56), (18)(26)(34)(57)
\[
  \resizebox{1\columnwidth}{!}{\includegraphics{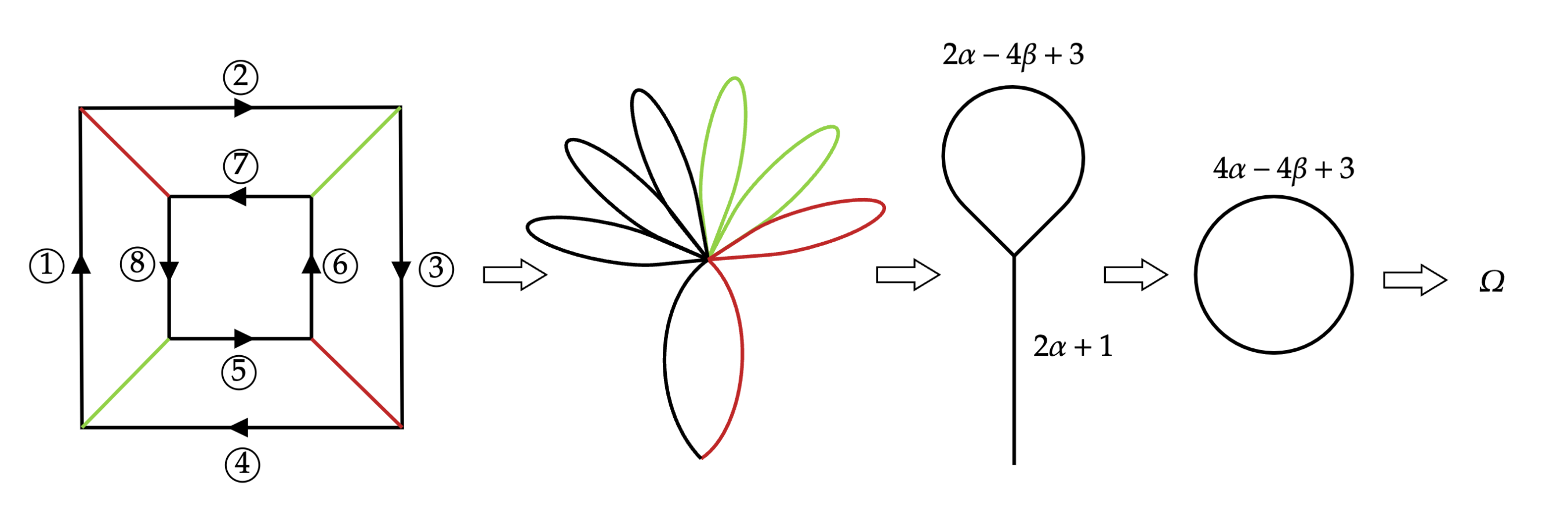}}
\]
6. (12)(35)(48)(67), (12)(37)(46)(58), (14)(26)(35)(78), (14)(28)(37)(56),
(15)(23)(46)(78), (15)(28)(34)(67), (17)(23)(48)(56), (17)(26)(34)(58)
\[
  \resizebox{1\columnwidth}{!}{\includegraphics{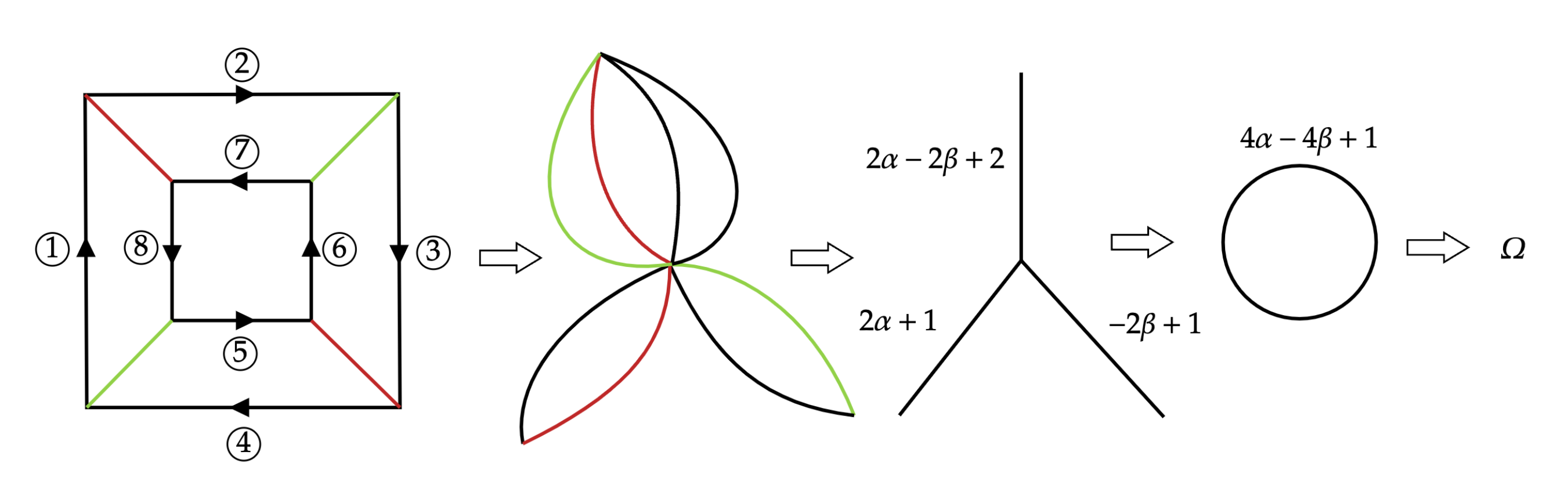}}
\]
7. (12)(36)(45)(78), (18)(27)(34)(56)
\[
  \resizebox{1\columnwidth}{!}{\includegraphics{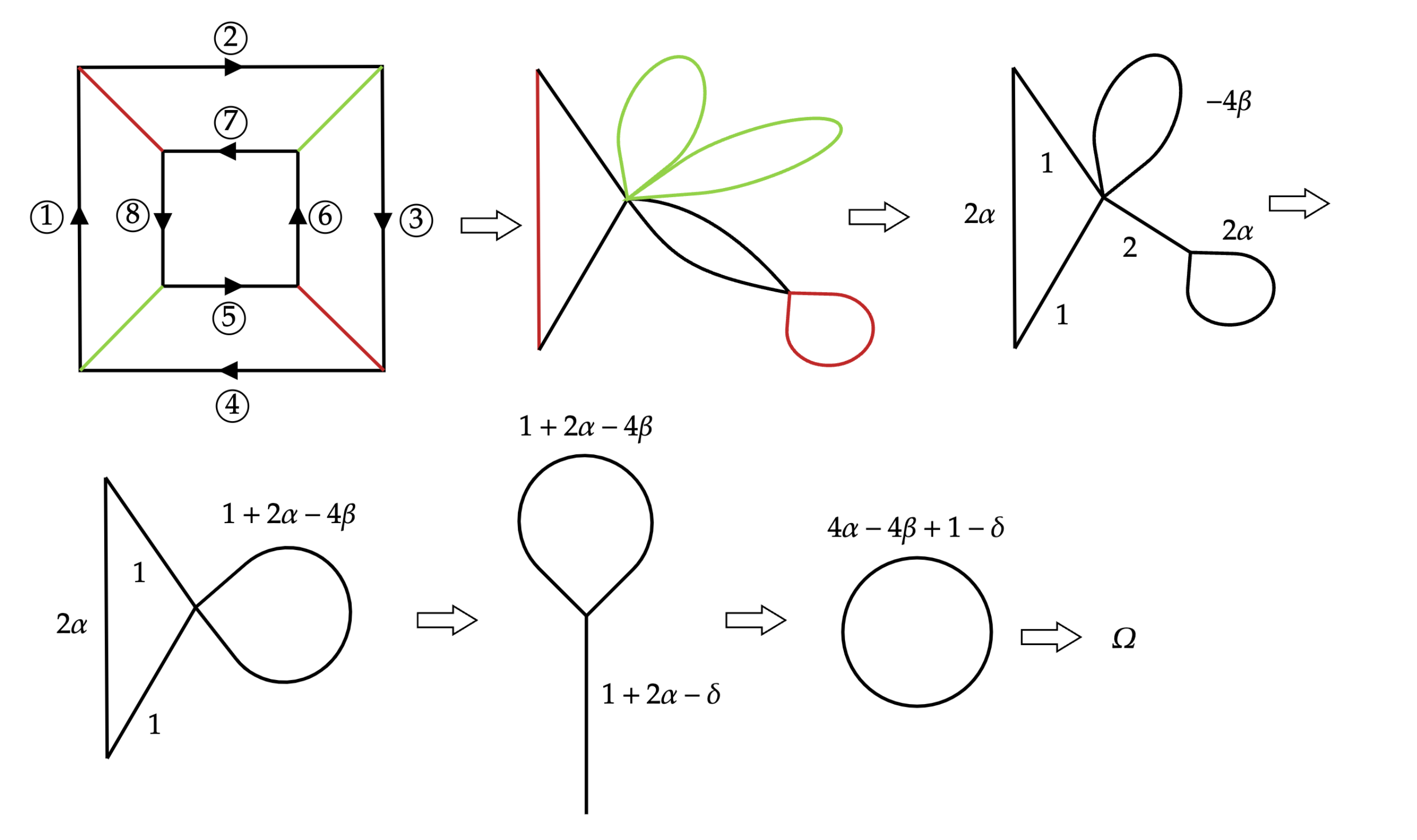}}
\]
8. (12)(36)(47)(58), (12)(38)(45)(67), (14)(25)(36)(78), (14)(27)(38)(56),
(16)(23)(45)(78), (16)(27)(34)(58), (18)(23)(47)(56), (18)(25)(34)(67)
\[
  \resizebox{1\columnwidth}{!}{\includegraphics{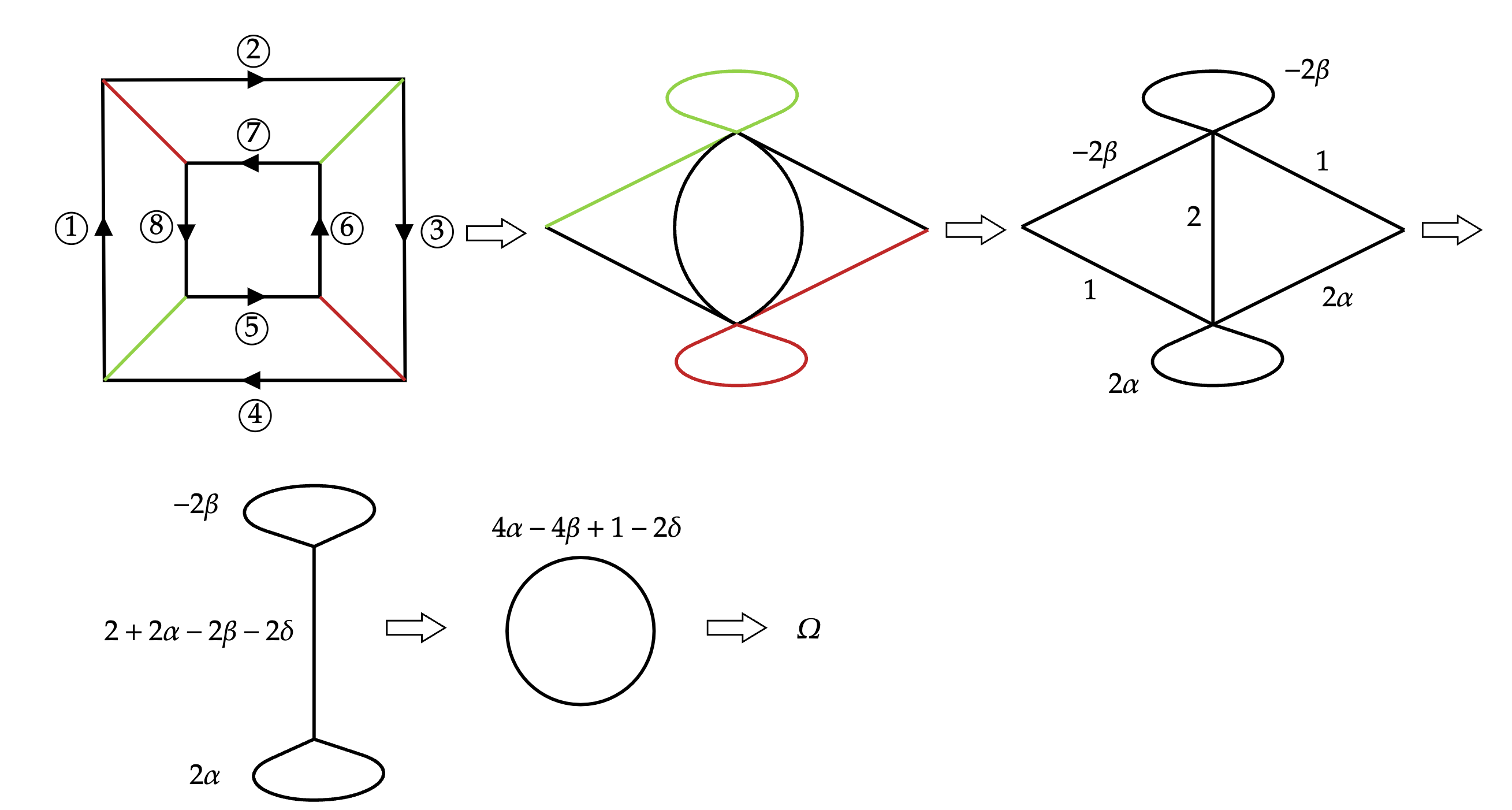}}
\]
9. (12)(37)(48)(56), (15)(26)(34)(78)
\[
  \resizebox{1\columnwidth}{!}{\includegraphics{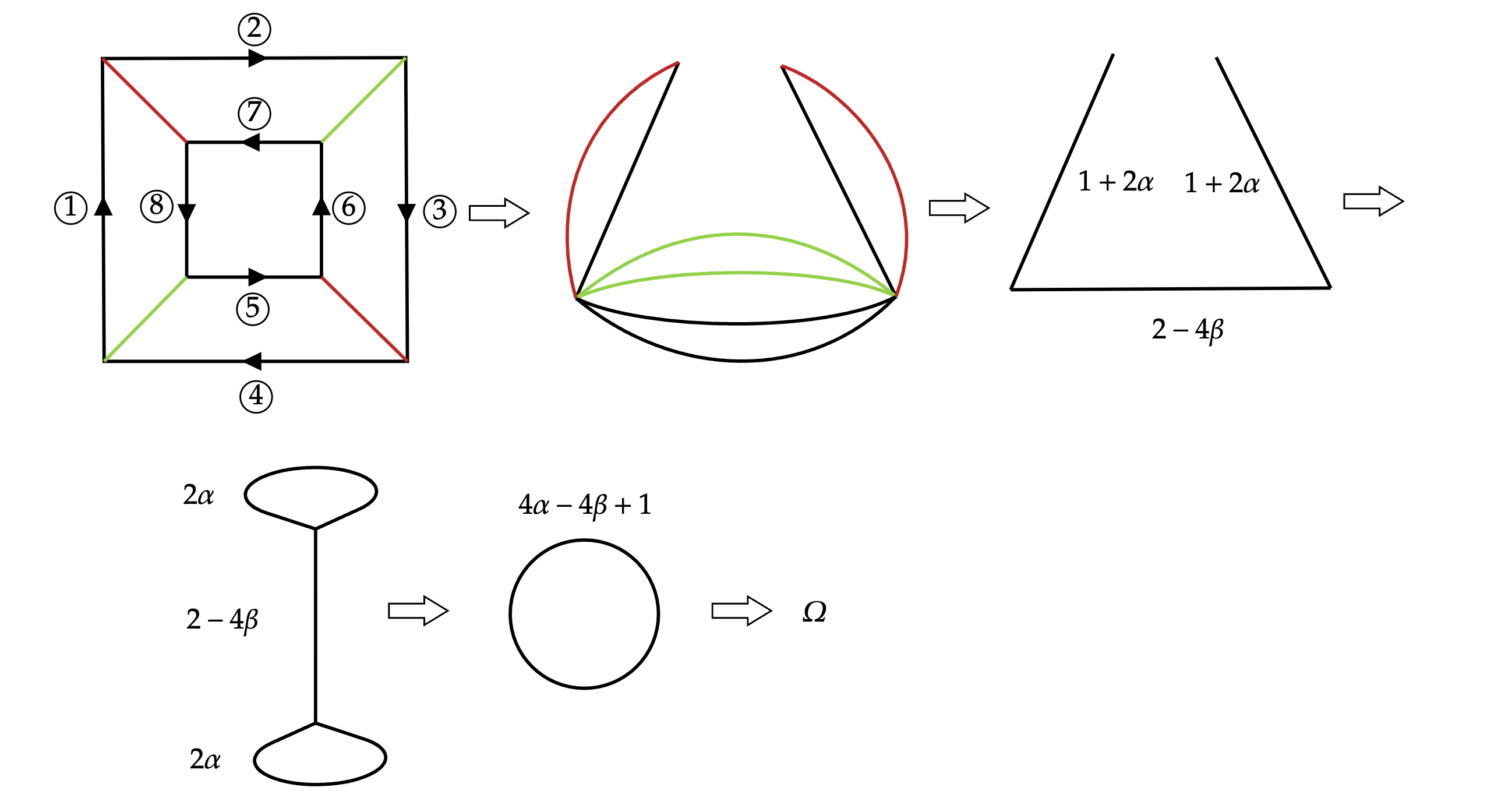}}
\]
10. (12)(38)(47)(56), (16)(25)(34)(78)
\[
  \resizebox{1\columnwidth}{!}{\includegraphics{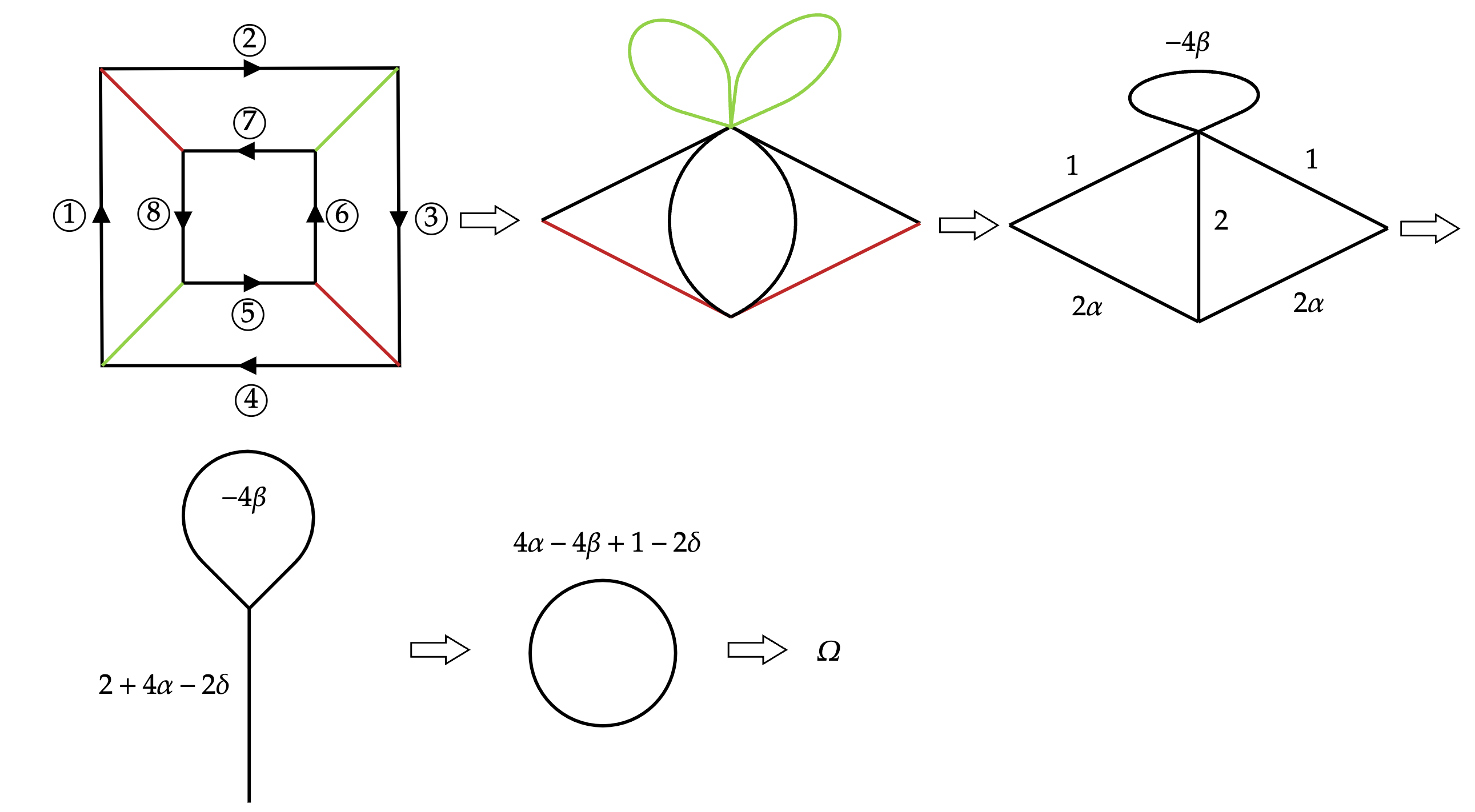}}
\]
11. (13)(24)(57)(68)
\[
  \resizebox{1\columnwidth}{!}{\includegraphics{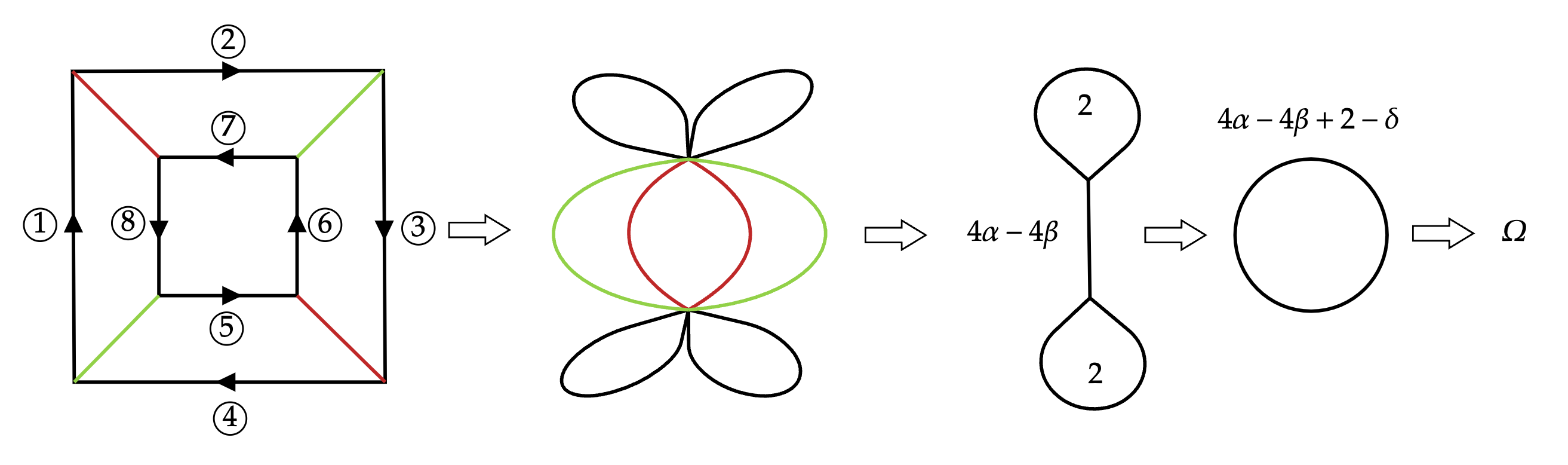}}
\]
12. (13)(24)(58)(67), (14)(23)(57)(68)
\[
  \resizebox{1\columnwidth}{!}{\includegraphics{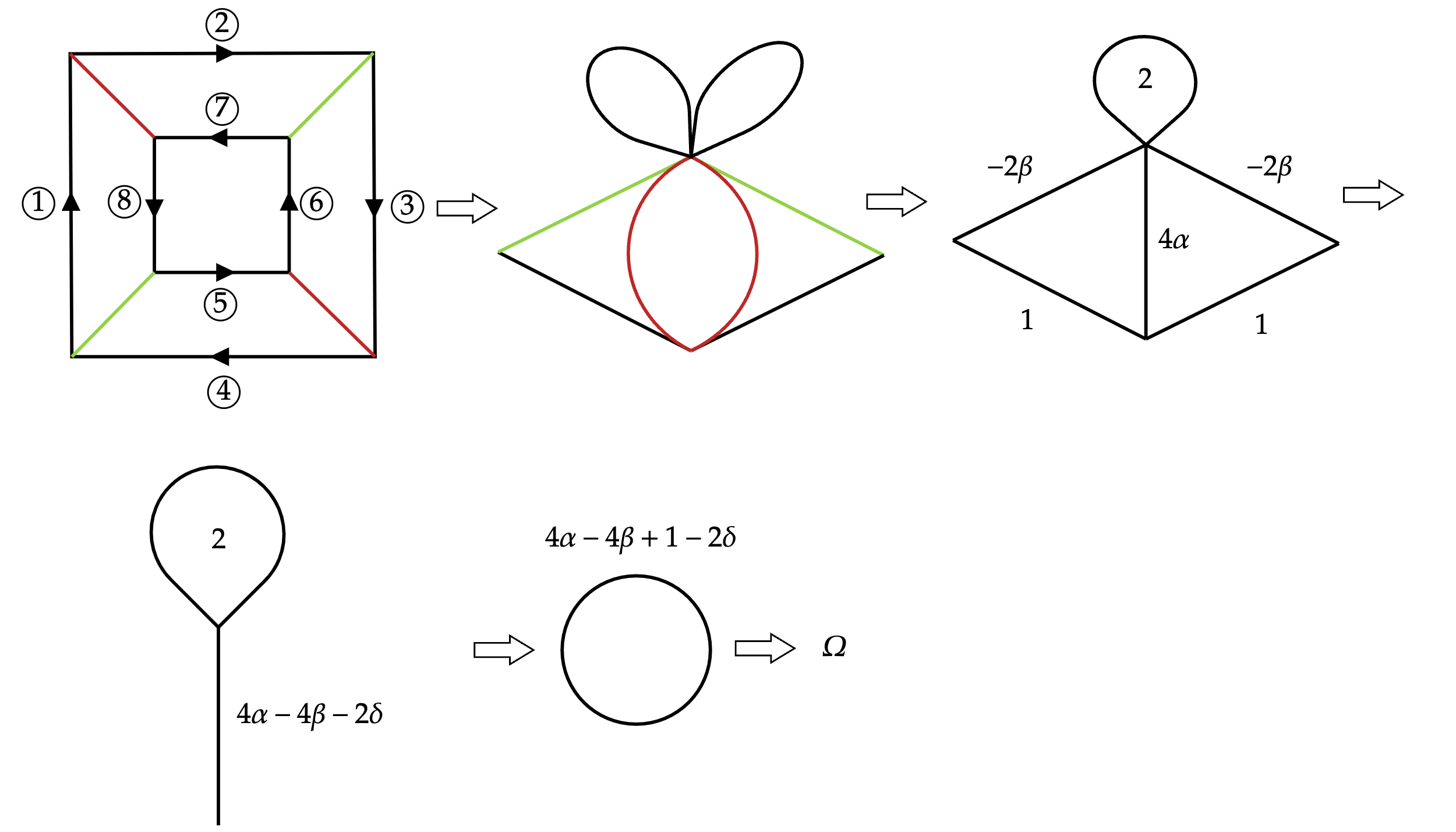}}
\]
13. (13)(25)(47)(68)
\[
  \resizebox{1\columnwidth}{!}{\includegraphics{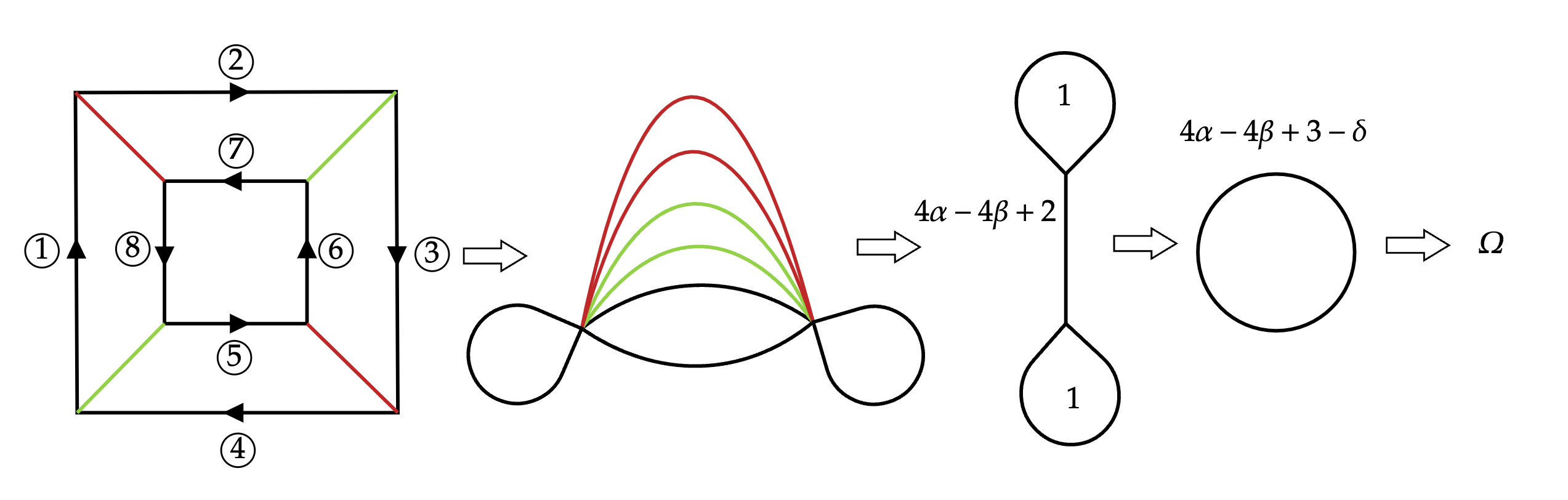}}
\]
14. (13)(25)(48)(67), (13)(26)(47)(58), (13)(27)(46)(58), (13)(28)(45)(67),
(14)(25)(37)(68), (14)(26)(38)(57), (14)(27)(35)(68), (14)(28)(36)(57),
(15)(23)(47)(68), (15)(24)(38)(67), (16)(23)(48)(57), (16)(24)(37)(58),
(17)(23)(45)(68), (17)(24)(36)(58), (18)(23)(46)(57), (18)(24)(35)(67)
\[
  \resizebox{1\columnwidth}{!}{\includegraphics{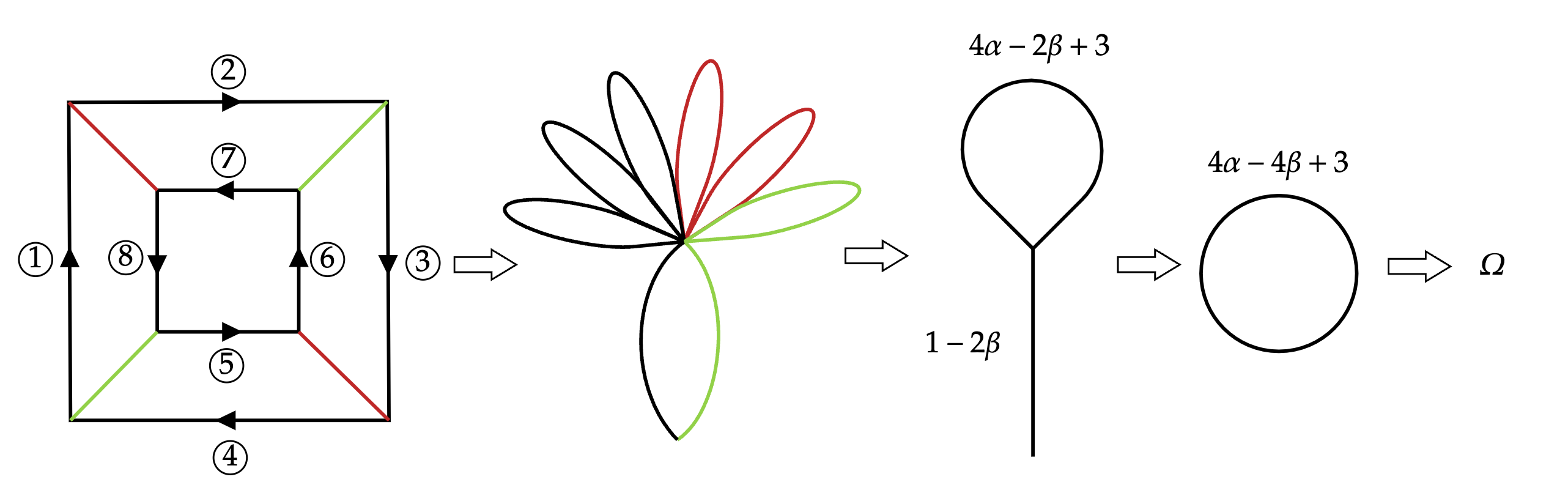}}
\]
15. (13)(26)(48)(57), (15)(24)(37)(68)
\[
  \resizebox{1\columnwidth}{!}{\includegraphics{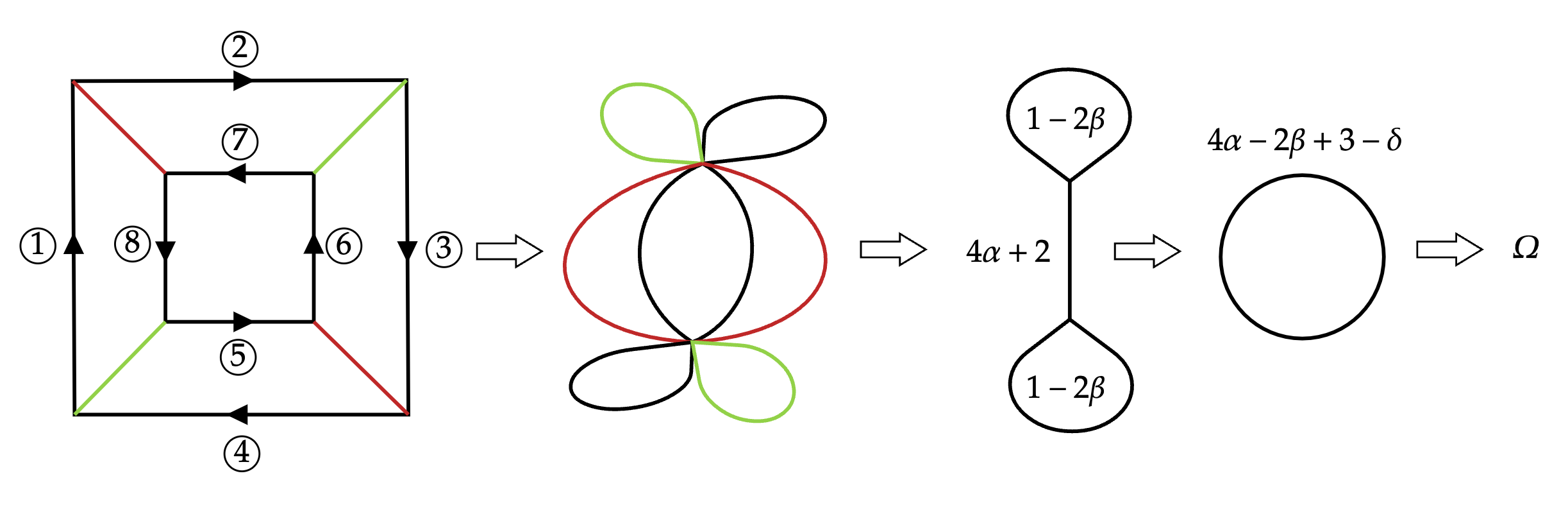}}
\]
16. (13)(27)(45)(68), (18)(24)(36)(57),
\[
  \resizebox{1\columnwidth}{!}{\includegraphics{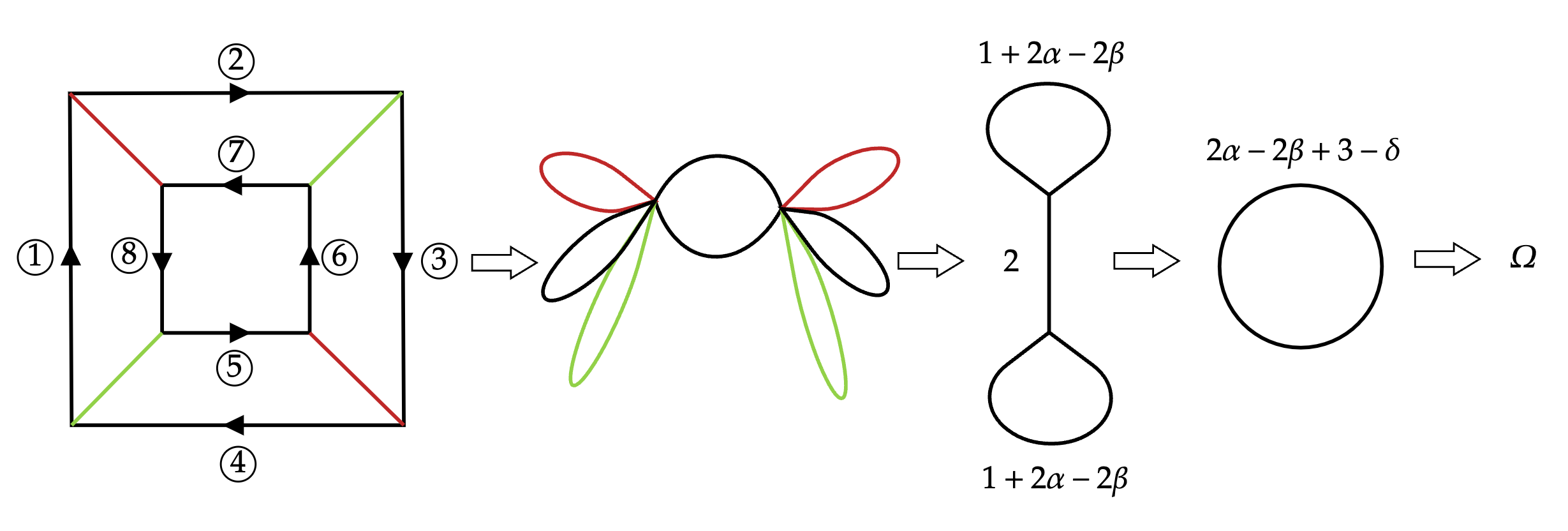}}
\]
17. (13)(28)(46)(57), (17)(24)(35)(68)
\[
  \resizebox{1\columnwidth}{!}{\includegraphics{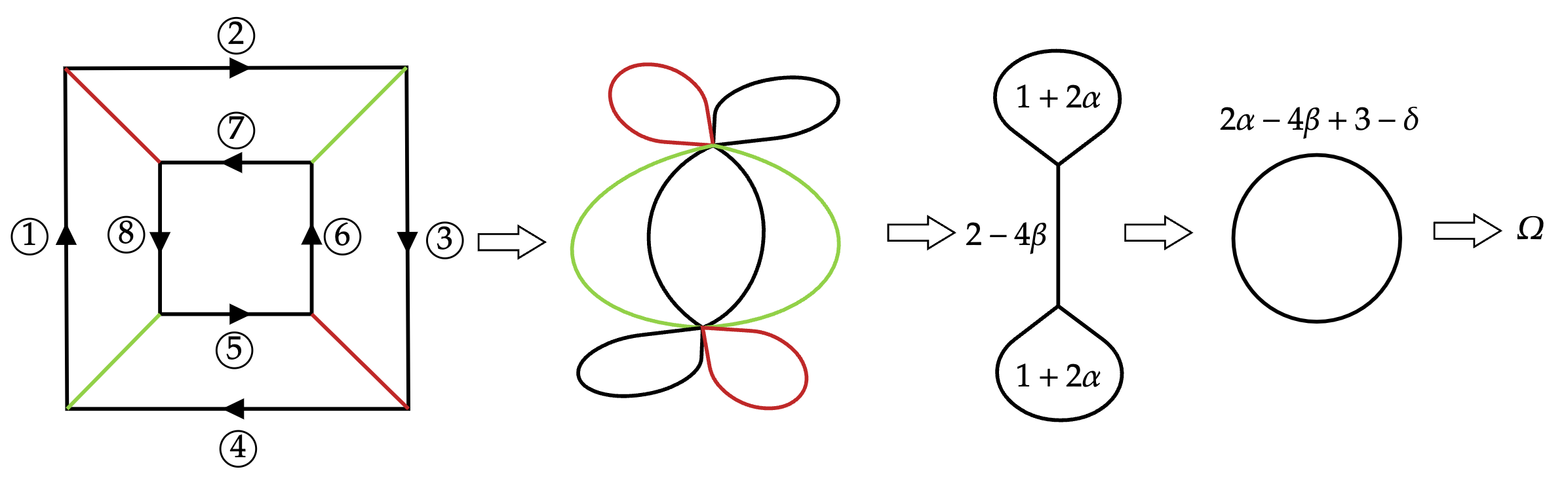}}
\]
18. (14)(23)(56)(78)
\[
  \resizebox{1\columnwidth}{!}{\includegraphics{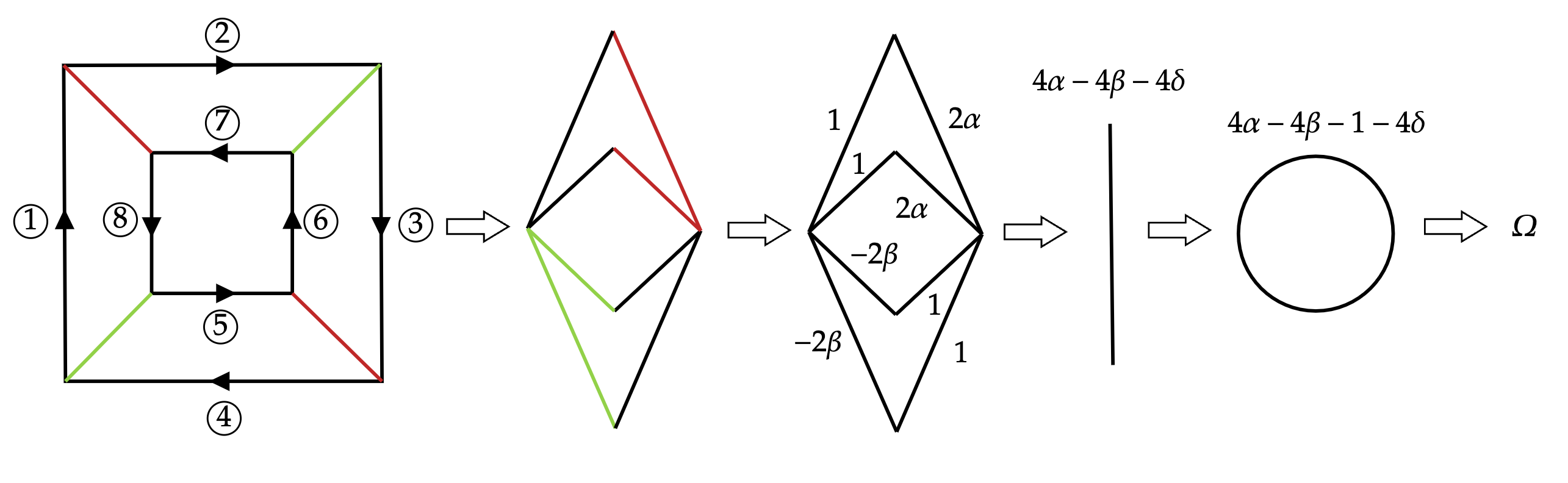}}
\]
19. (14)(23)(58)(67)
\[
  \resizebox{1\columnwidth}{!}{\includegraphics{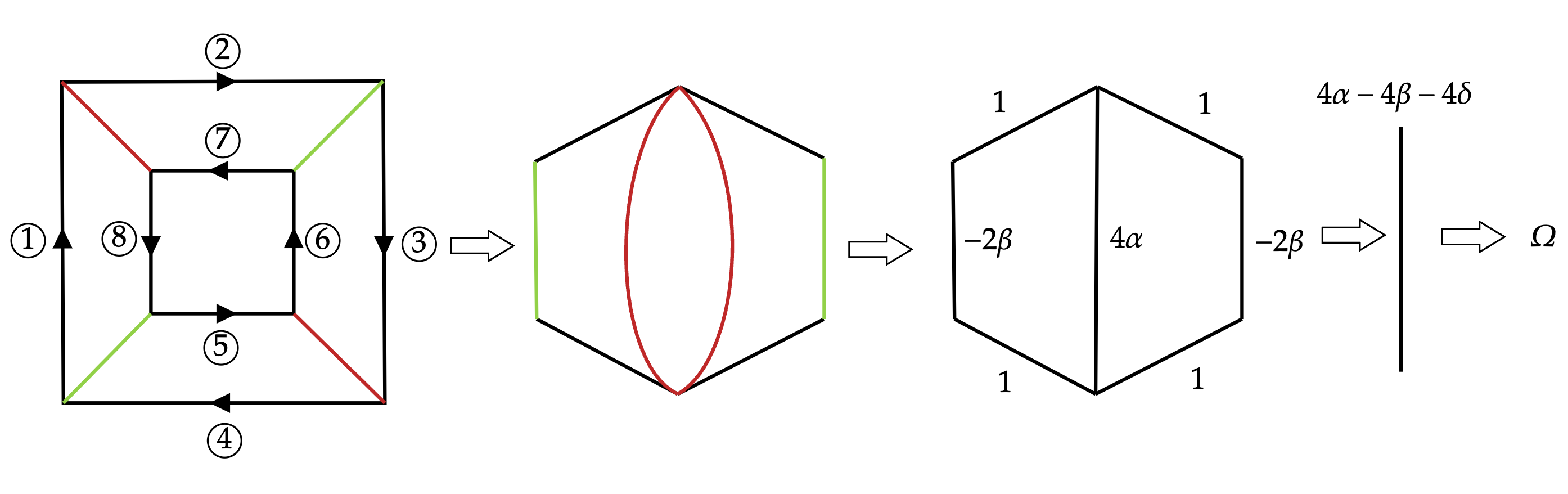}}
\]
20. (14)(25)(38)(67), (16)(23)(47)(58)
\[
  \resizebox{1\columnwidth}{!}{\includegraphics{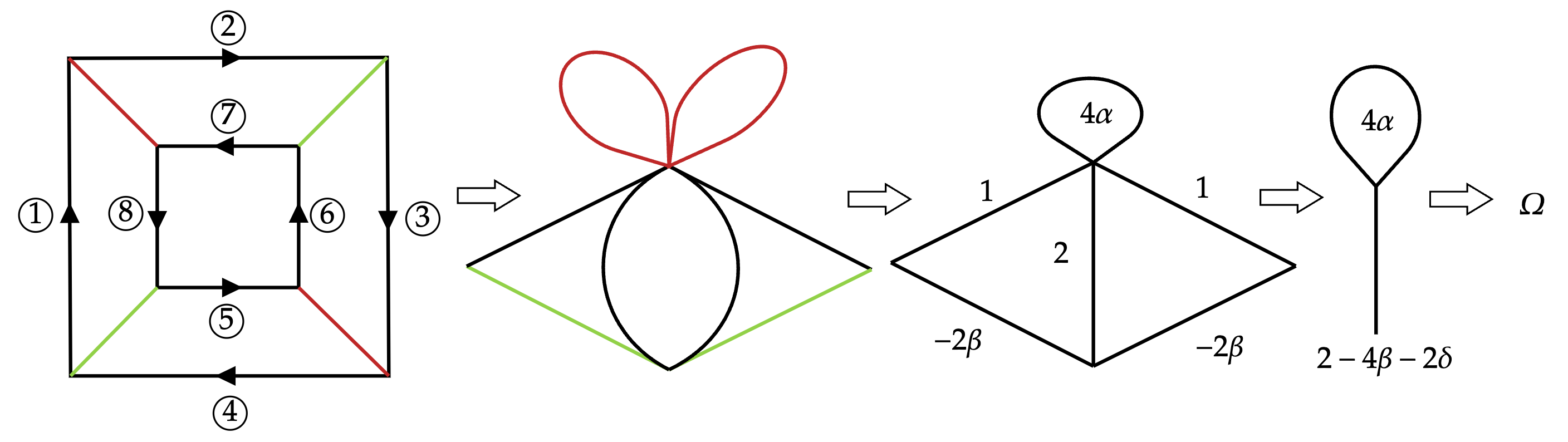}}
\]
21. (14)(26)(37)(58), (15)(23)(48)(67)
\[
  \resizebox{1\columnwidth}{!}{\includegraphics{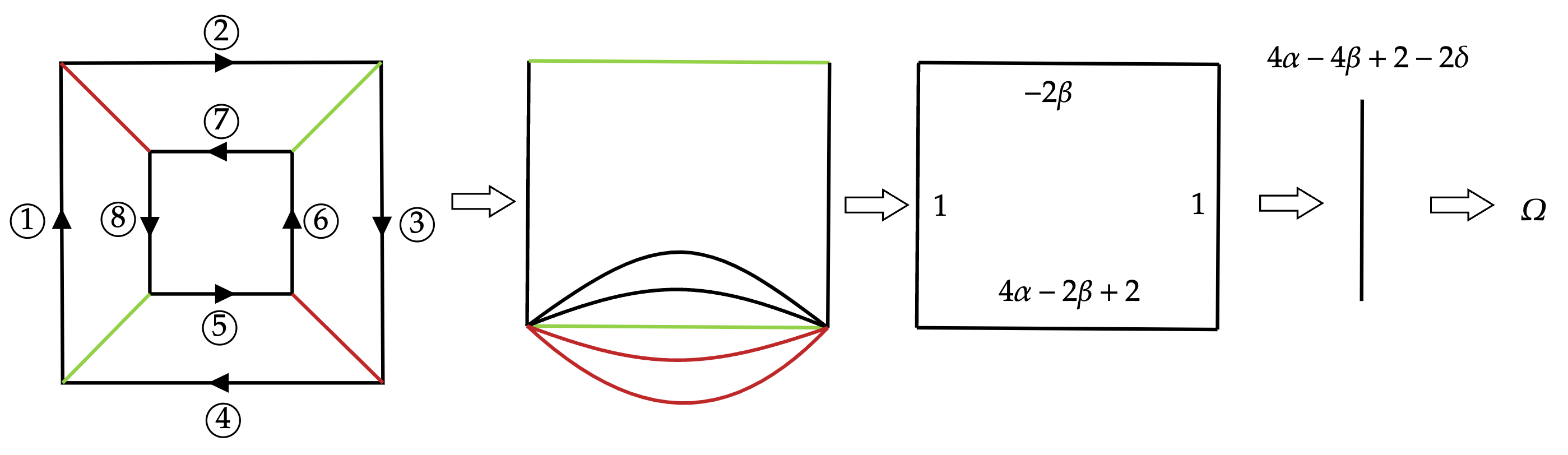}}
\]
22. (14)(27)(36)(58), (18)(23)(45)(67)
\[
  \resizebox{1\columnwidth}{!}{\includegraphics{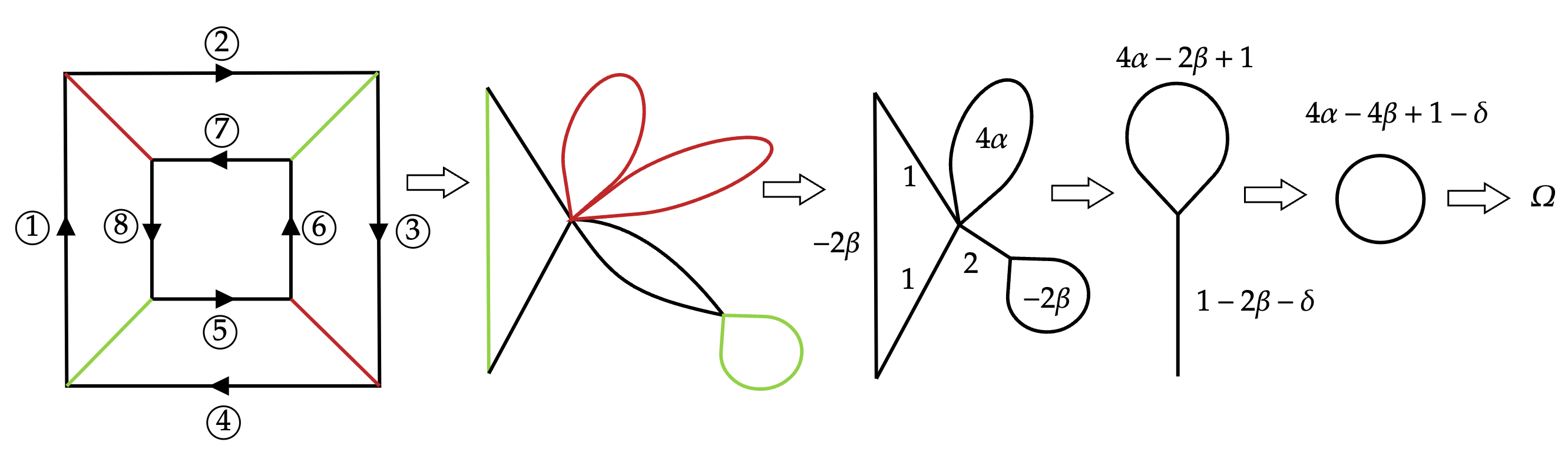}}
\]
23. (14)(28)(35)(67), (17)(23)(46)(58)
\[
  \resizebox{1\columnwidth}{!}{\includegraphics{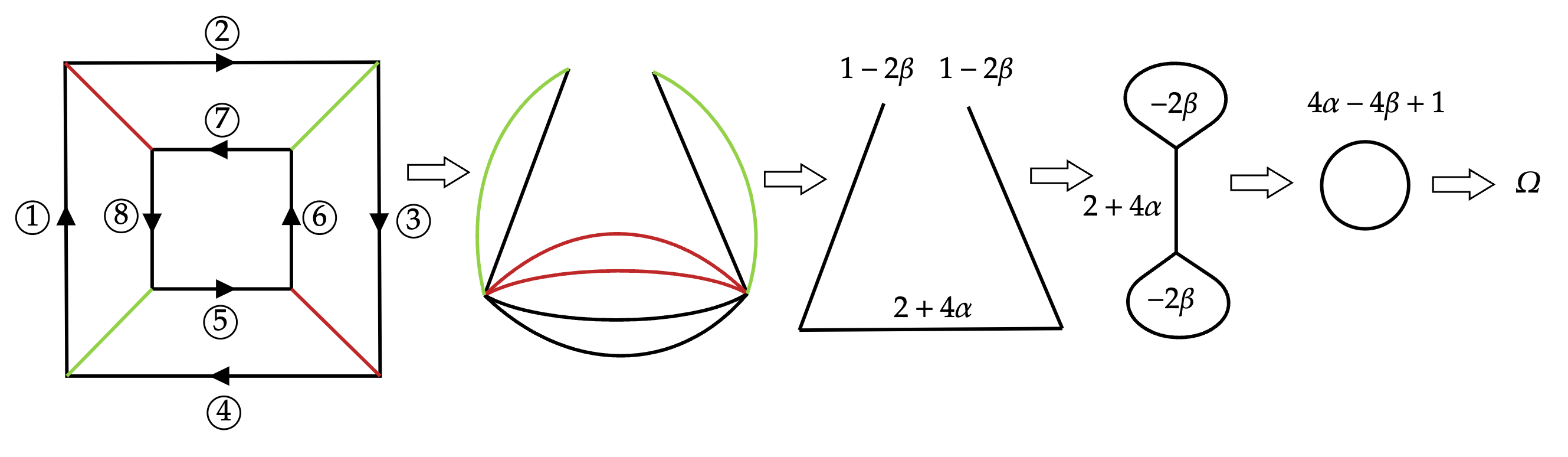}}
\]
24. (15)(26)(37)(48), (17)(28)(35)(46)
\[
  \resizebox{1\columnwidth}{!}{\includegraphics{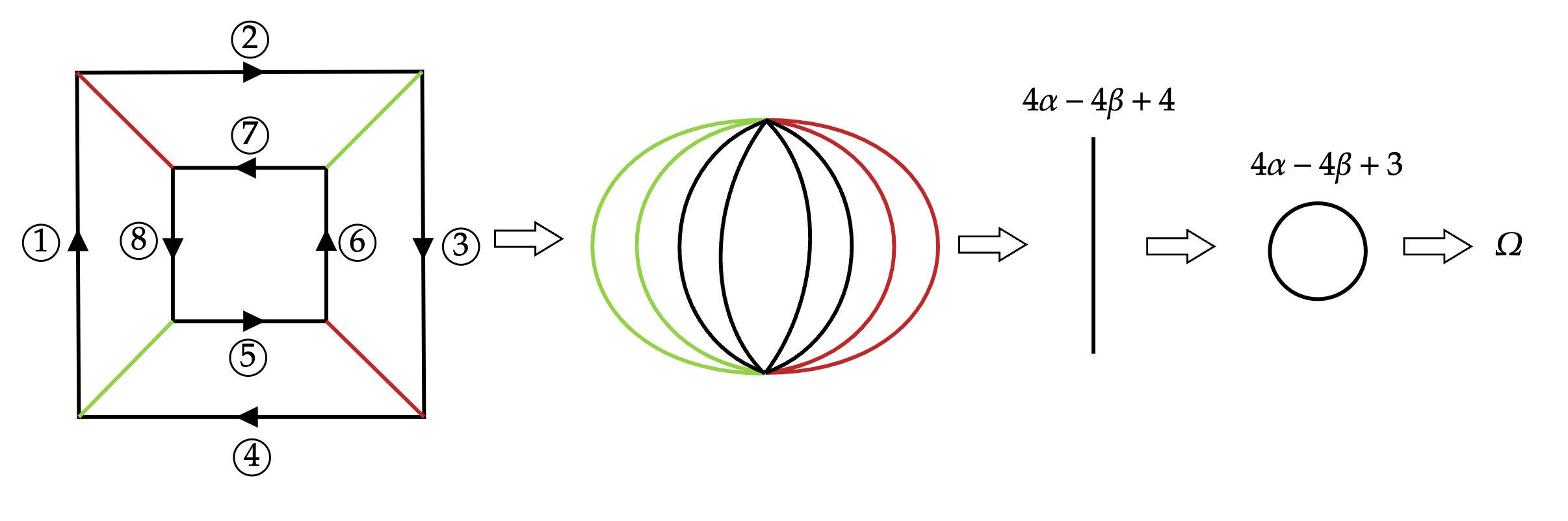}}
\]
25. (15)(26)(38)(47), (16)(25)(37)(48)
\[
  \resizebox{1\columnwidth}{!}{\includegraphics{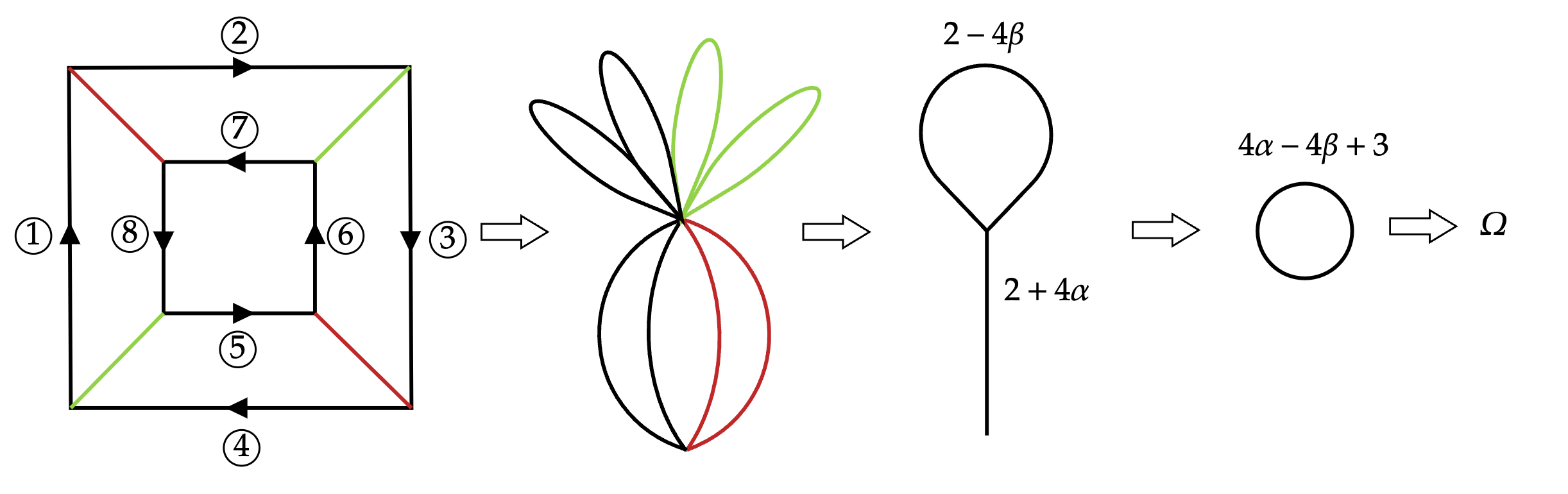}}
\]
26. (15)(27)(36)(48), (18)(26)(37)(45)
\[
  \resizebox{1\columnwidth}{!}{\includegraphics{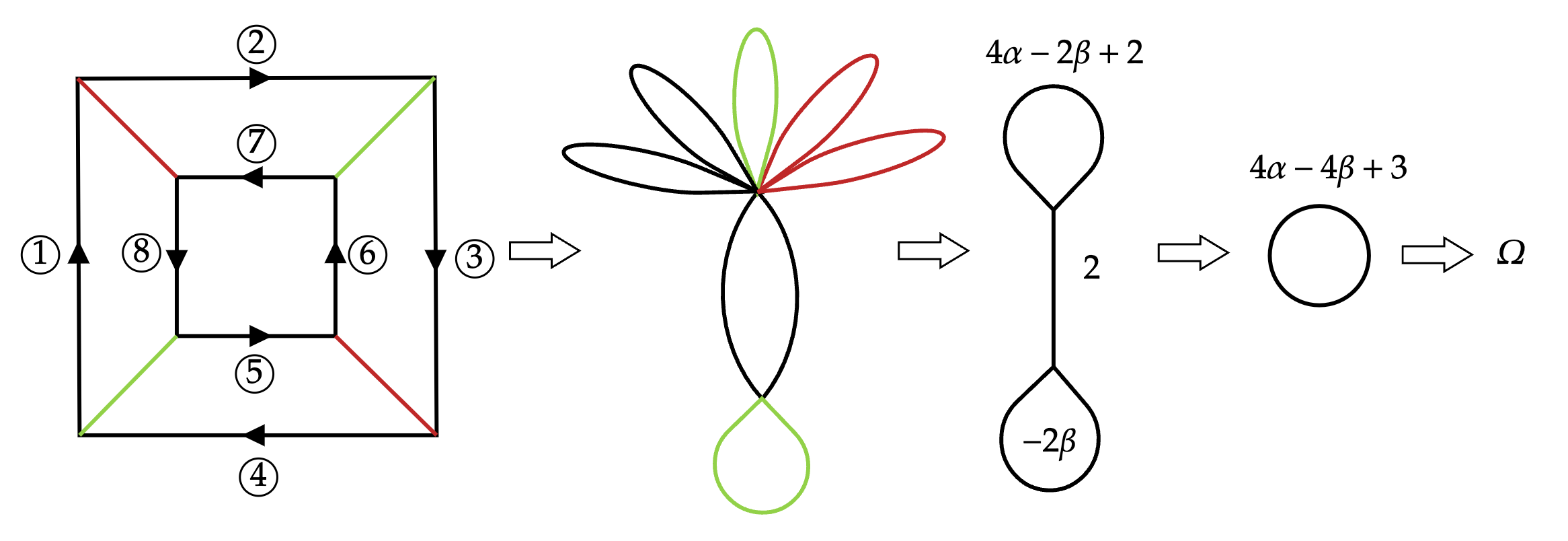}}
\]
27. (15)(27)(38)(46), (15)(28)(36)(47), (16)(27)(35)(48), (16)(28)(37)(45),
(17)(25)(36)(48), (17)(26)(38)(45), (18)(25)(37)(46), (18)(26)(35)(47)
\[
  \resizebox{1\columnwidth}{!}{\includegraphics{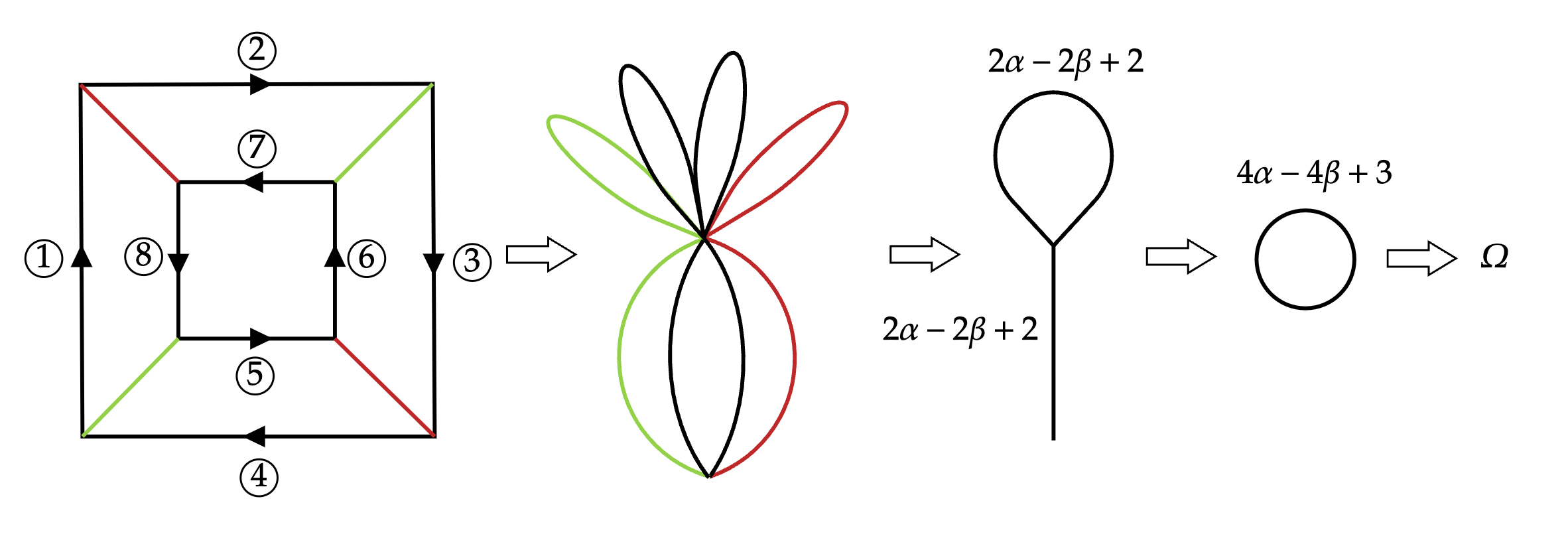}}
\]
28. (15)(28)(37)(46), (17)(26)(35)(48)
\[
  \resizebox{1\columnwidth}{!}{\includegraphics{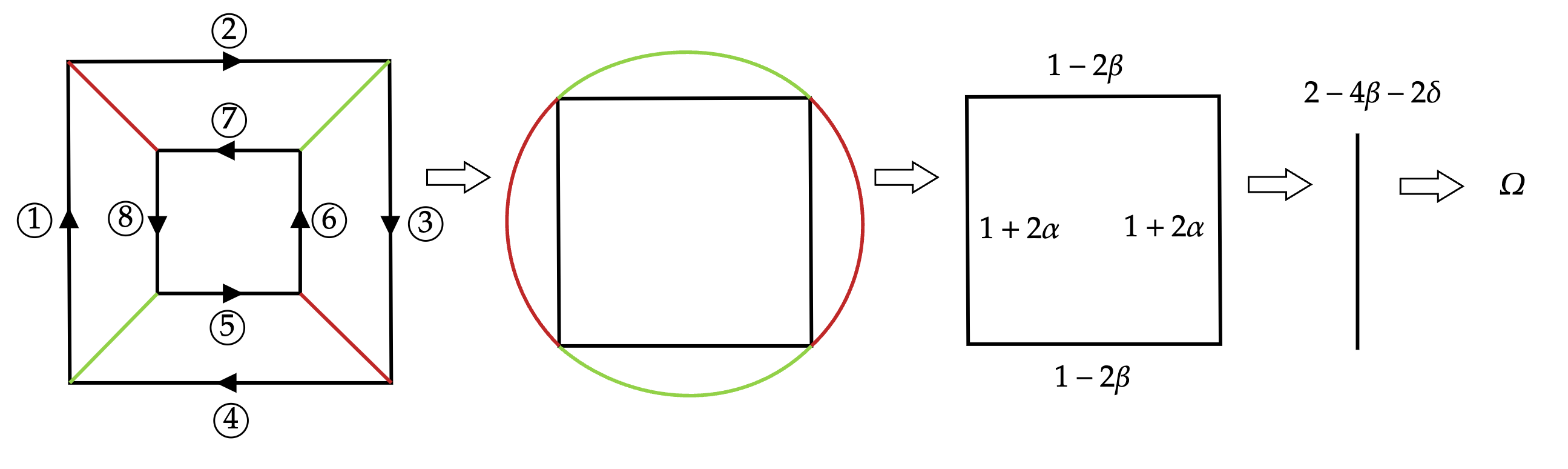}}
\]
29. (16)(24)(38)(57)
\[
  \resizebox{1\columnwidth}{!}{\includegraphics{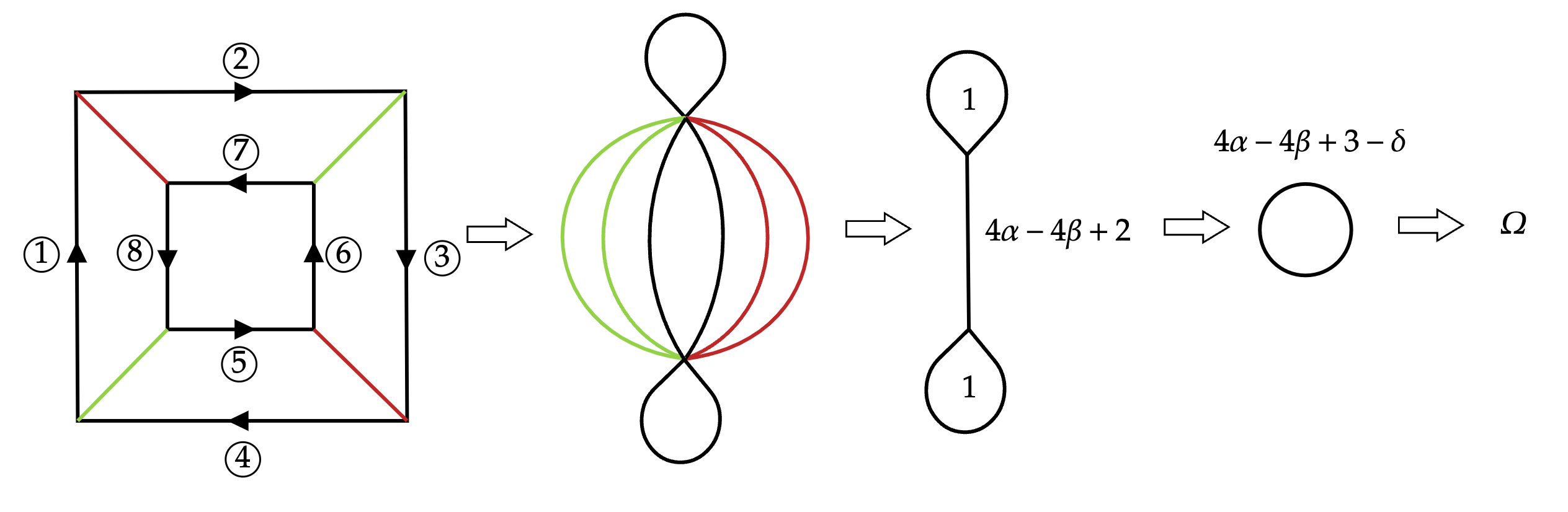}}
\]
30. (16)(25)(38)(47)
\[
  \resizebox{1\columnwidth}{!}{\includegraphics{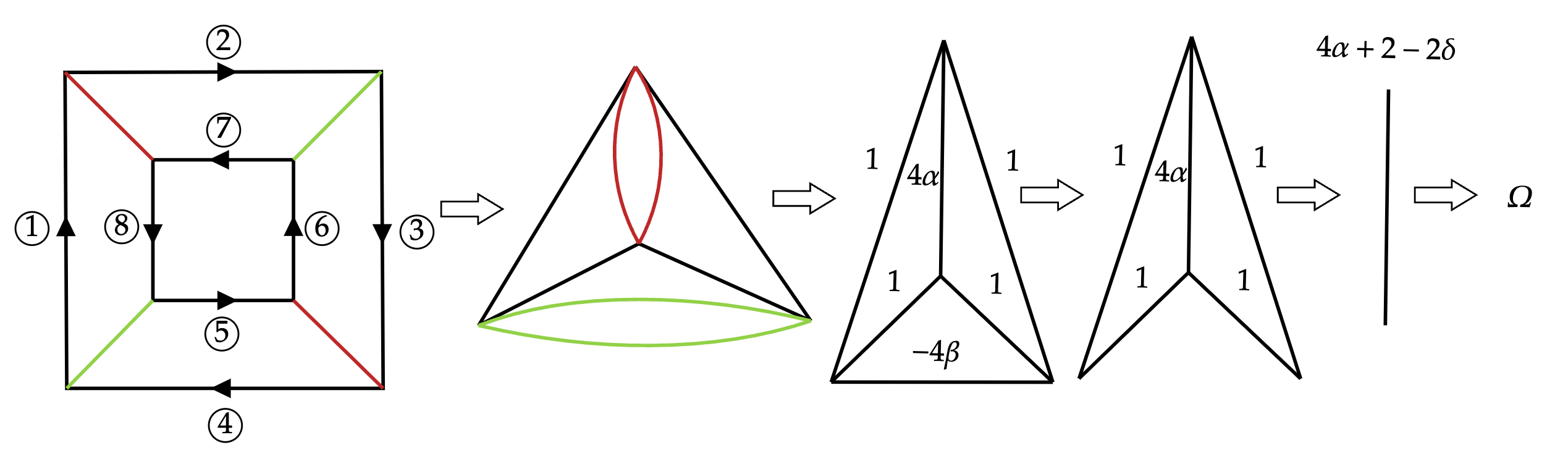}}
\]
In this case, we simply bound the edge with weight $- 4 \beta$ by 1, which is
equivalent to getting rid of this edge on the graph.

31. (16)(27)(38)(45), (18)(25)(36)(47)
\[
  \resizebox{1\columnwidth}{!}{\includegraphics{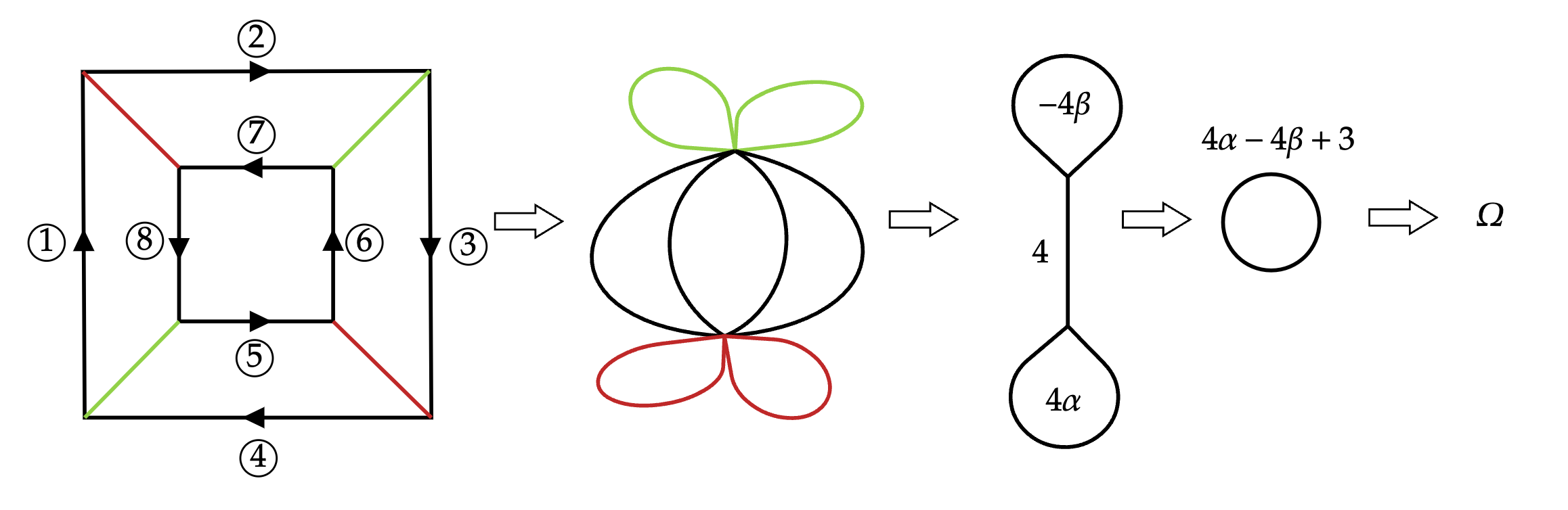}}
\]
32. (16)(28)(35)(47), (17)(25)(38)(46)
\[
  \resizebox{1\columnwidth}{!}{\includegraphics{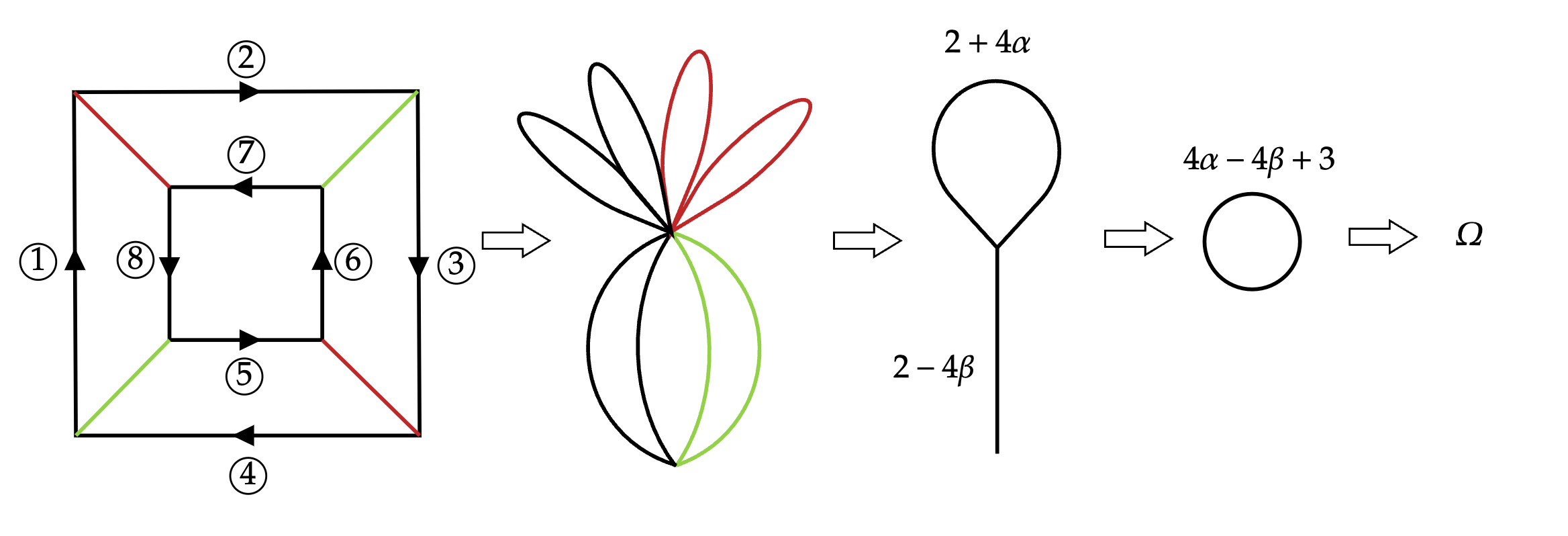}}
\]
33. (17)(28)(36)(45), (18)(27)(35)(46)
\[
  \resizebox{1\columnwidth}{!}{\includegraphics{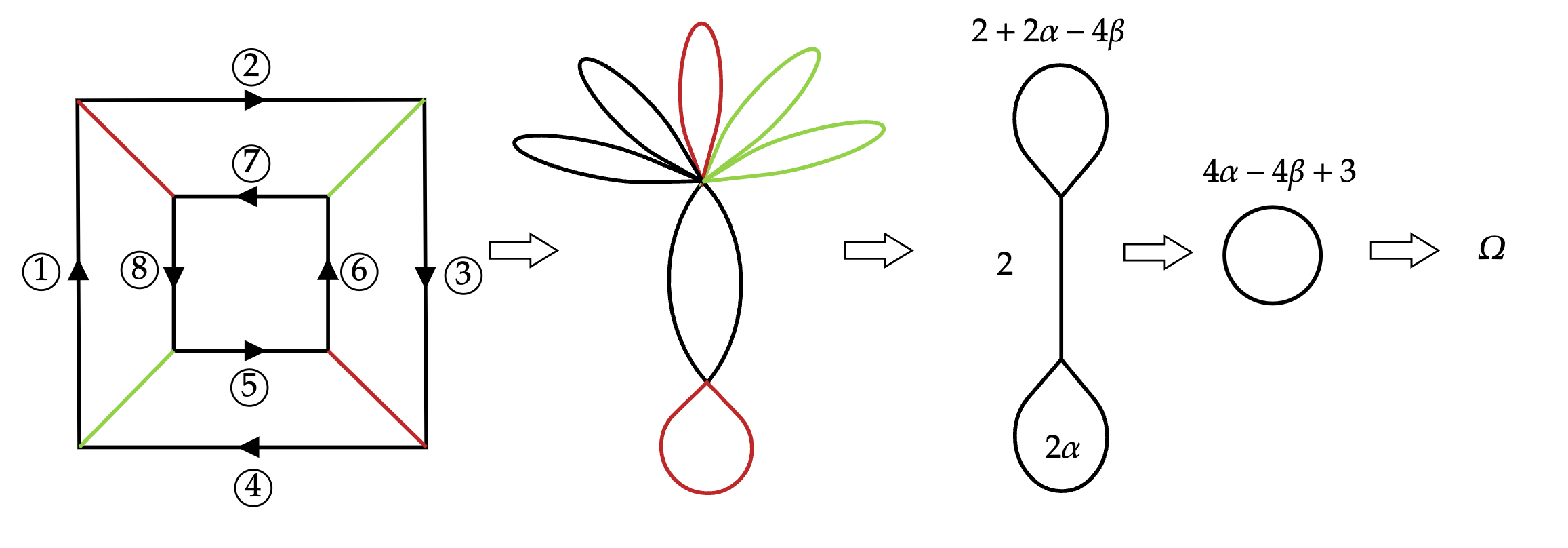}}
\]
34. (18)(27)(36)(45)
\[
  \resizebox{1\columnwidth}{!}{\includegraphics{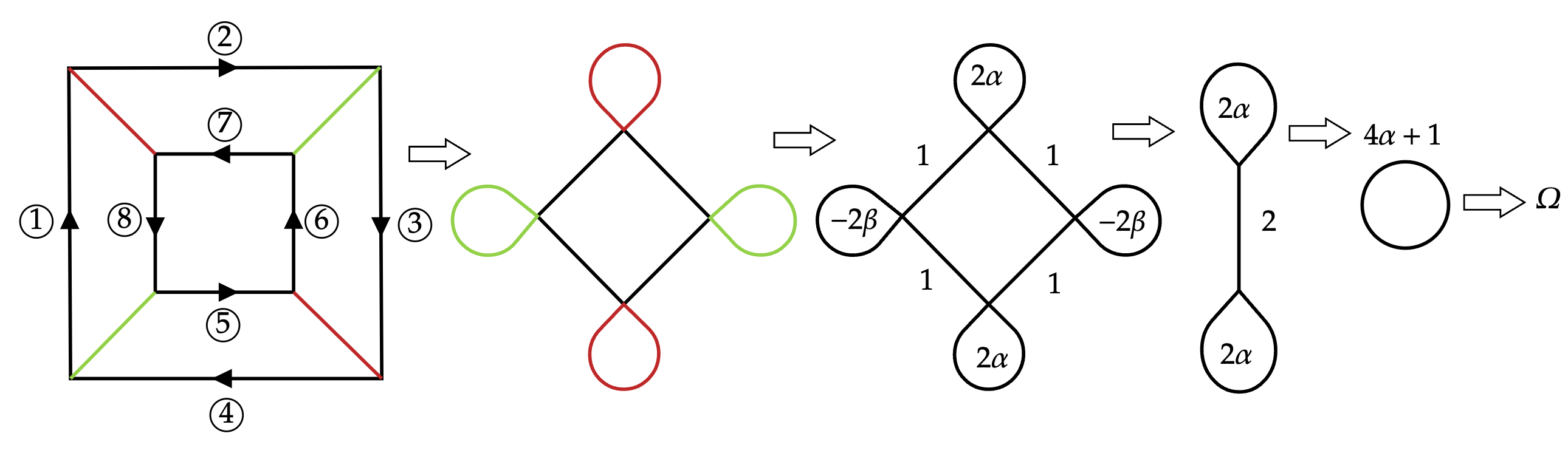}}
\]

\end{document}